\documentclass[twocolumn,preprintnumbers, amssymb,amsmath,aps,floatfix,nofootinbib,superscriptaddress,showpacs]{revtex4-1}

\usepackage{epsfig}
\usepackage{bm}
\usepackage{amssymb}
\usepackage{amsmath}
\usepackage{color}
\usepackage{subfigure}
\usepackage[colorlinks,
            linkcolor=blue,
            anchorcolor=black,
            citecolor=blue
            ]{hyperref}

\newcommand\lperp{\ensuremath{l_\perp}}

\newcommand\kperp{\ensuremath{k_\perp}}
\newcommand\pperp{\ensuremath{p_T}}
\newcommand\qperp{\ensuremath{q_\perp}}
\newcommand\uperp{\ensuremath{u_\perp}}

\newcommand\vperp{\ensuremath{v_\perp}}

\newcommand\bperp{\ensuremath{b_\perp}}
\newcommand\xperp{\ensuremath{x_\perp}}
\newcommand\yperp{\ensuremath{y_\perp}}

\newcommand\rperp{\ensuremath{r_\perp}}
\newcommand\rbperp{\ensuremath{r_{1\perp}}}
\newcommand\rcperp{\ensuremath{r_{2\perp}}}
\newcommand\Rperp{\ensuremath{R_\perp}}
\newcommand\Sperp{\ensuremath{S_\perp}}
\newcommand\alphas{\ensuremath{\alpha_s}}

\newcommand\qa{\ensuremath{ q_{1\perp}}}
\newcommand\qb{\ensuremath{ q_{2\perp}}}
\newcommand\qc{\ensuremath{ q_{3\perp}}}

\newcommand\dd{\ensuremath{\mathrm{d}}}

\newcommand\fcal{\mathcal{F}}

\allowdisplaybreaks[4]

\begin{document}

\title{Pursuing the Precision Study for Color Glass Condensate\\ in Forward Hadron Productions}
\author{Yu Shi} \email{yu.shi@sdu.edu.cn} 
\affiliation{Key Laboratory of Particle Physics and Particle Irradiation (MOE), Institute of frontier and interdisciplinary science, Shandong University, Qingdao, Shandong 266237, China}
\affiliation{Key Laboratory of Quark and Lepton Physics (MOE) and Institute of Particle Physics, Central China Normal University, Wuhan 430079, China}

\author{Lei Wang} \email{leiwang@mails.ccnu.edu.cn}
\affiliation{Key Laboratory of Quark and Lepton Physics (MOE) and Institute of Particle Physics, Central China Normal University, Wuhan 430079, China}

\author{Shu-Yi Wei}  \email{shuyi@sdu.edu.cn}
\affiliation{Key Laboratory of Particle Physics and Particle Irradiation (MOE), Institute of frontier and interdisciplinary science, Shandong University, Qingdao, Shandong 266237, China}
\affiliation{European Centre for Theoretical Studies in Nuclear Physics and Related Areas (ECT*)
and Fondazione Bruno Kessler, Strada delle Tabarelle 286, I-38123 Villazzano (TN), Italy}

\author{Bo-Wen Xiao}  \email{xiaobowen@cuhk.edu.cn}
\affiliation{School of Science and Engineering, The Chinese University of Hong Kong, Shenzhen 518172, China}

\begin{abstract}
With the tremendous accomplishments of RHIC and the LHC experiments and the advent of the future Electron-Ion Collider on the horizon, the quest for compelling evidence of the color glass condensate (CGC) has become one of the most aspiring goals in the high energy Quantum Chromodynamics research. Pursuing this question requires developing the precision test of the CGC formalism. By systematically implementing the threshold resummation, we significantly improve the stability of the next-to-leading-order calculation in CGC for forward rapidity hadron productions in $pp$ and $pA$ collisions, especially in the high $p_T$ region, and obtain reliable descriptions of all existing data measured at RHIC and the LHC across all $p_T$ regions. Consequently, this technique can pave the way for the precision studies of the CGC next-to-leading-order predictions by confronting them with a large amount of precise data. 
\end{abstract}

\maketitle

\textit{Introduction} The gluon saturation phenomenon\cite{Gribov:1984tu,Mueller:1985wy,McLerran:1993ni,McLerran:1993ka,arXiv:1002.0333, CU-TP-441a}, predicted by the small-$x$ framework, which is also known as the color glass condensate (CGC) formalism, has been an intriguing forefront research topic. A lot of experimental and theoretical research efforts around the globe have been devoted to this cutting-edge research frontier. Besides, in the upcoming era of the Electron-Ion Collider (EIC)\cite{Boer:2011fh, Accardi:2012qut, Proceedings:2020eah, AbdulKhalek:2021gbh}, probing the emergent properties of ultra-dense gluon has become one of the key fundamental questions that the EIC sets out to address. 

CGC is an effective formalism in Quantum Chromodynamics (QCD) which describes the novel non-linear dynamics of low-momentum gluons inside a hadron. These low momentum gluon degrees of freedom are generally referred to as the small-$x$ gluons, with $x$ being the longitudinal momentum fraction. First, color sources such as large-$x$ quarks and gluons inside fast-moving hadrons emit a large number of small-$x$ gluons\cite{Kuraev:1977fs,Balitsky:1978ic}. In the meantime, when its occupation number inside the hadron becomes sufficiently large, small-$x$ gluons start to overlap, recombine and then compress each other, eventually saturate. Usually, we introduce the saturation momentum $Q_s(x)$ at given $x$ to characterize the typical size of soft gluons. Due to the rise of the gluon density, $Q_s(x)$ increases at low-$x$ so that the corresponding gluon size becomes smaller in the transverse space and more gluons can fit into a confined transverse region. This non-linear dynamics can be captured by the evolution equation known as the BK-JIMWLK equation\cite{Balitsky:1995ub, Kovchegov:1999yj, JalilianMarian:1997jx, JalilianMarian:1997gr, Iancu:2000hn, Ferreiro:2001qy}.

In high-energy collisions, small-$x$ gluon degrees of freedom are unlocked and measured in terms of final state hadrons. To search for the experimental evidence of gluon saturation among existing data\cite{Arsene:2004ux, Adams:2006uz, Braidot:2010ig, Adare:2011sc,ALICE:2012xs, ALICE:2012mj, Hadjidakis:2011zz,ATLAS:2016xpn,LHCb:2021abm,LHCb:2021vww} and prepare for the future EIC precision studies, it is important to develop next-to-leading order (NLO) computations in the CGC formalism and achieve an accurate description of data collected from various kinematic regions. 

Among many different physical processes studied at RHIC and the LHC, the calculation and measurements of the single forward hadron production in proton-nucleus collisions (or deuteron-nucleus collisions at RHIC), $p \,(d)\,+A\to h( y \, , p_T) +X$, have attracted a great deal of attention\cite{Kovchegov:1998bi, Dumitru:2002qt, Dumitru:2005kb, Albacete:2010bs, Levin:2010dw, Fujii:2011fh, Albacete:2012xq, Albacete:2013ei, Altinoluk:2011qy, Chirilli:2011km, Chirilli:2012jd, Stasto:2013cha, Lappi:2013zma, vanHameren:2014lna, Stasto:2014sea, Altinoluk:2014eka, Watanabe:2015tja, Stasto:2016wrf, Iancu:2016vyg, Ducloue:2016shw, Ducloue:2017dit}. In the forward region, the projectile proton (or deuteron) can be viewed as a relatively dilute object that probes the ultra-dense gluon fields in the nuclear target\cite{Dumitru:2002qt, Chirilli:2011km, Chirilli:2012jd, Dominguez:2010xd, Dominguez:2011wm}. Experimentally, the evolution of the nuclear modification factor $R_{dAu}$ \cite{Arsene:2004ux, Adams:2006uz} from mid-rapidity to forward-rapidity regions is considered the evidence\cite{Kharzeev:2003wz, Kharzeev:2004yx, Albacete:2003iq, Iancu:2004bx, Albacete:2013ei} for the onset of gluon saturation. The measured $R_{dAu}$ is computed from the hadron spectra in deuteron $+$ gold collisions normalized by the spectra in $pp$ collisions times the number of binary collisions, and $R_{dAu}$ in forward rapidity regions is found to be suppressed. The physical interpretation is that the non-linear gluon dynamics can significantly reduce the gluon density at small-$x$ in nuclei compared to the proton (or deuteron) baseline.

In terms of the perturbative expansion, the corresponding cross-section can be schematically cast into
\begin{eqnarray}
&&\frac{\dd\sigma}{\dd y\dd^2p_T} =\int x_pf_a(x_p)\otimes D_a(z) \otimes \mathcal{F}^{x_g}_a(k_\perp) \otimes \mathcal{H}^{(0)} \notag \\
&& \quad \quad+\frac{\alpha_s }{2\pi} \sum_{a,b=q, g}\int xf_a(x)\otimes D_b(z) \otimes \mathcal{F}_{ab}^{x_g}\otimes \mathcal{H}_{ab}^{(1)}, \,\, \label{NLO1}
\end{eqnarray}
where the first term stands for the leading order (LO) contribution first computed in Refs.~\cite{Dumitru:2002qt, Dumitru:2005gt} and the second term represents the NLO corrections derived from one-loop diagrams. In our framework, the full NLO contribution includes the contributions computed in Ref.~\cite{Chirilli:2011km,Chirilli:2012jd} and the additional kinematic constraint corrections given in Ref.~\cite{Watanabe:2015tja}. The kinematic variables are defined as follows, $x_p=\frac{k_\perp}{\sqrt{s}}e^y$, $x_g=\frac{k_\perp}{\sqrt{s}}e^{-y}$, $p_T=zk_\perp$ with $k_\perp$ and $z$ being the parton transverse momentum and the longitudinal momentum fraction of produced hadron w.r.t. its original parton, respectively.

The LO production in various channels and the contribution together with running coupling effects have been calculated extensively in Refs.~\cite{Blaizot:2004wu, Blaizot:2004wv, Albacete:2010bs, Levin:2010dw, Fujii:2011fh, Albacete:2012xq, Lappi:2013zma, vanHameren:2014lna, Bury:2017xwd}, and part of the NLO contributions are studied in Refs~\cite{Dumitru:2005gt, Altinoluk:2011qy}. To obtain the full analytical expressions of the NLO corrections, one needs to evaluate all of the real and virtual one-loop diagrams and remove various types of divergences, as demonstrated in Ref.~\cite{Chirilli:2011km, Chirilli:2012jd}. First, we subtract the so-called rapidity divergences and absorb them into the evolution of the dipole scattering amplitude associated with the dipole gluon distribution $\mathcal{F}^{x_g}(k_\perp)$. This procedure reproduces the well-known BK equation\cite{Balitsky:1995ub,Kovchegov:1999yj} and allows us to resum the small-$x$ large logarithms systematically. Second, one can gather all the residual collinear divergences and remove them through the redefinition of collinear parton distribution functions (PDFs) $xf(x)$ or/and fragmentation functions (FFs) $D(z)$. Eventually, the resulting finite NLO corrections, which are simplified in the large $N_c$ limit and denoted as $\mathcal{H}_{ab}^{(1)}$ in Eq.~(\ref{NLO1}), can be numerically evaluated.

The direct evaluation of the complete NLO cross-section yields a good agreement with experimental data\cite{Arsene:2004ux, Adams:2006uz} from RHIC for forward rapidity hadron production in the low-$p_T$ region. However, the NLO result drastically turns negative in the high $p_T$ region\cite{Stasto:2013cha}. When the kinematic constraint corrections are included~\cite{Watanabe:2015tja}, the negative NLO cross-section issue can be slightly mitigated but not entirely resolved. In usual perturbative QCD calculations in the collinear factorization, similar issues occur as well for various processes. It indicates that large (and mostly negative) logarithms hidden in $\mathcal{H}_{ab}^{(1)}$ become important in the high $p_T$ region. In particular, in our case with the forward rapidity hadron production, the threshold logarithms cause the breakdown of the perturbative expansion, and they should be resummed in order to restore the predictive power of our calculation in the region of interest. 

The quest for positivity in this NLO CGC calculation has sparked a lot of interest. Over the last seven years, there have been a lot of studies\cite{Altinoluk:2014eka, Kang:2014lha, Stasto:2014sea, Watanabe:2015tja, Stasto:2016wrf, Ducloue:2016shw, Iancu:2016vyg, Ducloue:2017mpb, Ducloue:2017dit, Xiao:2018zxf, Liu:2019iml, Kang:2019ysm, Liu:2020mpy} dedicated to addressing the issue caused by the large negative NLO corrections. We believe that the threshold resummation is one of the feasible solutions to this issue, and the resummation technique developed in this work can also be useful in the study of other NLO calculations\cite{Mueller:2012bn, Hentschinski:2014esa, Benic:2016uku, Boussarie:2016bkq, Ducloue:2017ftk, Roy:2018jxq, Roy:2019cux, Roy:2019hwr, Roy:2019hwr, Iancu:2020mos, Caucal:2021ent} in CGC. In addition, there have been some further theoretical efforts\cite{Altinoluk:2014oxa, Altinoluk:2015gia, Chirilli:2018kkw, Altinoluk:2020oyd}, which go beyond the eikonal approximation and compute the next-to-eikonal corrections for this process. 

This paper is organized as follows. The following two sections are devoted to implementing the threshold resummation in the CGC framework for forward hadron productions and the corresponding numerical results, respectively. In the end, the conclusion and outlook are provided in Sec.~4. Finally, all the technical details are attached as the appendix. 

\textit{2. Implementation of the threshold resummation} To tackle the issue of the large negative corrections at NLO, we need to analytically extend the applicability of the NLO CGC calculation from the low-$p_T$ region to the high-$p_T$ region, thus obtain reliable numerical predictions for measurements at both RHIC and the LHC, and therefore better understand the transition from the ultra-dense regime to the dilute regime. First, to illustrate the origin of the threshold logarithms in the NLO corrections, let us discuss the appearance of the large NLO corrections that cause the issue in the sufficiently forward rapidity region when $p_T \gg Q_s$\cite{Stasto:2013cha, Watanabe:2015tja}. In fact, this indicates that the issue occurs when hard scatterings dominate in this region, where the corresponding events are approaching the kinematic threshold. Second, we identify and extract the large logarithms in the momentum space where the numerical computation of the NLO correction can be performed more efficiently. In the end, we introduce the resummation scheme, which allows us to take the higher-order large logarithms into account and restore the predictive power of the one-loop calculation for this process in the CGC framework. 

To see this clearly and intuitively, let us recall the kinematics at NLO\cite{Chirilli:2011km,Chirilli:2012jd} and define the hadron longitudinal momentum fraction $\tau=\frac{p_T}{\sqrt{s}}e^{y} $, which is equivalent to $\tau =x\xi z $ with $\xi$ being the remaining momentum fraction of a parton after emitting one gluon. In the forward rapidity region ($y>0$), as the hadron $p_T$ increases, $\tau$ starts to approach $1$. That is to say that we are approaching the threshold region where $x$, $z$, and $\xi$ are all forced to approach $1$. In this case, the phase space for the real gluon emission is severely limited since there is not much longitudinal momentum left for the radiation near the threshold. In contrast, there is no constraint imposed on the virtual graphs. As a result, after canceling singularities between real and virtual graphs, large logarithms appear in the NLO corrections. These large threshold logarithms are the culprits that upset the convergence of the $\alpha_s$ expansion in our NLO calculation. Two formulations of the threshold resummation within the CGC framework have been proposed earlier in Ref.~\cite{Xiao:2018zxf} and Refs.~\cite{Kang:2019ysm, Liu:2020mpy}, respectively. In this paper, our study follows closely with the former approach. 

Furthermore, let us describe the strategy used in our calculation to extract the above-mentioned threshold logarithms explicitly. Initially, the NLO corrections\cite{Chirilli:2011km, Chirilli:2012jd} were derived in the coordinate space where the physics interpretation for gluon saturation is manifest. However, due to the oscillating behavior of the complex phase factor in the coordinate space expression, it is challenging to evaluate them numerically, especially in the high $p_T$ region. Therefore, we later transform the complete NLO cross-sections, including the kinematic constraint corrections\cite{Watanabe:2015tja} into the momentum space, yielding much better numerical accuracy. In the coordinate space, we can identify two types of logarithms\cite{Sun:2013hua, Xiao:2018zxf} 
\begin{eqnarray}
\textrm{single log: } \ln \frac{k_\perp^2}{\mu_r^2}, \,\, \ln \frac{\mu^2}{\mu_r^2}; \,\quad  \textrm{double log: } \ln^2 \frac{k_\perp^2}{\mu_r^2},
\end{eqnarray}
where $\mu_r\equiv c_0/r_\perp$ with $r_\perp$ being the dipole size and $c_0=2e^{-\gamma_E}$. After integrating over $r_\perp$ in the coordinate space, these logarithms generate large contributions in the threshold region when $k_\perp$ (or $p_T$) becomes much larger than typical value of $\mu_r$. Therefore, in the momentum space, we need to introduce an auxiliary semi-hard scale $\Lambda$, much larger than the QCD scale $\Lambda_{\rm QCD}$, to extract these large logarithms for the resummation purpose. In the momentum space, the single and double logarithmic terms can be correspondingly cast into\cite{Mueller:2013wwa,Sun:2014gfa,Watanabe:2015tja}
\begin{eqnarray}
&&\textrm{single log: } \ln \frac{k_\perp^2}{\Lambda^2}+I_1(\Lambda) \quad \textrm{and} \quad \ln \frac{\mu^2}{\Lambda^2}+ I_1(\Lambda), \\
&&\textrm{double log: } \ln^2 \frac{k_\perp^2}{\Lambda^2} +I_2(\Lambda),
\end{eqnarray}
where $I_{1,2} (\Lambda)$ represent the residual matching functions. At one-loop order, our results are independent of the choice of the auxiliary scale $\Lambda$. The essential steps of the derivations can be found in the supplemental material. 

Usually, in the collinear factorization, the threshold logarithms are resummed in terms of the resummation of the ``plus'' distributions in the Mellin moment space\cite{Sterman:1986aj,Catani:1989ne, Catani:1996yz, deFlorian:2008wt}. The technique employed in the CGC framework is slightly different since the relevant gluon distribution is transverse momentum dependent. The threshold logarithms in forward hadron productions can be cast into two parts: the soft and the collinear parts. The soft part such as single and double logs of $\ln \frac{k_\perp^2}{\Lambda^2}$, associated with the soft gluon emission, can be resummed by the Sudakov factor $S_{\text{Sud}} ( k_\perp, \Lambda)$. As to the collinear part ($\ln \frac{\mu^2}{\Lambda^2}$), there are two similar approaches to deal with the corresponding resummation. The first method is to develop a renormalization group equation (whose solution is $\Delta (\Lambda^2, \mu^2, \omega\equiv \ln 1/\xi)$) in the momentum space to analytically resum logarithms of $\ln \frac{\mu^2}{\Lambda^2}$ combined with the above soft part in the threshold limit with $\xi \to 1$. This scheme is akin to the method first developed in the pioneering study\cite{Bosch:2004th,Becher:2006qw, Becher:2006nr,Becher:2006mr} for the deep-inelastic structure function using the soft-collinear effective theory.

Alternatively, since the above collinear logarithms are associated with the DGLAP splitting functions, they can be resummed with the help of the DGLAP evolution of the PDFs and FFs by resetting the factorization scale\cite{Xiao:2018zxf} from $\mu$ to the auxiliary scale $\Lambda$ in the LO resummed terms and then the resummed formula reads
\begin{eqnarray}
\sigma &=&\int xf_a(x,\Lambda)\otimes D_a(z, \Lambda) \otimes \mathcal{F}^{x_g}_a(k_\perp) \otimes \mathcal{H}^{(0)}  \otimes e^{-S_{\text{Sud}}}  \notag \\
&+&\frac{\alpha_s }{2\pi}   \sum_{a,b=q, g} \int xf_a(x)\otimes D_b(z) \otimes \mathcal{F}_{ab}^{x_g}\otimes \mathcal{H}_{ab}^{(1) } (\mu, \Lambda).
 \,\, \label{NLO2}
\end{eqnarray}
In fact, this choice of the factorization scale $\mu$ for LO cross-section is similar to the conventional practice of setting $\mu =\mu_b$ in the Collins-Soper-Sterman formalism\cite{Collins:1984kg}. These two resummation schemes are theoretically equivalent, and they yield similar numerical results. The resummation scheme is not unique, and one can certainly develop a similar scheme in the coordinate space as well.

Let us compare the resummed formulas as in Eq (\ref{NLO2}) to the original NLO results in Eq.~(\ref{NLO1}). Essentially, we take out the logarithmic term hidden in $ \mathcal{H}_{ab}^{(1)}$ from Eq.~(\ref{NLO1}), and extract the threshold logarithms which are resummed in Eq~(\ref{NLO2}). Then, the terms that are proportional to the residual matching functions $I_{1,2} (\Lambda)$ are put back into the new NLO coefficient $ \mathcal{H}_{ab}^{(1) } (\mu, \Lambda)$. Initially, Eq.~(\ref{NLO1}) only depends on the factorization scale $\mu$. After the implementation of the threshold resummation, Eq~(\ref{NLO2}) now depends on the choice of the factorization scale $\mu$ and the auxiliary  scale $\Lambda$. Both scale dependences cancel to the one-loop order (NLO), and the residual scale dependences, which start from the two-loop order in this process, are due to the truncation of the perturbative expansions. 

In principle, the cross-section would be independent of both $\mu$ and $\Lambda$ if all-order results were included. In practice, we can estimate the size of higher-order corrections by varying these two scales. Furthermore, to minimize the higher-order corrections, the ``natural" choice of these two scales should be adopted. In the collinear part, the hard scale $Q$ ($\sim 2k_\perp$ when $k_\perp$ is sufficiently large) sets the scale for the factorization scale $\mu$. As to the semi-hard auxiliary scale $\Lambda$, the ``natural" choice should be $\mu_r=c_0/r_\perp$, which depends on the typical value of $r_\perp$ when $r_\perp$ is integrated over. Following Refs.~\cite{Collins:1984kg, Parisi:1979se, Qiu:2000ga}, we use the saddle point approximation to locate the dominant region of the $r_\perp$ integral, thus estimate the physical value of $\Lambda$ via the running coupling prescription
\begin{equation}
\Lambda^2 \approx \textrm{max}\left\{ \Lambda_{\rm QCD}^2\left[\frac{(1-\xi)k_\perp^2}{\Lambda_{\rm QCD}^2}\right]^{\frac{C_R}{C_R+N_c \beta_0}}, Q_s^2\right\}, \label{lambdavalue}
\end{equation}
where $\beta_0 = \frac{11}{12} - \frac{n_f}{6 N_c}$. $C_R$ is the Casimir factor, which gives $C_F$ and $C_A$ for the quark and gluon channel, respectively. In the gluon channel, the saturation momentum $Q_s^2$ is increased by a factor of $N_c/C_F$ as compared to the quark channel. We set $\Lambda^2 =Q_s^2$ when the saturation effect is strong, while $(1-\xi)k_\perp^2 \sim (1-\tau)p_T^2$ becomes the dynamical scale near the threshold region\cite{Becher:2006nr,Becher:2006mr}.

\begin{figure}[!h]
\includegraphics[width=0.40\textwidth]{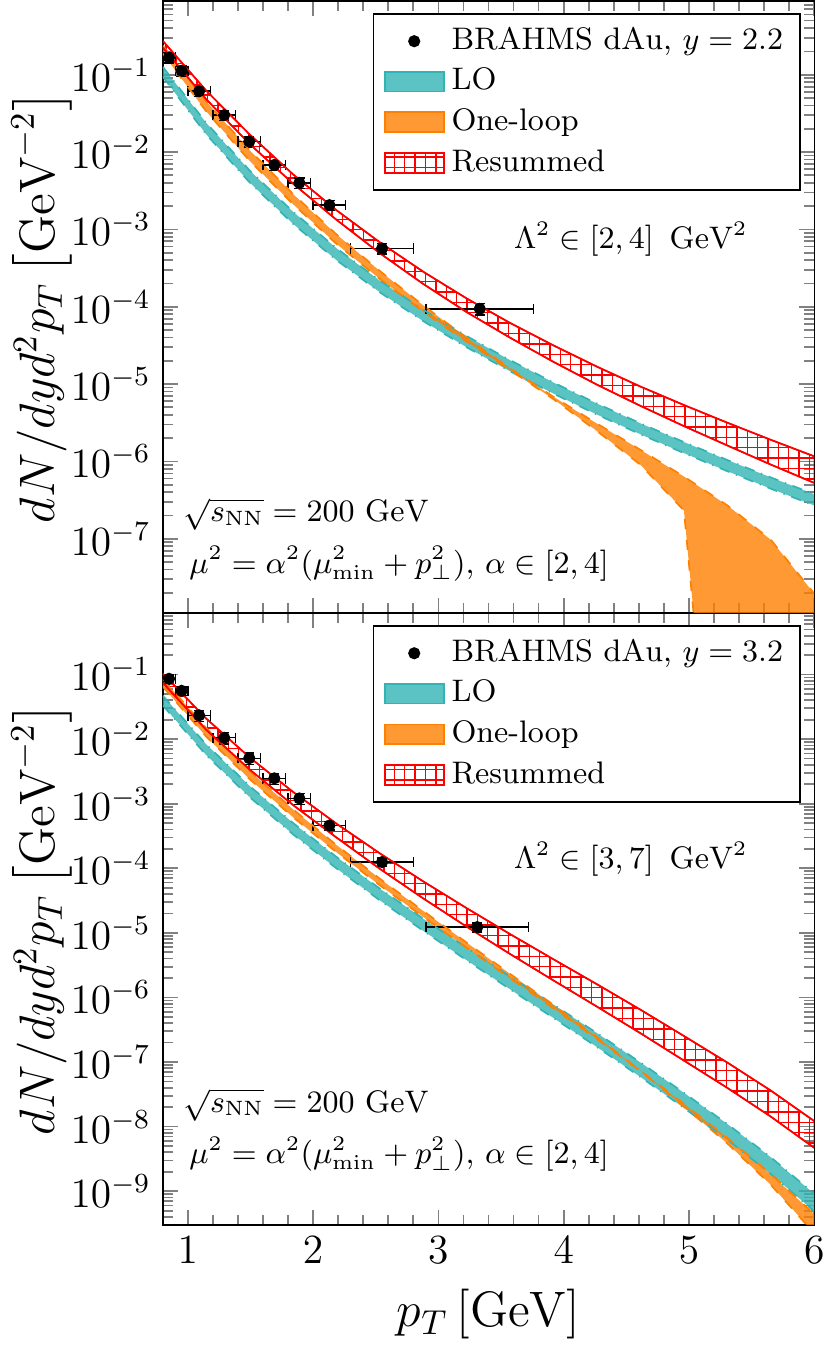}
\caption{Theoretical results computed in the CGC framework compared with the BRAHMS data \cite{Arsene:2004ux}. Many additional plots are provided at the end of the appendix.}  
\label{fig:rhic}\end{figure}

\begin{figure*}
\centering
\includegraphics[width=0.95\textwidth]{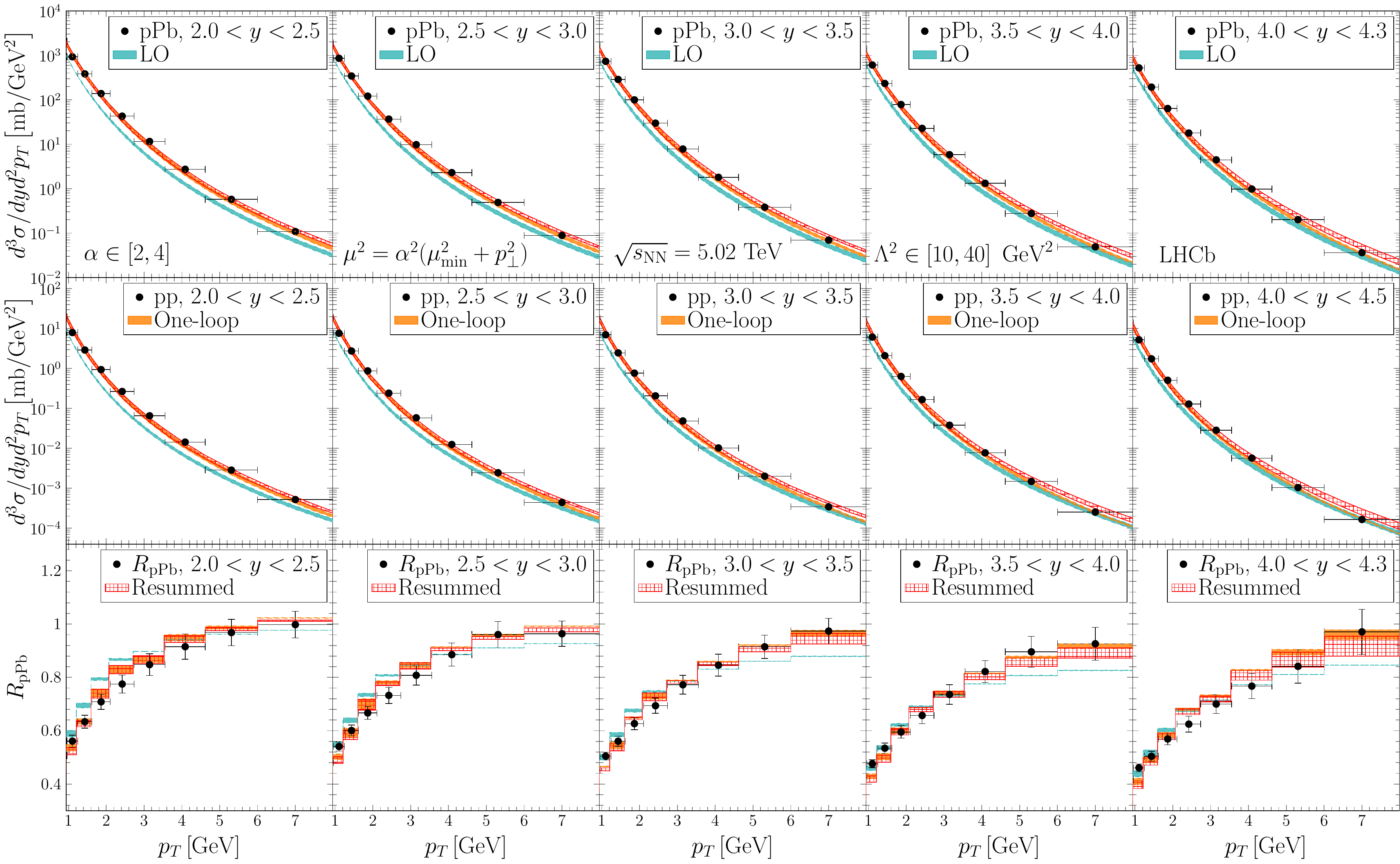}
\caption{Comparisons of the $pPb$, $pp$ and $R_{pPb}$ data\cite{LHCb:2021vww} from LHCb with CGC calculations in five forward rapidity bins.}
\label{fig:lhcb}
\end{figure*}

\textit{3. Numerical results} In the numerical evaluation, we use the NLO MSTW PDFs \cite{Martin:2009iq} and NLO DEHSS FFs \cite{deFlorian:2014xna} together with the one-loop running coupling. For the dipole gluon distribution $\mathcal{F}^{x_g}_a(k_\perp)$, we use the modified McLerran-Venugopalan model\cite{McLerran:1993ni,McLerran:1993ka,Golec-Biernat:1998zce,Stasto:2000er,Mueller:1999wm,Gelis:2001da} with parameters given by the Set h in Table 1 of Ref.~\cite{Albacete:2010sy} as the initial condition, solve the running-coupling BK equation numerically in the coordinate space\cite{Golec-Biernat:2001dqn, Balitsky:2006wa, Kovchegov:2006vj, Gardi:2006rp, Balitsky:2007feb, Albacete:2007yr, Albacete:2010sy,Berger:2010sh, Fujii:2013gxa, Iancu:2020jch}, and then obtain the numerical inputs in the momentum space via Fourier transform. As shown in Ref.~\cite{Lappi:2013zma}, the numerical results are sensitive to the above inputs especially the initial condition for $\mathcal{F}^{x_g}_a(k_\perp)$. To universally describe the data from RHIC and the LHC, it is important to choose proper initial conditions, include the NLO corrections, and implement the threshold resummation near the kinematic threshold. 

As shown in Fig.~\ref{fig:rhic}, with the proper choices of the $\Lambda^2$ scales, the improved NLO CGC calculations with the implementation of the threshold resummation, which are labeled in red gridded bands, agree with the data collected at RHIC and the LHC in both low and high $p_T$ regions. Similar to Ref.~\cite{deFlorian:2008wt}, the edges of the various bands were computed by varying $\Lambda^2$ in the appropriate ranges and $\mu^2=\alpha^2\left(\mu_{\text{min}}^2+p_T^2\right)$ with $\alpha =2 \sim 4$. To ensure that $\mu^2$ is not too small in the low $p_T$ region, a minimum value $\mu_{\text{min}} =2 \text{ GeV}$ is used. In the high $p_T$ region, the factorization scale is set by the hard scale $Q$, which is estimated to be at least twice the parton transverse momentum $k_T$. Therefore, the proper value of $\mu$ should be larger than $2p_T$ in this region. 

Compared to the one-loop results marked in orange bands, the resummed results, which are depicted in red grids, are roughly unchanged in the low-$p_T$ region. In fact, when $\Lambda$ is set to the value around $\mu \sim k_\perp$, the resummation formulation naturally reduces to the one-loop result since the threshold logarithms become small in this limit. Meanwhile, the resummation significantly improves the stability of the NLO calculation for the high-$p_T$ spectrum with the values of the auxiliary scale $\Lambda^2$ prescribed by Eq.~(\ref{lambdavalue}). 

In addition, we compare our calculation with the latest data measured by the LHCb collaboration\cite{LHCb:2021vww} in Fig.~\ref{fig:lhcb}. In the forward rapidity regions, LHCb measured the prompt charged particle production in $pPb$ and $pp$ collisions at $5\, \textrm{TeV}$ in five rapidity ranges around $y=2.25, \, 2.75, \, 3.25, \, 3.75$ and $4.2$. Within the same framework, we obtain a good agreement with the hadron spectra measured in both $pPb$ and $pp$ collisions for all rapidity windows. The impact of the resummation at the LHCb regime is less pronounced than that at RHIC since the kinematic range of this measurement is still far away from the threshold boundary.

Strictly speaking, our NLO calculation can not be directly applied to forward $pp$ collisions since we assume that the target is much larger than the proton projectile. This assumption allows us to integrate over the impact parameter and obtain the transverse area $S_\perp$ of the target nucleus. For $pp$ collisions, the above assumption is no longer justified, and thus $S_\perp^{pp}$ becomes less under control. Interestingly, we find that our results agree with the hadron spectra measured in $pp$ collisions if we choose $S_\perp^{pp}=2\pi R_p^2$ with $R_p$ the proton radius.

Eventually, this allows us to calculate the nuclear modification factor, which is defined as
\begin{equation}
R_{pPb} =\frac{1}{A} \frac{\dd ^2 \sigma_{pPb}/\dd p_T \dd y}{\dd^2 \sigma_{pp}/\dd p_T \dd y}. 
\end{equation}
The suppression of this factor $R_{pPb}$ as shown in Fig.~\ref{fig:lhcb} reflects the onset of the gluon saturation phenomenon. As we increase the rapidity or decrease the transverse momentum, more suppression in $R_{pPb}$ can be observed as the indication of strengthening of the saturation effect. In the high $p_T$ region, $R_{pPb}$ approaches unity as the saturation effect attenuates. 

\textit{4. Conclusion} By incorporating the threshold resummation in the CGC formalism, we extend the applicability regime of the CGC NLO calculation for forward hadron productions to the large transverse momentum region. Furthermore, the resummation allows us to reliably compute the hadron spectra and corresponding nuclear modification factor from low $p_T$ to high $p_T$ regions, and thus enables us to quantitatively understand the transition from the gluon saturation regime to the dilute regime. This study, which may serve as a benchmark example for other NLO CGC calculations, demonstrates that the NLO phenomenology is essential to test the CGC formalism and collect compelling evidence for the onset of gluon saturation. Lastly, the resummation formulation developed in this paper can also shed light on other higher-order calculations in the CGC framework.

\textit{Acknowledgments} We thank Tuomas Lappi, Xiaohui Liu, Feng Yuan and David Zaslavsky for useful inputs and discussions. This work is partly supported by the Natural Science Foundation of China (NSFC) under Grant Nos. 11575070 and by the university development fund of CUHK-Shenzhen under Grant No. UDF01001859.

\clearpage
\section*{Supplemental Material}

As the supplemental material of the paper, we provide all the technical details attached below, which include the following ten sections. 

\begin{enumerate}
\item First, we present the summary of the leading-order (LO) and next-to-leading order (NLO) cross-section for hadron productions in $pA$ collisions in the forward rapidity region in both the coordinate space and momentum space in Sec.~\ref{sec:full-one-loop}. Since the multi-dimensional numerical integrations (ranging from one to eight-dimensional integrations) of the NLO cross-section are pretty demanding, we have to adopt several technical procedures to accelerate the computation and improve the numerical accuracy significantly. For example, it is essential to note that the numerical evaluation of the momentum space expressions is much faster than that of the coordinate space expressions. This improvement is the primary reason that we Fourier transform all the terms of the LO and NLO cross-sections from the coordinate space into the analytic expressions in the momentum space. In addition, to prepare for the threshold resummation discussed below, we also identify the appearance of logarithms in the NLO corrections. 

\item With the help of the density plots of the hadron momentum fraction $\tau$ at RHIC and LHC energies, we illustrate and discuss the kinematics near the threshold region in detail in Sec.~\ref{sec:tau}. These plots allow one to visualize the regions of $y$ and $p_T$ where the threshold logarithms become important and the boundaries due to the small-$x$ kinematic constraint. 

\item There are two types of threshold logarithms in forward hadron productions: the collinear and the soft logarithms. Using two complimentary methods, we can resum the collinear part and obtain similar numerical results. First, in Sec.~\ref{sec:dglap-resum}, we show that the collinear logarithms can be resummed with the help of the DGLAP equations by resetting the factorization scale $\mu$ to the auxiliary semi-hard scale $\Lambda$ in the parton distribution functions (PDFs) and fragmentation functions (FFs) of the resummed contribution. Alternatively, we also show that one can directly solve the DGLAP equation in the threshold ($\xi \to 1$) limit and analytically resum the collinear threshold logarithms by using the so-called forward threshold jet function $\Delta \left(\mu^2, \Lambda^2, \omega \right)$ in Sec.~\ref{sec:rge-resum}. The latter method is equivalent to the renormalization group approach developed in the soft-collinear effective theory (SCET). Furthermore, we demonstrate that these two resummation approaches are numerically equivalent.

\item Following the above discussion, the soft part of the threshold logarithms is then resummed via the corresponding Sudakov factor. Besides, the remainder of the finite terms due to the mismatch between the running coupling and the fixed coupling cases are then redefined as the Sudakov matching term and implemented as part of the NLO hard factor. The resummation of the soft logarithms is summarized and presented in Sec.~\ref{sec:sudakov}.

\item For the reader's convenience, we summarize the full resummed expressions after implementing the threshold resummation in Sec.~\ref{sec:resummed-full-expression}. The numerical outcomes of the resummed expressions are labeled ``resummed" results in our plots.  

\item Sec.~\ref{kin-choice} is devoted to discussing the ``natural" choice of the semi-hard auxiliary scale $\Lambda$. Based on the kinematics and the partonic interactions, we first provide an intuitive way to show that this auxiliary scale $\Lambda^2$ is related to the saturation momentum $Q_s^2$ and the scale $(1-\xi) k_\perp^2$ when the coupling constant $\alpha_s$ is fixed. Then, using the saddle point approximation, we analytically show that one can reproduce the previous results for the choice of $\Lambda^2$ in the fixed coupling case. In addition, for the running coupling case, we further identify the typical value of $\Lambda^2$ determined by the saddle point of the resummation integral in coordinate space. According to the quantitative estimate of the typical value for $\Lambda^2$, we thus summarize the corresponding $\Lambda^2$ values used in the numerical evaluation at various rapidity bins and collision energies.

\item In Sec.~\ref{sec:rcbk}, we describe in detail the dipole gluon distributions used in this NLO calculation. First, we adopt the initial condition for the dipole scattering amplitude $S^{(2)}_{x_g} (r_\perp)$ that is widely used in CGC calculations, and then numerically evolve it with the running-coupling Balitsky-Kovchegov (rcBK) equation. Through the numerical Fourier transform, this numerical solution of the scattering amplitude is converted into the transverse momentum-dependent dipole gluon distribution, and it provides us the input for small-$x$ gluons with $x_g \leq 10^{-2}$.

\item Also, we briefly discuss the issue of the correlated uncertainties when we compute the nuclear modification factor from the ratio of the cross-section computed from the $pA$ and $pp$ collisions in Sec.~\ref{sec:theo-uncertainty}. In principle, the uncertainties of these two cross-sections, obtained by varying the factorization scale $\mu^2$ and the auxiliary scale $\Lambda^2$, are correlated in theory calculations since they are computed from the same CGC formalism. 

\item In addition, to bridge our theoretical calculation and experimental data reported by different collaborations at RHIC and the LHC, we adopt a systematic conversion between the hadron multiplicities computed in our framework and the measured cross-sections (multiplicities) for various hadrons. All of our numerical results are obtained from the rcBK solution described above and calculated with a uniform set of parameters. We provide the details of the relation between our theoretical calculation and various experimental measurements in Sec.~\ref{sec:relating}.

\item In Sec.~\ref{sec:additional-plots}, we present many additional plots and show the detailed comparisons of our numerical calculations with all of the available RHIC and the LHC data measured in forward $pp$ and $pPb$ collisions, and further discuss the range of validity of the CGC calculation at both RHIC and the LHC kinematic regimes. 

\end{enumerate}

\section{The cross-sections at one-loop order}
\label{sec:full-one-loop}

\begin{figure}[h]
\centering
\includegraphics[width=0.3\textwidth]{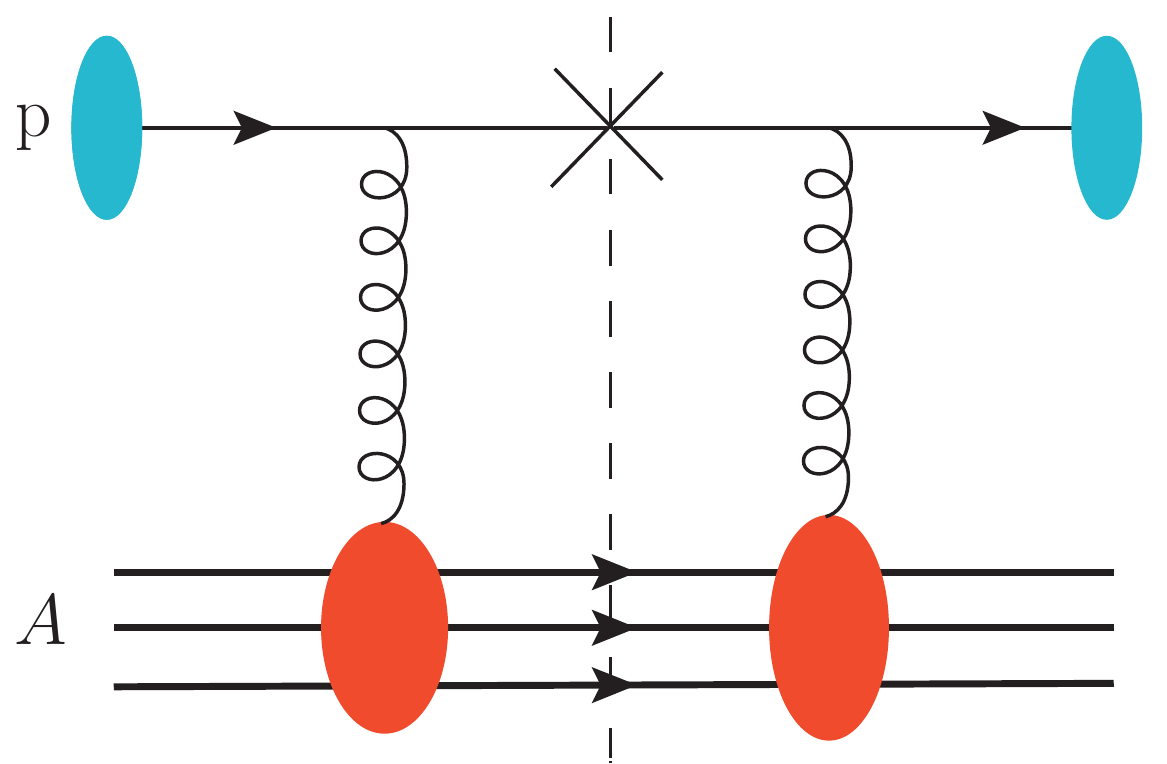}
\caption{LO diagram for the $q\to q$ channel in $pA$ collisions.}
\label{LO-diagram}
\end{figure}

As shown in Fig.~\ref{LO-diagram}, forward single inclusive hadron productions in $pA$ collisions at leading order (LO) in the CGC formalism can be modeled as follows: A collinear parton (either a quark or a gluon) from the projectile proton interacts with the dense gluon fields in the nuclear target before it fragments into an observed hadron with the transverse momentum $\pperp$ at the rapidity $y$. In high energy limit, the LO cross-section is cast into\cite{Dumitru:2002qt} 
\begin{eqnarray}
\frac{\dd \sigma^{\rm LO}}{ \dd y \dd^2 \pperp}&=& \int_\tau^1\frac{\dd z}{z^2}\sum_f x_p q_f(x_p)\mathcal{F}^{x_g}(k_\perp) D_{h/q}(z) \notag \\
&&+ \int_\tau^1\frac{\dd z}{z^2} x_pg(x_p)\tilde{\mathcal{F}}^{x_g}(k_\perp)D_{h/g}(z),
\end{eqnarray}
where $q_f (x_p)$ and $g(x_p)$ represent the quark and gluon collinear PDF with the longitudinal momentum fraction $x_p$, respectively. $D_{h/q, g}  (z)$ is the corresponding FF which describes the probability density of a quark or a gluon fragmenting into a hadron with the momentum fraction $z$. These quantities all depend on the so-called factorization scale $\mu$ due to the DGLAP evolution of the PDF and FF. The dependence on $\mu$ in the PDF and FF can be compensated and thus reduced by higher order corrections in the hard factor. $\mathcal{F}^{x_g} (k_\perp)$ and $\tilde{\mathcal{F}}^{x_g} (k_\perp)$ stand for the Fourier transforms of the dipole scattering amplitude in the fundamental and adjoint representations, respectively. If we define the quark dipole amplitude as $S^{(2)}_{x_g} (\rperp)$, then $\mathcal{F}^{x_g} (k_\perp) \equiv  S_\perp\int \frac{\dd^2\rperp}{(2\pi)^2} e^{-i\kperp\cdot \rperp} S^{(2)}_{x_g} (\rperp)$. $S_\perp$ is defined as the effective transverse area of the target nucleus obtained after averaging over the impact parameter. It is usually convenient to define $F(\eta, \kperp) \equiv \mathcal{F}^{x_g} (k_\perp) / \Sperp = \int \frac{\dd^2\rperp}{(2\pi)^2} e^{-i\kperp\cdot \rperp} S^{(2)}_{x_g} (\rperp)$ in solving the rcBK evolution equation with $\eta = \ln\frac{x_0}{x_g}$ and $x_0 = 0.01$. In principle, $F(\eta,\kperp)$ and $S^{(2)}_{x_g} (\rperp)$ depend on $x_g$ or $\eta$. We suppress the $x_g$ or $\eta$ dependence for simplicity in the following sections. The adjoint representation of the dipole amplitude yields $\tilde{\mathcal{F}}^{x_g} (k_\perp)$ accordingly. These two dipole amplitudes encode the strength of the multiple interactions between the quark/gluon projectiles and the dense gluon fields in the small-$x$ regime inside the nuclear target. At LO, the transverse momentum $k_\perp$ of the final state measured parton is determined by the transverse momentum that the incoming quark/gluon receives due to the multiple interaction. The LO kinematics imply $x_p=k_\perp e^y/\sqrt{s}$, $x_g=k_\perp e^{-y}/\sqrt{s}$ and $p_T=z k_\perp $. $\sqrt{s}$ is the total energy in the center-of-mass frame for $pp$ collisions, and it is identified as the center-of-mass energy per nucleon pair $\sqrt{s_{\text{NN}}}$ in $dAu$ or $pPb$ collisions in our calculation. The advantage of measuring hadron productions in the forward rapidity region is that the active parton from the proton projectile is from the large $x$ region while target gluon fields deep in the low-$x$ region are probed.

By considering the diagrams of the real emission of an additional gluon and the corresponding virtual contributions in this process as illustrated in Fig.~\ref{one-loop-diagrams}, we can compute hadron production at the one-loop order. At this order, both the collinear and rapidity divergence appear in the one-loop contributions. Although the results of the one-loop diagrams are unique, there are freedoms for divergence subtractions. For example, one can choose either the modified minimal subtraction ($\overline{\textrm{MS}}$) scheme or other schemes when one removes collinear divergences from one-loop contributions. In our calculation, we adopt the $\overline{\textrm{MS}}$ scheme in order to implement the widely used PDFs and FFs in numerical calculations. Similarly, there are various proposals of scheme choices for the subtraction of the rapidity logarithm\cite{Kang:2014lha, Ducloue:2016shw, Iancu:2016vyg, Ducloue:2017mpb, Ducloue:2017dit, Liu:2019iml, Kang:2019ysm, Liu:2020mpy}. In this paper, we follow the scheme choice adopted in Refs.~\cite{Chirilli:2011km, Chirilli:2012jd, Watanabe:2015tja}, and we only subtract the exact amount of the logarithm $\alpha_s \ln\frac{1}{x_g}$ from the one-loop contributions according to the LO kinematics. Therefore, the corresponding amount of the rapidity evolution that is put into the rcBK equation is then $\Delta \eta =\ln\frac{1}{x_g}$. In light of recent new developments of other factorization schemes, such as the one proposed in Refs.~\cite{Iancu:2016vyg, Ducloue:2016shw, Ducloue:2017dit, Ducloue:2017ftk}, it will be interesting to explore the numerical performance of these schemes and further improve the NLO calculations in CGC framework. Nevertheless, we leave this for a future study. 

After subtracting all the divergences under the chosen schemes, the NLO corrections are free of any singularities and they can be evaluated numerically. Based on the calculation presented in Refs.~\cite{Chirilli:2011km, Chirilli:2012jd, Watanabe:2015tja}, the results for the NLO cross-sections are summarized below. 

\begin{widetext} 

\subsection{$q \to q$ channel}
\begin{figure}[htp]
\centering
\includegraphics[width=0.22\textwidth]{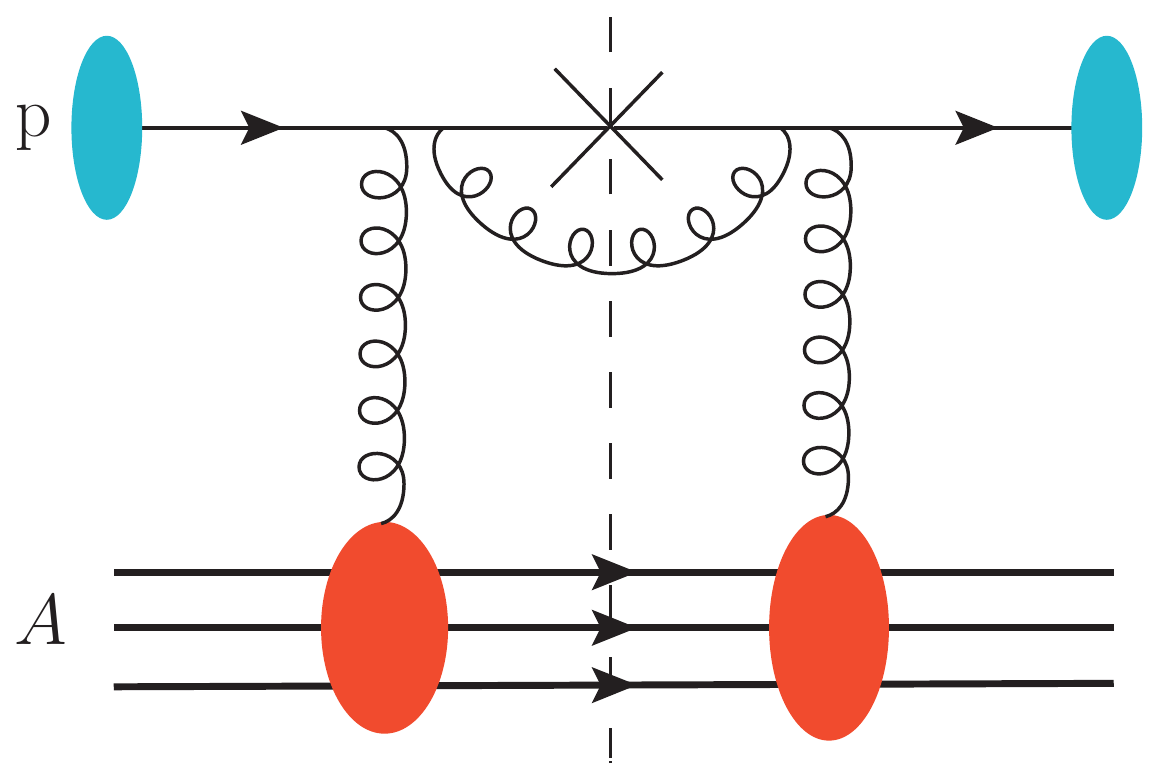}
\includegraphics[width=0.22\textwidth]{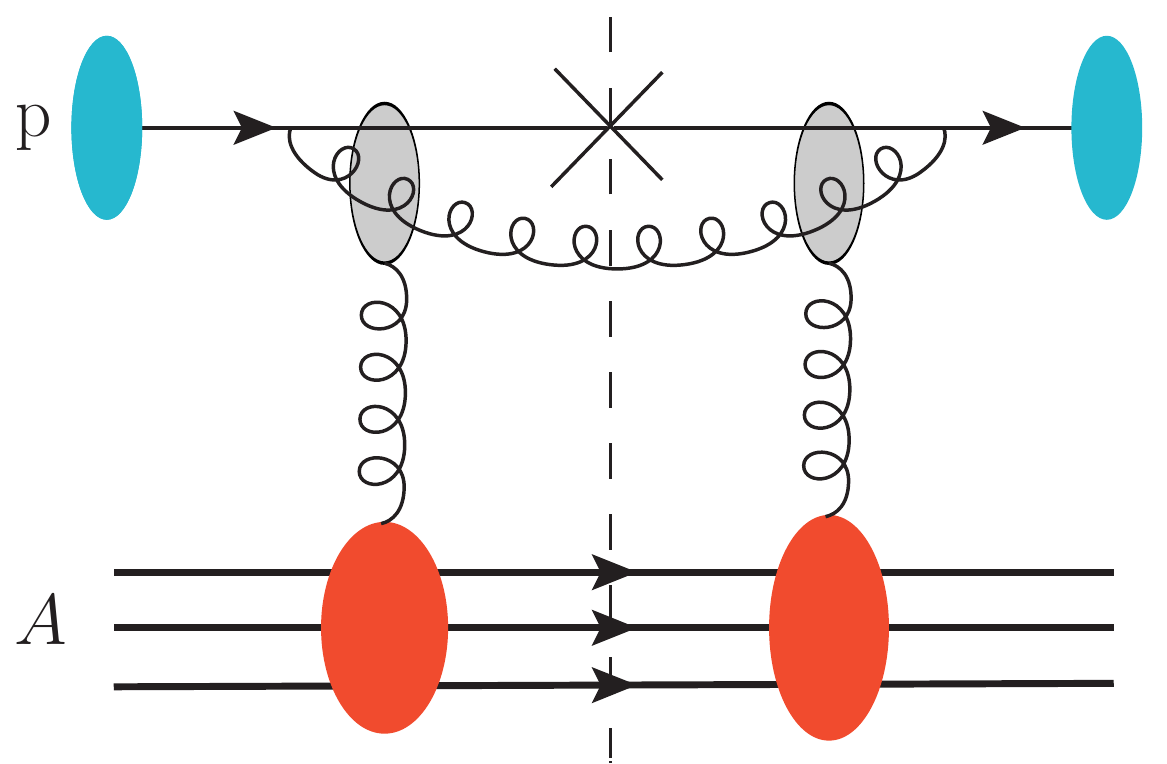}
\includegraphics[width=0.22\textwidth]{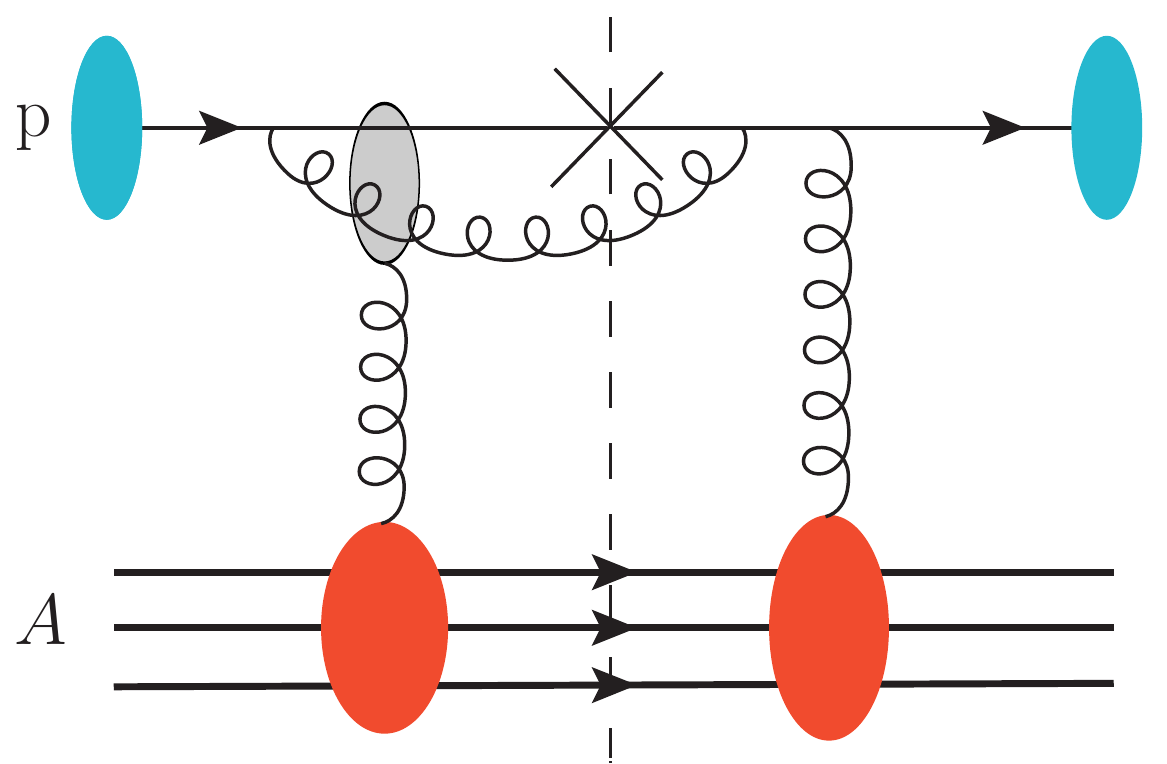}
\includegraphics[width=0.22\textwidth]{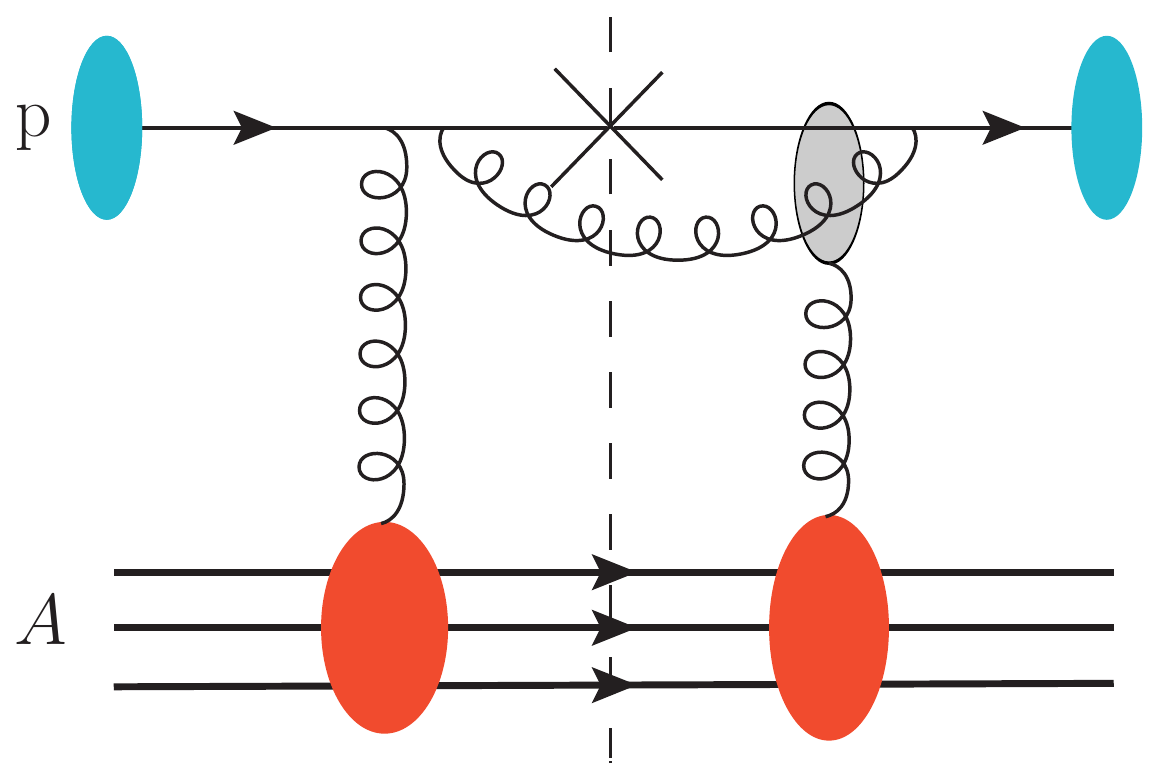}
\includegraphics[width=0.22\textwidth]{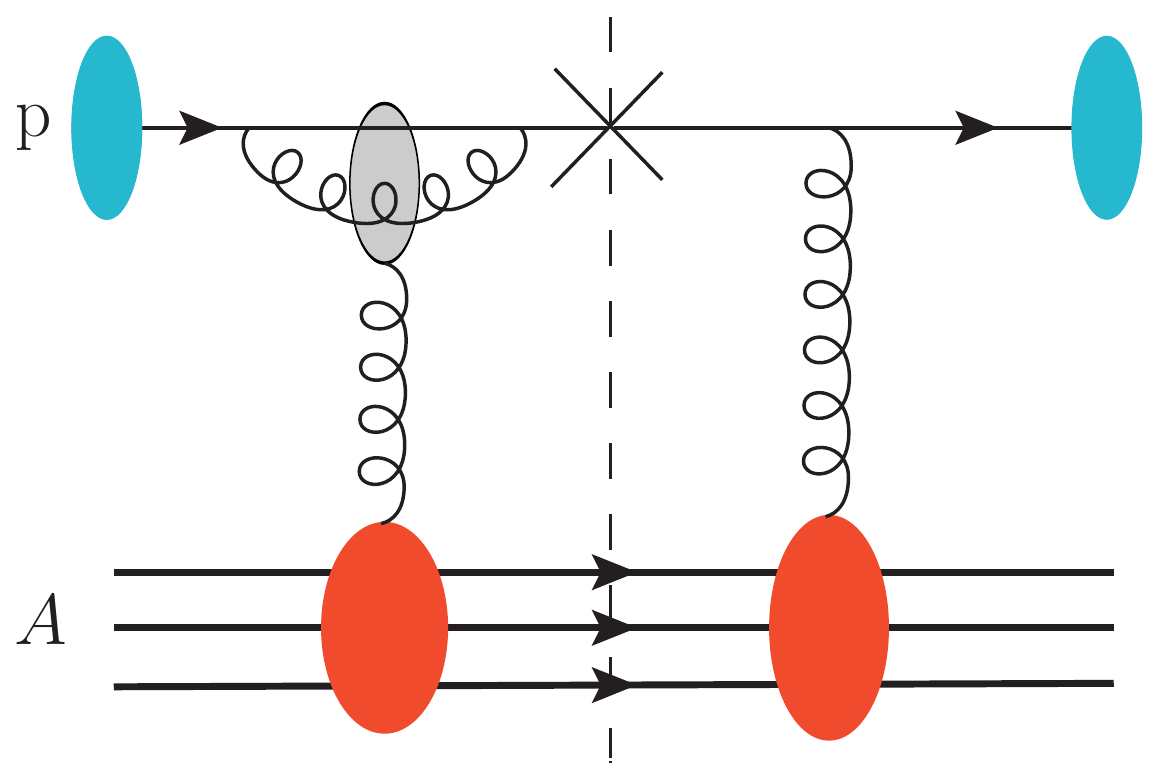}
\includegraphics[width=0.22\textwidth]{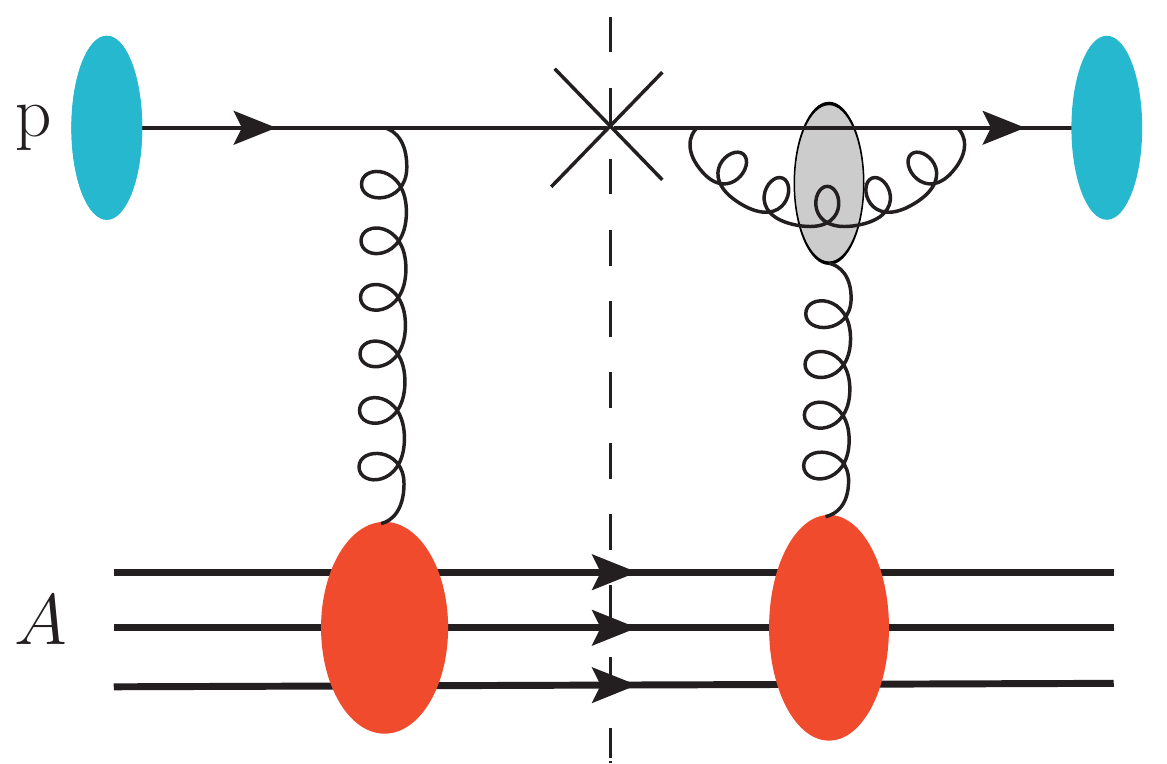}
\includegraphics[width=0.22\textwidth]{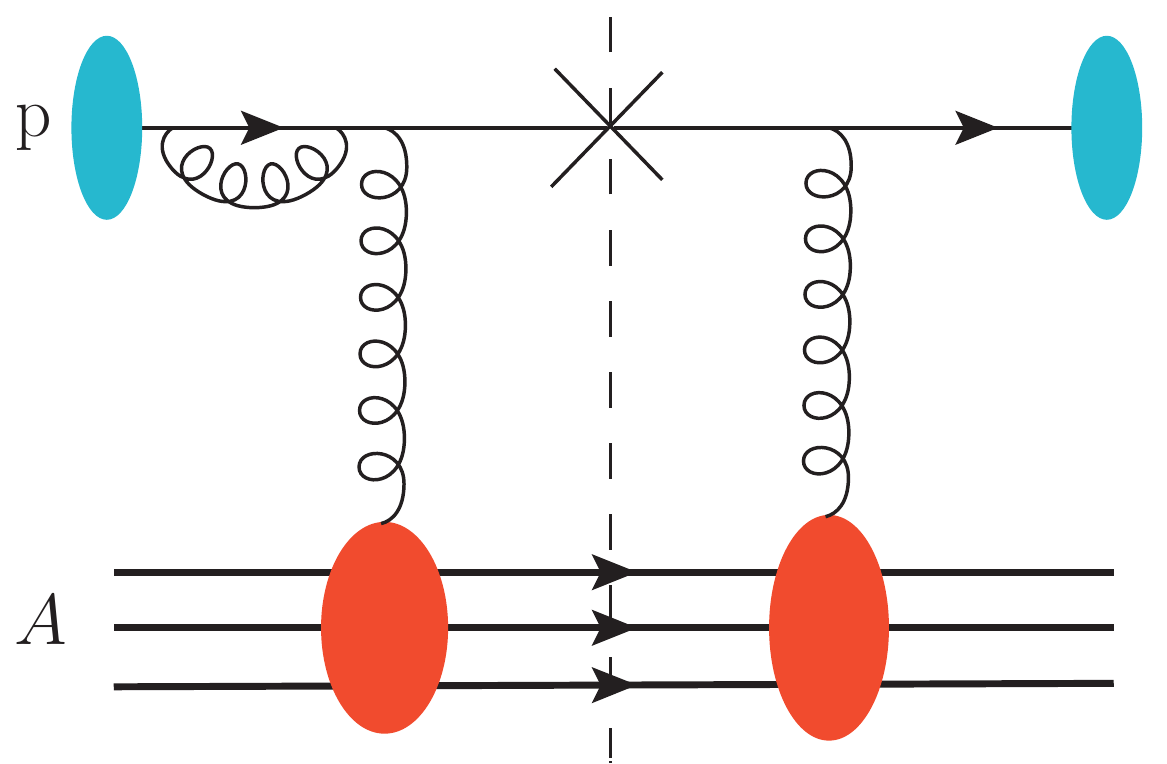}
\includegraphics[width=0.22\textwidth]{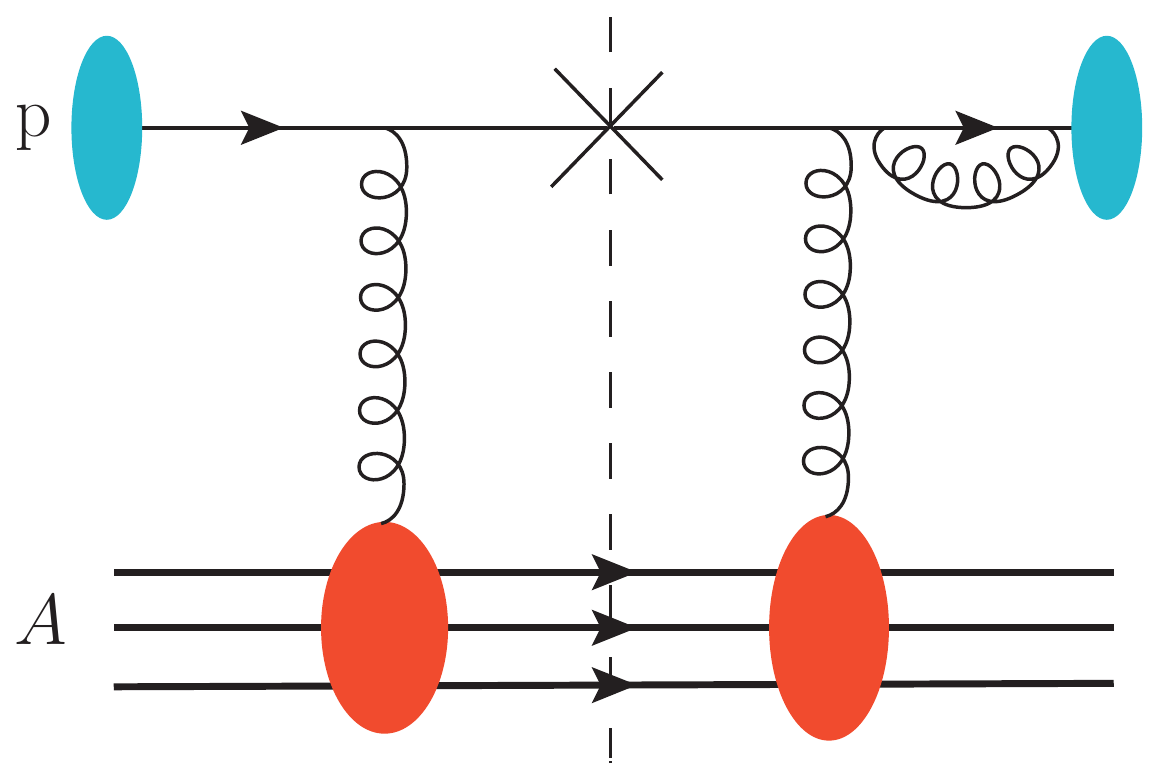}
\caption{The NLO real and virtual diagrams for the $q\to q$ channel in $pA$ collisions. Here the grey blobs indicate where non-linear multiple interactions occur between the vertical gluons and the quark-gluon pair.}
\label{one-loop-diagrams}
\end{figure}
The cross-section in the coordinate space has been obtained in Refs.~\cite{Chirilli:2011km, Chirilli:2012jd} with two additional terms presented in Ref.~\cite{Watanabe:2015tja}. To be self-contained, we summarize the final results here as the starting point. In the large $N_c$ limit, the complete one-loop cross-section for the $q \to q$ channel is divided into the following parts
\begin{align}
\frac{\dd \sigma_{qq}}{\dd y \dd^2 \pperp}
= 
\frac{\dd \sigma_{qq}^{\rm LO}}{\dd y \dd^2 \pperp}
+
\frac{\dd \sigma_{qq}^{\rm NLO}}{\dd y \dd^2 \pperp}
= 
\frac{\dd \sigma_{qq}^{\rm LO}}{\dd y \dd^2 \pperp}
+
\sum_{i=a}^e
\frac{\dd \sigma_{qq}^{i}}{\dd y \dd^2 \pperp},
\end{align}
where the LO and NLO parts read
\begin{align} 
\frac{\dd \sigma_{qq}^{\rm LO}}{\dd y \dd^2 \pperp}
= &
\Sperp\int_\tau^1 \frac{\dd z}{z^2} x_p q(x_p,\mu^2) D_{h/q}(z,\mu^2) \int \frac{\dd^2\rperp}{(2\pi)^2} e^{-i\kperp \cdot \rperp} S^{(2)}(\rperp),
\\ 
\frac{\dd \sigma_{qq}^{a}}{\dd y \dd^2 \pperp}
= &
\frac{\alpha_s}{2\pi} \Sperp C_F
\int_\tau^1 \frac{\dd z}{z^2} \int_{\tau/z}^1 \dd\xi x q(x,\mu^2) D_{h/q}(z,\mu^2) \mathcal{P}_{qq} (\xi)
\int \frac{\dd^2\rperp}{(2\pi)^2} 
\ln \frac{c_0^2}{\rperp^2 \mu^2}
\nonumber\\
& \times \left( e^{-i\kperp \cdot \rperp} + \frac{1}{\xi^2} e^{-i\frac{1}{\xi}\kperp \cdot \rperp} \right) S^{(2)}(\rperp),
\\
\frac{\dd \sigma_{qq}^{b}}{\dd y \dd^2 \pperp}
= &
- 3 \frac{\alpha_s}{2\pi} \Sperp C_F\int_\tau^1 \frac{\dd z}{z^2} x_p q(x_p,\mu^2) D_{h/q}(z,\mu^2) \int \frac{\dd^2\rperp}{(2\pi)^2} e^{-i\kperp \cdot \rperp}S^{(2)}(\rperp)  \ln \frac{c_0^2}{\rperp^2\kperp^2}  \, ,
\\
\frac{\dd \sigma_{qq}^{c}}{\dd y \dd^2 \pperp}
= &
-8\pi\frac{\alpha_s}{2\pi} \Sperp C_F 
\int_\tau^1 \frac{\dd z}{z^2} \int_{\tau/z}^1 \dd\xi x q(x,\mu^2) D_{h/q}(z,\mu^2) \int \frac{\dd^2\uperp\dd^2\vperp}{(2\pi)^4} e^{-i\kperp \cdot (\uperp - \vperp)}  e^{-i \frac{1-\xi}{\xi}\kperp \cdot \uperp} 
\nonumber\\
&
\times \frac{1+\xi^2}{(1-\xi)_+} \frac{1}{\xi} \frac{\uperp \cdot \vperp}{\uperp^2 \vperp^2}
S^{(2)}(\uperp)S^{(2)}(\vperp),
\\
\frac{\dd \sigma_{qq}^{d}}{\dd y \dd^2 \pperp}
= & 
8\pi\frac{\alpha_s}{2\pi} \Sperp C_F 
\int_\tau^1 \frac{\dd z}{z^2} x_p q(x_p,\mu^2) D_{h/q}(z,\mu^2) \int \frac{\dd^2\uperp\dd^2\vperp}{(2\pi)^4}  e^{-i \kperp \cdot (\uperp - \vperp)} S^{(2)}(\uperp)S^{(2)}(\vperp)
\nonumber\\
& \times 
\int_0^1 \dd\xi' \frac{1+\xi'^2}{(1-\xi')_+}
\left(
e^{-i(1-\xi') \kperp \cdot \vperp} \frac{1}{\vperp^2} - \delta^2 (\vperp) \int \dd^2 \rperp' e^{i\kperp \cdot \rperp'}\frac{1}{\rperp'^2} 
\right),
\\
\frac{\dd \sigma_{qq}^{e}}{\dd y \dd^2 \pperp}
= & \frac{\alpha_s}{\pi^2} \Sperp \frac{N_c}{2} \int_\tau^1 \frac{\dd z}{z^2} x_p q(x_p,\mu^2) D_{h/q}(z,\mu^2) \int \frac{\dd^2 \uperp\dd^2\vperp}{(2\pi)^2} e^{-i\kperp\cdot (\uperp-\vperp)} [S^{(2)}(\uperp)S^{(2)}(\vperp) - S^{(2)}(\uperp-\vperp)] 
\nonumber\\
& \times \Bigg[
\frac{1}{\uperp^2} \ln \frac{\kperp^2 \uperp^2}{c_0^2} +
\frac{1}{\vperp^2} \ln \frac{\kperp^2 \vperp^2}{c_0^2} - \frac{2\uperp\cdot\vperp}{\uperp^2 \vperp^2} \ln \frac{\kperp^2 |\uperp| |\vperp|}{c_0^2}
\Bigg],
\end{align}
with the kinematic variables $x_p=\tau/z$, $x= x_p/\xi$, and $\kperp=\pperp/z$. The coordinate variables are defined as follows $\rperp=\xperp-\yperp$, $\uperp=\xperp-\bperp$, $\vperp=\yperp-\bperp$. For convenience, we also denote $c_0=2e^{-\gamma_E}$ with $\gamma_E$ the Euler constant and the splitting function $\mathcal{P}_{qq} (\xi) = \frac{1+\xi^2}{(1-\xi)_+} + \frac{3}{2} \delta (1-\xi)$. For simplicity, $S^{(2)}(\rperp)$ is assumed to be only a function of $|\rperp|$ while the impact parameter dependence is neglected throughout this calculation. Therefore, one can simply define $\Sperp=\int \dd ^2\Rperp$ as the effective transverse area of the target nucleus after integrating over the impact parameter $\Rperp$. The first four terms $\sigma_{qq}^{a-d}$ among the NLO corrections are first derived in Refs.~\cite{Chirilli:2011km, Chirilli:2012jd} and the last term $\sigma_{qq}^{e}$ is due to the kinematic constraint as illustrated in Ref.~\cite{Watanabe:2015tja}. As we show in the discussion in Sec.~\ref{kin-choice}, there are two logarithms $\ln \frac{1}{x_g}$ and $\ln \frac{k_\perp^2}{q_\perp^2}$ arising from the rapidity integral when we consider the kinematic constraint. The first logarithm $\ln \frac{1}{x_g}$ is corresponding to the rapidity divergence when the center of mass energy $\sqrt{s}$ is taken to be $\infty$, and it is resummed through the BK evolution equation. In our scheme choice, we keep the second logarithm $\ln \frac{k_\perp^2}{q_\perp^2}$ in the NLO hard factor and this eventually gives rise to the last term $\sigma_{qq}^{e}$ as shown in Ref.~\cite{Watanabe:2015tja}. 

In arriving the above expressions, we have taken the large $N_c$ limit and assumed the Gaussian approximation for color charge distributions inside the target nucleus. Then, we can safely neglect the NLO corrections which are suppressed by $1/N_c^2$, and we also simplify multiple point correlation functions and write them in terms of products of dipole amplitudes $S^{(2)}$ as shown in the last three terms, i.e., $\sigma_{qq}^{c}$, $\sigma_{qq}^{d}$, and $\sigma_{qq}^{e}$. Since we do not distinguish between $N_c/2$ and $C_F$ in the large $N_c$ limit, we have replaced the color factors in $\sigma_{qq}^{c}$, $\sigma_{qq}^{d}$ by $C_F$ and we will change the color factor $\frac{N_c}{2}$ in $\sigma_{qq}^{e}$ to $C_F$ in the following discussions.

Although the physical interpretation of each NLO correction is manifest in the above coordinate space expressions\cite{Chirilli:2011km, Chirilli:2012jd}, it is challenging to evaluate some of the NLO corrections accurately in numerical computations, especially in the LHC kinematic regime. To achieve better numerical performance, we adopt an analytical procedure including the following three steps of manipulations: 1. Fourier transform; 2. Combining terms that are cancelling each other; 3. Shifting coordinates. 

\subsubsection{Fourier Transform}
Due to the oscillatory behavior of the phase factor $e^{-ik_\perp\cdot \rperp}$, which can be translated into a Bessel function $J_0(|k_\perp| |\rperp|)$ after averaging over the azimuthal angle, it is notoriously difficult to numerically calculate the cross-section in the coordinate space especially in the large $k_\perp$ region. To achieve a much better numerical performance, we analytically transform all of the above coordinate space expressions to the momentum space. This step is vital in the numerical evaluation of the NLO corrections since we need to perform up to eight-dimensional numerical integrations with high precision. 

The Fourier transform of the $\sigma_{qq}^{\rm LO}$ term is straightforward, while the transforms of other terms are less trivial. For example, let us consider the Fourier transform of the $\sigma_{qq}^{a}$ and $\sigma_{qq}^{b}$ terms. Since the splitting function ${\cal P}_{qq}(\xi) = \frac{1+\xi^2}{(1-\xi)_+} + \frac{3}{2} \delta(1-\xi)$ contains two terms, we can rewrite $\sigma_{qq}^{a}$ as
\begin{align}
\frac{\dd \sigma_{qq}^{a}}{\dd y \dd^2 \pperp}
= &
\frac{\alpha_s}{2\pi} \Sperp C_F
\int_\tau^1 \frac{\dd z}{z^2} \int_{\tau/z}^1 \dd\xi x q(x,\mu^2) D_{h/q}(z,\mu^2) \frac{1+\xi^2}{(1-\xi)_+}
\int \frac{\dd^2\rperp}{(2\pi)^2} \ln \frac{c_0^2}{\rperp^2 \mu^2}
\nonumber\\
& \times \left( e^{-i\kperp \cdot \rperp} + \frac{1}{\xi^2} e^{-i\frac{1}{\xi}\kperp \cdot \rperp} \right) S^{(2)}(\rperp)
\nonumber\\
+ &
3\frac{\alpha_s}{2\pi} \Sperp C_F
\int_\tau^1 \frac{\dd z}{z^2} x_p q(x_p,\mu^2) D_{h/q}(z,\mu^2) 
\int \frac{\dd^2\rperp}{(2\pi)^2} e^{-i\kperp \cdot \rperp} S^{(2)}(\rperp)
\ln \frac{c_0^2}{\rperp^2 \mu^2}.
\label{eq:sigqqa-0920}
\end{align}
We then combine the second term in Eq.~(\ref{eq:sigqqa-0920}), which is proportional to $ \ln \frac{c_0^2}{\rperp^2\mu^2} $, together with $\sigma_{qq}^{b}$, which is proportional to $- \ln \frac{c_0^2}{\rperp^2\kperp^2} $, and obtain the following contribution
\begin{align}
3\frac{\alpha_s}{2\pi} \Sperp C_F
\int_\tau^1 \frac{\dd z}{z^2} x_p q(x_p,\mu^2) D_{h/q}(z,\mu^2) 
\int \frac{\dd^2\rperp}{(2\pi)^2} e^{-i\kperp \cdot \rperp} S^{(2)}(\rperp)
\ln \frac{\kperp^2}{\mu^2}.
\label{eq:sigqqb-0920}
\end{align}
The Fourier transform of this term is then straightforward. For the remaining terms of $\sigma_{qq}^{a}$ (i.e., the first term of Eq.~(\ref{eq:sigqqa-0920})), the derivation is a bit more involved. We use the following identities
\begin{align}
\int\frac{\dd^2\rperp}{(2\pi)^2} e^{-ik_\perp\cdot \rperp} S^{(2)}(\rperp)\ln \frac{c_{0}^{2}}{\rperp^{2}\mu ^{2}}=&\frac{1}{\pi}\int\frac{\dd^2\lperp}{\lperp^2}\left[F(\kperp-\lperp)-J_0\left(\frac{c_0}{\mu}|\lperp|\right)F(\kperp)\right]\nonumber\\
=&\frac{1}{\pi}\int\frac{\dd^2\lperp}{\lperp^2}\left[F(\kperp-\lperp)-\frac{\Lambda^2}{\Lambda^2+\lperp^2}F(\kperp)\right]+F(\kperp)\ln\frac{\Lambda^2}{\mu^2},
\label{Fourier1}
\end{align}
where we introduce a convenient notation for the dipole gluon distribution $F(\kperp) \equiv \int \frac{\dd^2\rperp}{(2\pi)^2} e^{-i\kperp\cdot \rperp} S^{(2)} (\rperp)$ (note that $\mathcal{F}^{x_g} (k_\perp) \equiv  S_\perp F(\kperp)$ as previously defined). The second term arises from the integral identity
\begin{equation}
\frac{1}{\pi}\int \frac{\dd^2 l_\perp}{l_\perp^2} \left[ \frac{\Lambda^2}{\Lambda^2+\lperp^2} -J_0\left(\frac{c_0}{\mu}|\lperp|\right) \right] =\ln\frac{\Lambda^2}{\mu^2}.
\end{equation}
To optimize the Monte Carlo integrations and extract the large logarithm, we add and subtract a regulator term $\frac{1}{\pi\lperp^2}\frac{\Lambda^2}{\Lambda^2+\lperp^2}F(\kperp)$ and eventually obtain two terms shown in Eq.~(\ref{Fourier1}). The choice of the regulating counter term is not unique. It is straightforward to find that one gets the same result if one replaces $\frac{\Lambda^2}{\Lambda^2+\lperp^2}$ by $J_0\left(\frac{c_0}{\Lambda}|\lperp|\right)$ in the above equation. Nevertheless, it appears that the counter term $\frac{\Lambda^2}{\Lambda^2+\lperp^2}$ is better behaved in numerical evaluations since it does not have the oscillatory behavior as $J_0\left(\frac{c_0}{\Lambda}|\lperp|\right)$ does.

As shown in the above mathematical manipulation, an auxiliary scale $\Lambda^2$ has been introduced and it is clear that Eq.~(\ref{Fourier1}) does not depend on the value of the auxiliary scale $\Lambda^2$. The second term of Eq.~(\ref{Fourier1}) is finite as long as $\Lambda \neq 0$ and it is ready for resummation as it is proportional to $F(\kperp)$. With proper choices of $\Lambda^2$, one can extract the corresponding logarithms and efficiently evaluate the remaining first term in numerical computations. 

Furthermore, this auxiliary scale can receive the physical interpretation as the semi-hard scale related to the saturation momentum or the semi-hard scale in the threshold resummation. For example, when $k_\perp^2 \sim Q_s^2$, a reasonable choice of $\Lambda^2$ is expected to be $Q_s^2$. In this case, on the LHS of Eq.~(\ref{Fourier1}), the $r_\perp$ integral is dominated by the region where $r_\perp \sim 1/Q_s$. Meanwhile, the desired logarithm $\ln \frac{Q_s^2}{\mu^2}$ naturally arises from the second term once we set $\Lambda^2= Q_s^2$, while the first term on the RHS of Eq.~(\ref{Fourier1}) should be small due to the cancellation between those two terms inside the square brackets. When $k_\perp^2 \gg Q_s^2$ near the threshold regime, $\Lambda^2$ can act as the semi-hard scale separating the hard momentum exchange and the soft momentum emission. In the latter case, the ``natural" choice of $\Lambda^2$ will be discussed in detail later.

As for $\sigma_{qq}^{c}$, the Fourier transform gives
\begin{align}
\frac{\dd \sigma_{qq}^{c}}{\dd y\dd^2\pperp} & = \frac{\alpha_s}{2\pi^2} C_F \Sperp
\int_\tau^1 \frac{\dd z}{z^2} \int_{\tau/z}^1 \dd\xi xq(x,\mu^2) D_{h/q}(z,\mu^2)
\int \dd^2 \qa \dd^2\qb  F(\qa-\kperp/\xi) F(\qb-\kperp) \nonumber\\
&\times \frac{1}{\xi} \frac{1+\xi^2}{(1-\xi)_+} \frac{-2\qa\cdot\qb}{\qa^2\qb^2}.
\end{align}
In the step discussed in the next subsection, we will combine $\sigma_{qq}^{c}$ with the first term on the RHS of Eq.~(\ref{Fourier1}) to expedite the numerical integration when $k_\perp$ is large. 

In addition, to deal with the $\sigma_{qq}^{d}$ term, we need to employ the following identity
\begin{align}
\int \frac{\dd^2 \rperp}{(2\pi)^2} e^{-i \qa \cdot \rperp} \frac{1}{\rperp^2} = \lim_{\rho \to 0} \int \frac{\dd^2 \rperp}{(2\pi)^2} e^{-i \qa \cdot \rperp} \frac{1}{\rperp^2 +\rho^2} 
= \frac{1}{2\pi} \lim_{\rho \to 0} K_0 (\rho \qa).
\end{align}
where we have introduced an infinitesimal parameter $\rho$ to regulate the divergence. The two terms in $\sigma^{d}$ correspond to the following kernel
\begin{align}
\frac{1}{2\pi} \lim_{\rho \to 0} [K_0 (\rho (\qa - \xi' \kperp)) - K_0 (\rho \kperp)] = -\frac{1}{4\pi} \ln\frac{(\qa - \xi' \kperp)^2}{\kperp^2},
\end{align}
which leads to a finite contribution after cancelling the logarithmic divergences between them. Therefore, the Fourier transform of $\sigma_{qq}^{d}$ becomes $\sigma_{qq}^4$ as shown in Eq.~(\ref{sigmaqq4}).

The Fourier transform of $\sigma_{qq}^{e}$ is the most difficult one. The analytical techniques have been outlined in Ref.~\cite{Watanabe:2015tja}, thus only a brief summary of the steps is provided here. By employing the following relations
\begin{align}
& 
\frac{1}{\rperp^2} \ln \frac{\kperp^2 \rperp^2}{c_0^2} = \frac{1}{8\pi} \int \dd^2\qa e^{i\qa \cdot \rperp} \ln^2 \frac{\kperp^2}{\qa^2},
\\
&
\frac{\vec r_\perp}{\rperp^2} \ln \frac{\kperp^2 \rperp^2}{c_0^2} = \frac{1}{2\pi} \int \dd^2\qa e^{i\qa\cdot \rperp} \frac{i\vec q_{1\perp}}{\qa^2} \ln \frac{\kperp^2}{\qa^2},
\end{align}
we can convert the terms in $\sigma_{qq}^{e}$ into the following three terms
\begin{align}
\frac{\dd \sigma_{qq}^{e}}{\dd y \dd^2 \pperp} = & \int_\tau^1 \frac{\dd z}{z^2} x_p q(x_p,\mu^2) D_{h/q} (z,\mu^2) [L_{q1} (\kperp) + L_{q2} (\kperp) + L_{q3} (\kperp)]
\\
L_{q1} (\kperp) = & -\frac{2\alphas }{\pi^2}C_F S_\perp\int \frac{\dd^2\uperp \dd^2\vperp }{(2\pi)^2} e^{-i\kperp \cdot (\uperp-\vperp)} S^{(2)}(\uperp -\vperp)\left[\frac{1}{\uperp^2} \ln\frac{\kperp^2 \uperp^2}{c_0^2} -\frac{\uperp \cdot \vperp}{\uperp^2\vperp^2} \ln\frac{\kperp^2 \vert\uperp\vert\vert\vperp\vert}{c_0^2}\right]\notag \\
= & \lim_{\rho\to 0}\frac{\alphas }{\pi}C_F S_\perp\int \frac{\dd^2\rperp}{(2\pi)^2} e^{-i\kperp \cdot \rperp} S^{(2)}(\rperp )\left[-\ln^2\frac{\kperp^2}{\rho^2}+4\int_\rho^\infty\frac{\dd\qa}{\qa}\ln \frac{\kperp^2}{\qa^2}J_0(\qa\rperp)\right]\notag \\
=& -\frac{\alphas  }{2\pi^2}C_F S_\perp\int \frac{\dd^2\rperp  }{2\pi} e^{-i\kperp \cdot \rperp} S^{(2)}(\rperp) \ln^2\frac{\kperp^2 \rperp^2}{c_0^2}, \label{l1}\\
L_{q2} (\kperp)=&\frac{2\alpha_s}{\pi^2} \Sperp C_F \int \frac{\dd^2 \uperp\dd^2\vperp}{(2\pi)^2} e^{-i\kperp\cdot (\uperp-\vperp)} S^{(2)}(\uperp)S^{(2)}(\vperp)\frac{1}{\uperp^2} \ln \frac{\kperp^2 \uperp^2}{c_0^2}\nonumber\\
 =&\frac{\alphas}{\pi}C_F S_\perp F(\kperp)\int\dd^2\qa F(\kperp-\qa)\ln^2\frac{\kperp^2}{\qa^2}, \\
L_{q3} (\kperp) =&-\frac{2\alpha_s}{\pi^2} \Sperp C_F \int \frac{\dd^2 \uperp\dd^2\vperp}{(2\pi)^2} e^{-i\kperp\cdot (\uperp-\vperp)} S^{(2)}(\uperp)S^{(2)}(\vperp)\frac{\uperp\cdot\vperp}{\uperp^2 \vperp^2} \ln \frac{\kperp^2 |\uperp| |\vperp|}{c_0^2}\nonumber\\
=&-\frac{2\alpha_s}{\pi^2} \Sperp C_F\int\dd^2\qa\dd^2\qb F(\kperp-\qa)F(\kperp-\qb)\frac{\qa\cdot\qb}{\qa^2\qb^2}\ln\frac{\kperp^2}{\qa^2}, 
\end{align}
where we have introduced $\rho$ as an infrared cutoff to regularize the integration in Eq.~(\ref{l1}). While the last two terms are already in the momentum space, Eq.~(\ref{l1}) needs an extra step of manipulation. To perform the Fourier transform of the double logarithmic factor $\ln^2 \frac{\kperp^2 \rperp^2}{c_0^2}$, we utilize the following relations inspired by the Sudakov double logarithm calculation\cite{Mueller:2013wwa, Watanabe:2015tja} in the dimensional regularization 
\begin{align}
& \left( \frac{\mu^2 e^{\gamma_E}}{4\pi} \right)^\epsilon \int \frac{\dd^{2-2\epsilon} \qa}{(2\pi)^{2-2\epsilon} \qa^2} e^{-i\qa\cdot\rperp} \ln \frac{\kperp^2}{\qa^2} = \frac{1}{4\pi} 
\left[
\frac{1}{\epsilon^2} - \frac{1}{\epsilon} \ln \frac{\kperp^2}{\mu^2} + \frac{1}{2} \ln^2 \frac{\kperp^2}{\mu^2} - \frac{1}{2} \ln^2 \frac{\kperp^2 \rperp^2}{c_0^2} - \frac{\pi^2}{12}
\right],
\\
& \left( \frac{\mu^2 e^{\gamma_E}}{4\pi} \right)^\epsilon \int \frac{\dd^{2-2\epsilon} \qa}{(2\pi)^{2-2\epsilon} \qa^2} \ln \frac{\kperp^2}{\qa^2} \theta(\kperp^2-\qa^2) = \frac{1}{4\pi} 
\left[
\frac{1}{\epsilon^2} - \frac{1}{\epsilon} \ln \frac{\kperp^2}{\mu^2} + \frac{1}{2} \ln^2 \frac{\kperp^2}{\mu^2} - \frac{\pi^2}{12}
\right].
\end{align}
By taking the difference of these two terms, we obtain
\begin{align}
\ln^2 \frac{\kperp^2 \rperp^2}{c_0^2} = 8\pi \int \frac{\dd^2 \qa}{(2\pi)^2} \frac{1}{\qa^2} \ln\frac{\kperp^2}{\qa^2} [\theta(\kperp^2-\qa^2) - e^{-i\qa\cdot \rperp}].
\end{align}
With the above relations, we can easily find
\begin{align}
\int \frac{\dd^2 \rperp}{(2\pi)^2} e^{-i\kperp\cdot\rperp} S^{(2)}(\rperp) \ln^2 \frac{\kperp^2 \rperp^2}{c_0^2} = \frac{2}{\pi} \int \frac{\dd^2\qa}{\qa^2} \ln \frac{\kperp^2}{\qa^2} [\theta(\kperp^2-\qa^2) F(\kperp) - F(\kperp-\qa)].
\end{align}
Finally, the Fourier transform of $\sigma_{qq}^e$ becomes $\sigma_{qq}^5$ as given in Eq.~(\ref{sigmaqq5}).

\subsubsection{Combinations of terms}
Using the normalization of $F(\qa)$, we can rewrite the first term of Eq.~(\ref{Fourier1}) as
\begin{align}
\frac{1}{\pi} \int \dd^2\qa\dd^2\qb \frac{1}{\qb^2} F(\qa) \left[ F(\qb-\kperp)-\frac{\Lambda^2}{\Lambda^2+\qb^2} F(\kperp)\right].
\end{align}
$\sigma_{qq}^{a}$ contains two Fourier transforms. Combining these two terms from $\sigma_{qq}^{a}$ with $\sigma_{qq}^{c}$, we obtain $\sigma_{qq}^{3}$ in the momentum space with the following kernel 
\begin{align}
\int \dd^2 \qa \dd^2\qb & \left[ 
\frac{(\qb-\qa/\xi)^{2}}{\qa^2 \qb^2} F(\qa-\kperp/\xi) F(\qb-\kperp)
\right.
\nonumber\\
& \left.
- \frac{1}{\qa^2} \frac{\Lambda^{2}}{\Lambda^{2}+\qa^2}
F(\qb)F(\kperp)
- \frac{1}{\xi^2 \qb^2 } \frac{\Lambda^{2}}{\Lambda^{2}+\qb^2}
F(\kperp/\xi) F(\qa)
\right]. \label{eq:integrand-before-shift}
\end{align}
In the above expression, the poles at $|\qa| = 0$ in the first and second terms cancel each other and those at $|\qb|=0$ in the first and third terms also cancel each other. Therefore, the above formula is free from any divergences. However, the integration region of $|\qa|$ and $|\qb|$ is from 0 to $+\infty$, which is impossible to implement exactly in the Monte Carlo similation. In practice, we always numerically integrate in the finite region and assume that the contribution from outside is negligible. The validity of this technique requires that the integrand falls off much faster than $\frac{1}{V}$ with $V$ being the volume of integration region. The dominant contribution of the first term in Eq.~(\ref{eq:integrand-before-shift}) arises from both $\qa\sim 0$, $\qb \sim 0$ and $\qa \sim \kperp/\xi$ and $\qb\sim \kperp$, while that of the second and third terms comes only from $\qa\sim 0$, $\qb \sim 0$. Each term results in a large contribution while the sum of them is relatively small. The cancellation of large contributions does not occur at the integrand level since the dominant contributions of these three terms arise from different $|\qa|$ and $|\qb|$ regions. Therefore, if one directly evaluates Eq.~(\ref{eq:integrand-before-shift}), it may consume enormous amount of computing resources in order to obtain accurate numerical results. Besides, the final results may also be sensitive to the upper cuts of $|\qa|$ and $|\qb|$ in the numerical implementation, if these cuts are not sufficiently large enough. This issue can be significantly mitigated by shifting the momentum coordinates as discussed in the following subsection. 

\subsubsection{Shifting Coordinates}
The above-mentioned numerical challenges can be solved by coordinate shifts. Defining $\qa' = \qa - \kperp/\xi$ and $\qb' = \qb - \kperp$, we switch to the following optimized expression for numerical evaluation
\begin{align}
\int \dd^2\qa \dd^2\qb 
& \left[ \frac{(\qb-\qa/\xi)^{2}}{(\kperp+\qa)^2 (\kperp/\xi+\qb)^2} F(\qa)F(\qb) \right.
\nonumber\\
& - \frac{1}{(\kperp+\qa)^2} \frac{\Lambda^{2}}{\Lambda^{2}+(\kperp+\qa)^2}
F(\qb)F(\kperp)
\nonumber \\
& 
\left. - \frac{1}{(\kperp +\xi\qb)^2 } \frac{\Lambda^{2}}{\Lambda^{2}+(\kperp/\xi+\qb)^2}
F(\kperp/\xi) F(\qa) \right]. \label{eq:integrand-after-shift}
\end{align}
It is straightforward to show that Eq.~(\ref{eq:integrand-after-shift}) and Eq.~(\ref{eq:integrand-before-shift}) are equivalent. The dominant contributions of these three terms come from the small $|\qa|$ and small $|\qb|$ region. The cancellation now occurs at the integrand level. Therefore, it is much more efficient to numerically evaluate Eq.~(\ref{eq:integrand-after-shift}) rather than Eq.~(\ref{eq:integrand-before-shift}), although they are analytically equivalent.

\subsubsection{The full NLO/one-loop corrections in the momentum space}
At the end of the day, we arrive at the results of the one-loop cross-section in the momentum space in the large $N_c$ limit after following the above three-step procedure
\begin{align}
\frac{\dd\sigma_{qq}}{\dd y \dd^2\pperp} = 
\frac{\dd \sigma^{\rm LO}_{qq}}{\dd y\dd^{2}\pperp} 
+ \frac{\dd \sigma^{\rm NLO}_{qq}}{\dd y\dd^{2}\pperp} 
=
\frac{\dd \sigma^{\rm LO}_{qq}}{\dd y\dd^{2}\pperp} 
+ \sum_{i=1}^5 \frac{\dd \sigma^{i}_{qq}}{\dd y\dd^{2}\pperp} \, ,
\end{align}
where the new corresponding LO and NLO contributions are cast into
\begin{align}
\frac{\dd \sigma^{\rm LO}_{qq}}{\dd y\dd^2\pperp} = & \Sperp  \int_\tau^1 \frac{\dd z}{z ^2} x_pq(x_p,\mu^2)D_{h/q}(z,\mu^2)  F(\kperp),
\\
\frac{\dd\sigma_{qq}^1}{\dd y\dd^{2}\pperp} 
=&
\frac{\alpha_{s}}{2\pi}C_{F}\Sperp
 \int_{\tau}^{1} \frac{\dd z}{z^{2}}\int_{\tau/z}^{1}\dd\xi
 x q(x,\mu^{2})D_{h/q}(z,\mu^{2})
\mathcal{P}_{qq}(\xi) 
\ln\frac{\Lambda^{2}}{\mu^{2}}\left(F(\kperp)+\frac{1}{\xi^{2}}F(\kperp/\xi)\right),
\label{eq:sigqq1fir}
 \\
\frac{\dd\sigma_{qq}^2}{\dd y\dd^{2}\pperp} 
=&
3\frac{\alphas}{2\pi} C_{F} \Sperp
\int_{\tau}^{1} \frac{\dd z}{z^{2}}
x_p q(x_p,\mu^{2})D_{h/q}(z,\mu^{2})
\ln\frac{\kperp^{2}}{\Lambda^2}F(\kperp),
\\
\frac{\dd\sigma_{qq}^3}{\dd y\dd^{2}\pperp}
=& \frac{\alpha_{s}}{2\pi^2} C_F \Sperp \int_{\tau}^{1} \frac{\dd z}{z^{2}}\int_{\tau/z}^{1}\dd\xi\int\dd^{2}\qa\dd^{2}\qb
 x q(x,\mu^{2})D_{h/q}(z,\mu^{2})
\frac{1+\xi^2}{(1-\xi)_+} \mathcal{T}_{qq}^{(1)}(\xi,\qa,\qb,\kperp), \label{t-momentum}
\\
\frac{\dd\sigma_{qq}^4}{\dd y\dd^{2}\pperp}
=& 
- \frac{\alpha_{s}}{\pi} C_F \Sperp
 \int_{\tau}^{1} \frac{\dd z}{z^{2}}\int_{0}^{1}\dd\xi'\int\dd^{2}\qa 
 x_p q(x_p,\mu^{2})D_{h/q}(z,\mu^{2})
\frac{1+{\xi'}^{2}}{\left(1-\xi'\right)_+}\ln\frac{(\qa-\xi'\kperp)^{2}}{\kperp^{2}}
F(\qa)F(\kperp),  \label{sigmaqq4}
\\
\frac{\dd\sigma_{qq}^5}{\dd y\dd^{2}\pperp}
=& 
\frac{2\alpha_s}{\pi^2} C_F \Sperp 
\int_\tau^1 \frac{\dd z}{z^2} \int \dd^2 \qa x_p q(x_p,\mu^2) D_{h/q} (z,\mu^2) 
\frac{1}{\qa^2} \ln \frac{\kperp^2}{\qa^2} [F(\kperp-\qa) - \theta(\kperp^2-\qa^2) F(\kperp)] 
\nonumber\\
+ & 
\frac{\alpha_s}{\pi} C_F \Sperp 
\int_\tau^1 \frac{\dd z}{z^2} \int \dd^2 \qa x_p q(x_p,\mu^2) D_{h/q} (z,\mu^2) 
F(\qa) F(\kperp) \ln^2 \frac{\kperp^2}{(\kperp - \qa)^2} \nonumber \\
- & \frac{2\alpha_s}{\pi^2} C_F \Sperp 
\int_\tau^1 \frac{\dd z}{z^2} \int \dd^2 \qa \int \dd^2 \qb x_p q(x_p,\mu^2) D_{h/q} (z,\mu^2) 
F(\qa) F(\qb) \ln \frac{\kperp^2}{(\kperp-\qa)^2} \nonumber\\
& \times \frac{(\kperp-\qa)\cdot (\kperp-\qb)}{(\kperp-\qa)^2 (\kperp-\qb)^2} .  \label{sigmaqq5}
\end{align}
In Eq.~(\ref{t-momentum}), we use the shorthand notation $\mathcal{T}_{qq}^{(1)}(\xi,\qa,\qb,\kperp)$ defined as 
\begin{align}
\mathcal{T}_{qq}^{(1)}(\xi,\qa,\qb,\kperp)= 
& \frac{(\qb-\qa/\xi)^{2}}{(\kperp+\qa)^2 (\kperp/\xi+\qb)^2} F(\qa)F(\qb)
\nonumber\\
& - \frac{1}{(\kperp+\qa)^2} \frac{\Lambda^{2}}{\Lambda^{2}+(\kperp+\qa)^2}
F(\qb)F(\kperp)
\nonumber \\
& 
- \frac{1}{(\kperp +\xi\qb)^2 } \frac{\Lambda^{2}}{\Lambda^{2}+(\kperp/\xi+\qb)^2}
F(\kperp/\xi)F(\qa).
\end{align}
The $\Lambda$ dependence in $\sigma_{qq}^1$, $\sigma_{qq}^2$ and $\sigma_{qq}^3$ completely cancels when they are summed together. Therefore, the above complete one-loop cross-section for the $q\to q$ channel is $\Lambda$-independent. This result together with the one-loop contributions from the other three channels is numerically evaluated and denoted as the ``one-loop" results in plots.

\subsubsection{Identifying the double logarithmic contribution}

In the high $k_\perp$ regime, it was found that the threshold type logarithm is one of the dominant contributions to the one-loop results. To set the stage for the resummation in the threshold regime, let us identify the double logarithm hidden in $\sigma_{qq}^5$. Since there are partial cancellations among terms in $\sigma^{5}$, we need to adopt a rather nuanced approach to analytically extract the double logarithmic term. 

\begin{figure}[!h]
\includegraphics[width=0.4\textwidth]{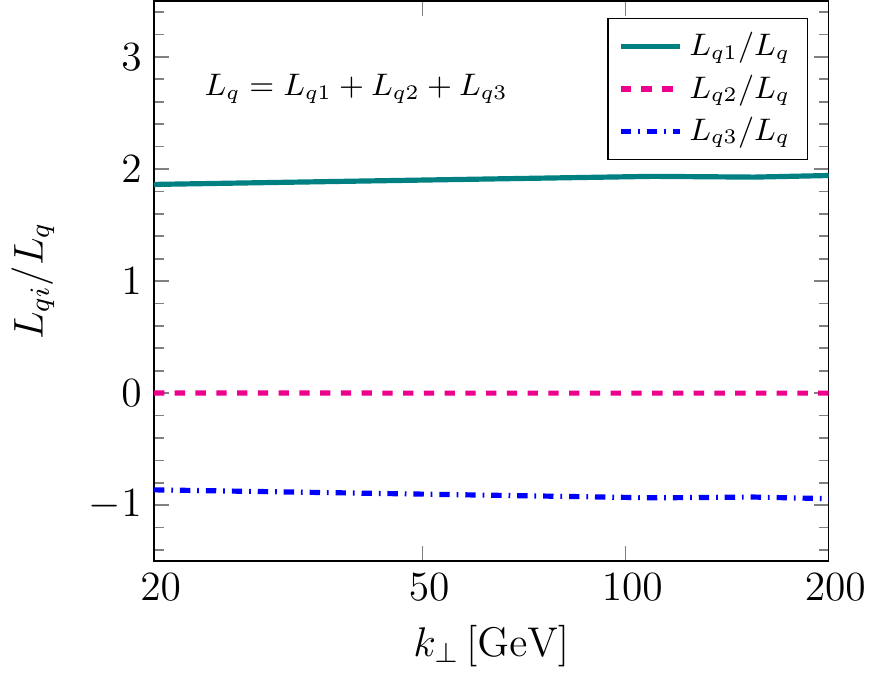}
\caption{Ratios of $L_{q1}$, $L_{q2}$, $L_{q3}$ to $L_{q}=L_{q1}+L_{q2}+L_{q3}$.}
\label{fig:lq}
\end{figure}

First of all, as shown in Ref.~\cite{Watanabe:2015tja}, one can analytically study the high $k_\perp$ limit of all three terms in $\sigma^{5}$ by using the GBW model and setting $S^{(2)}(k_\perp) = \exp\left(-Q_s^2r^2/4\right) $ with fixed $Q_s^2$. In this special case, the first term ($L_{q1}$) and the third term ($L_{q3}$) in $\sigma^5$ are proportional to $\frac{8Q_s^2}{\kperp^4}$ and $-\frac{4Q_s^2}{\kperp^4}$ at the large-$\kperp$ limit, respectively. The second term ($L_{q2}$) is found to be exponentially suppressed. Furthermore, as shown in Fig.~\ref{fig:lq}, we have numerically evaluated $\sigma_{qq}^5$ with the rcBK solution as the input and checked that $\sigma_{qq}^5$ is roughly half of the first term in the high $k_\perp$ limit. In addition, due to the kinematic constraint, the gluon emissions from both the initial and final state quark are required to be soft in the forward threshold limit. Thus, according to the counting rule~\cite{Mueller:2013wwa,Sun:2014gfa} for the double logarithmic contributions arising from the soft regime, the corresponding coefficient of the double logarithmic term should be $2\frac{\alpha_s}{4\pi}C_F$, since both initial state and final state gluons radiations can contribute and each of them contributes to $\frac{\alpha_s}{4\pi}C_F$. 

Therefore, to analytically extract the double logarithmic term, we can decompose the $\theta$-function term of $\sigma_{qq}^5$ as follows
\begin{equation}
\frac{2}{\pi}\int\frac{\dd^{2}\qa}{\qa^{2}}\ln\frac{\kperp^{2}}{\qa^{2}}\left[\theta(\Lambda ^2 -\qa^2)-\theta(\kperp^2-\qa ^2 ) \right] F(\kperp)
-
\frac{2}{\pi}\int\frac{\dd^{2}\qa }{\qa^{2}}\ln\frac{\kperp^{2}}{ \qa^{2}}\theta(\Lambda^2-\qa ^2)F(\kperp)
 ,
\end{equation}
where the first term leads to a double logarithm. As discussed above, we rewrite $\sigma_{qq}^5$ as
\begin{align}
\frac{\dd\sigma_{qq}^5}{\dd y\dd^{2}\pperp}
=
\frac{\dd\sigma_{qq}^{5a}}{\dd y\dd^{2}\pperp}
+
\frac{\dd\sigma_{qq}^{5b}}{\dd y\dd^{2}\pperp},
\end{align}
where,
\begin{align}
\frac{\dd\sigma_{qq}^{5a}}{\dd y\dd^{2}\pperp}
=& - 
\frac{\alpha_s}{2\pi} C_F \Sperp 
\int_\tau^1 \frac{\dd z}{z^2} x_p q(x_p,\mu^2) D_{h/q} (z,\mu^2) F(\kperp)
\ln^2 \frac{\kperp^2}{\Lambda^2},
\\
\frac{\dd\sigma_{qq}^{5b}}{\dd y\dd^{2}\pperp}
=& 
\frac{2\alpha_s}{\pi^2} C_F \Sperp 
\int_\tau^1 \frac{\dd z}{z^2} \int \dd^2 \qa x_p q(x_p,\mu^2) D_{h/q} (z,\mu^2) 
\frac{1}{\qa^2} \ln \frac{\kperp^2}{\qa^2} [F(\kperp-\qa) -  \theta(\Lambda^2-\qa^2) F(\kperp)] 
\nonumber\\
- &
\frac{\alpha_s}{2\pi} C_F \Sperp 
\int_\tau^1 \frac{\dd z}{z^2} x_p q(x_p,\mu^2) D_{h/q} (z,\mu^2) F(\kperp)
\ln^2 \frac{\kperp^2}{\Lambda^2}  \nonumber\\
+ & 
\frac{\alpha_s}{\pi} C_F \Sperp 
\int_\tau^1 \frac{\dd z}{z^2} \int \dd^2 \qa x_p q(x_p,\mu^2) D_{h/q} (z,\mu^2) 
F(\qa) F(\kperp) \ln^2 \frac{\kperp^2}{(\kperp - \qa)^2} \nonumber \\
- & \frac{2\alpha_s}{\pi^2} C_F \Sperp 
\int_\tau^1 \frac{\dd z}{z^2} \int \dd^2 \qa \int \dd^2 \qb x_p q(x_p,\mu^2) D_{h/q} (z,\mu^2) 
F(\qa) F(\qb) \ln \frac{\kperp^2}{(\kperp-\qa)^2} \nonumber\\
& \times \frac{(\kperp-\qa)\cdot (\kperp-\qb)}{(\kperp-\qa)^2 (\kperp-\qb)^2} .
\end{align}
In the large $k_\perp$ limit, with proper choice of $\Lambda^2$, $\sigma^{5a}$ represents the Sudakov double logarithm contribution. In the mean time, $\sigma^{5b}$ is put back into the NLO hard factor as part of $\sigma_{\rm NLO~matching}$.

\subsection{$g \to g$ channel}
 
The computation for the $g\to g$ channel is akin to the calculation we have done above for the $q\to q$ channel. In the large $N_c$ limit, the one loop cross-section for the $g\to g$ channel in the coordinate space reads
\begin{align}
\frac{\dd \sigma_{gg}}{\dd y \dd^2 \pperp}
= 
\frac{\dd \sigma_{gg}^{\rm LO}}{\dd y \dd^2 \pperp}
+
\frac{\dd \sigma_{gg}^{\rm NLO}}{\dd y \dd^2 \pperp}
= 
\frac{\dd \sigma_{gg}^{\rm LO}}{\dd y \dd^2 \pperp}
+
\sum_{i=a}^f
\frac{\dd \sigma_{gg}^{i}}{\dd y \dd^2 \pperp},
\end{align}
where the LO and NLO cross-sections are given by 
\begin{align}
\frac{\dd \sigma_{gg}^{\rm LO}}{\dd y \dd^2 \pperp}
= &
\Sperp\int_\tau^1 \frac{\dd z}{z^2} x_p g(x_p,\mu^2) D_{h/g} (z,\mu^2) \int \frac{\dd^2\rperp}{(2\pi)^2} e^{-i\kperp\cdot \rperp} S^{(2)} (\rperp) S^{(2)} (\rperp),
\\
\frac{\dd \sigma_{gg}^a}{\dd y \dd^2 \pperp}
= &
\frac{\alpha_s}{2\pi} \Sperp N_c \int_\tau^1 \frac{\dd z}{z^2} \int_{\tau/z}^1 \dd\xi xg(x,\mu^2) D_{h/g}(z,\mu^2) \mathcal{P}_{gg} (\xi) \int \frac{\dd^2\rperp}{(2\pi)^2} S^{(2)} (\rperp) S^{(2)} (\rperp)
\nonumber \\
& \times \Big( e^{-i\kperp\cdot \rperp} + \frac{1}{\xi^2} e^{-i\frac{1}{\xi} \kperp \cdot \rperp} \Big) \ln \frac{c_0^2}{\rperp^2\mu^2},
\\
\frac{\dd \sigma_{gg}^b}{\dd y \dd^2 \pperp}
= &
-\frac{\alpha_s}{2\pi} 4\beta_0 \Sperp N_c \int_\tau^1 \frac{\dd z}{z^2} x_p g(x_p,\mu^2) D_{h/g} (z,\mu^2) \int \frac{\dd^2\rperp}{(2\pi)^2} e^{-i\kperp\cdot \rperp} 
S^{(2)} (\rperp) S^{(2)} (\rperp) \ln \frac{c_0^2}{\rperp^2 \kperp^2},
\\
\frac{\dd \sigma_{gg}^c}{\dd y \dd^2 \pperp}
= &
8\pi \Sperp N_f T_R \frac{\alpha_s}{2\pi} \int_\tau^1 \frac{\dd z}{z^2} x_p g(x_p,\mu^2) D_{h/g} (z,\mu^2) \int \frac{\dd^2\uperp\dd^2\vperp}{(2\pi)^4} e^{-i\kperp\cdot\vperp} S^{(2)} (\uperp) S^{(2)} (\vperp)
\nonumber\\
& \times \int_0^1 \dd\xi' [\xi'^2 + (1-\xi')^2] \left[ \frac{e^{-i\xi' \kperp\cdot (\uperp-\vperp)}}{(\uperp-\vperp)^2} -\delta^2 (\uperp-\vperp) \int \dd^2 \rperp' \frac{e^{i\kperp\cdot \rperp' }}{\rperp'^2 } \right],
\\
\frac{\dd \sigma_{gg}^d}{\dd y \dd^2 \pperp}
= & 
-16\pi \Sperp N_c \frac{\alpha_s}{2\pi} 
\int_\tau^1 \frac{\dd z}{z^2} \int_{\tau/z}^1 \dd\xi xg(x,\mu^2) D_{h/g} (z,\mu^2) \int\frac{\dd^2\uperp\dd^2\vperp}{(2\pi)^4} 
e^{-i\kperp\cdot(\uperp-\vperp)} e^{-i\frac{1}{\xi}\kperp\cdot\vperp} 
\nonumber\\
&
\times S^{(2)} (\uperp) S^{(2)} (\vperp) S^{(2)} (\uperp-\vperp) \frac{[1-\xi(1-\xi)]^2}{(1-\xi)_+}
\frac{1}{\xi^2} \frac{(\uperp-\vperp) \cdot\vperp}{(\uperp-\vperp)^2\vperp^2},
\\
\frac{\dd \sigma_{gg}^e}{\dd y \dd^2 \pperp}
= & 
16\pi \Sperp N_c \frac{\alpha_s}{2\pi} \int_\tau^1 \frac{\dd z}{z^2} x_p g(x_p,\mu^2) D_{h/g} (z,\mu^2)  \int\frac{\dd^2\uperp\dd^2\vperp}{(2\pi)^4} S^{(2)} (\uperp) S^{(2)} (\vperp) S^{(2)} (\uperp-\vperp)
\nonumber\\
&
\times e^{-i\kperp\cdot(\uperp-\vperp)} \int_0^1 \dd\xi' \left[ \frac{\xi'}{(1-\xi')_+} + \frac{\xi' (1-\xi')}{2} \right]
\left[ \frac{e^{-i\xi'\kperp\cdot\vperp}}{\vperp^2} - \delta^2 (\vperp) \int \dd^2 \rperp' \frac{e^{i\kperp\cdot \rperp' }}{\rperp'^2 }  \right],
\\
\frac{\dd \sigma_{gg}^f}{\dd y \dd^2 \pperp}
= & 
\frac{\alpha_s}{\pi^2} \Sperp N_c \int_\tau^1 \frac{\dd z}{z^2} x_p g(x_p,\mu^2) D_{h/g} (z,\mu^2) \int \frac{\dd^2\uperp\dd^2\vperp}{(2\pi)^2} e^{-i\kperp\cdot(\uperp-\vperp)} [S^{(2)}(\uperp)S^{(2)}(\vperp)-S^{(2)}(\uperp-\vperp)] 
\nonumber\\
& 
\times S^{(2)}(\uperp-\vperp) 
\Bigg[ 
  \frac{1}{\uperp^2} \ln \frac{\kperp^2\uperp^2}{c_0^2} 
+ \frac{1}{\vperp^2} \ln \frac{\kperp^2\vperp^2}{c_0^2}- \frac{2\uperp\cdot \vperp}{\uperp^2 \vperp^2} \ln \frac{\kperp^2 |\uperp||\vperp|}{c_0^2}
\Bigg],
\end{align}
where $\mathcal{P}_{gg} (\xi) = 2 \left[\frac{\xi}{(1-\xi)_+}+\frac{1-\xi}{\xi}+\xi(1-\xi)\right] + 2\beta_0 \delta(1-\xi)$ and $\beta_0=\frac{11}{12} - \frac{2 N_f T_R}{6N_c}$.

Due to the same reasons discussed in the last subsection, we also need to Fourier transform the above equations to the momentum space analytically. Most of the techniques have been presented in the last subsection except for the following one
\begin{align}
\int \frac{\dd^2\rperp}{(2\pi)^2} e^{-i\kperp\cdot \rperp} S^{(2)} (\rperp) S^{(2)} (\rperp)
= & \int \frac{\dd^2\rperp}{(2\pi)^2} e^{-i\kperp\cdot \rperp} \int \dd^2 \qa e^{i\qa \cdot \rperp} F(\qa) \int \dd^2 \qb e^{i\qb\cdot \rperp} F(\qb)
\nonumber \\
= & \int \dd^2 \qa F(\qa) F(\kperp-\qa),
\end{align}
which transforms the gluon dipole into the momentum space. In addition, the $\sigma_{gg}^f$ term can be written as
\begin{align}
\frac{\dd \sigma_{gg}^f}{\dd y \dd^2 \pperp}
= & 
\int_\tau^1 \frac{\dd z}{z^2} x_p g(x_p,\mu^2) D_{h/g} (z,\mu^2) [L_{g1}(\kperp) + L_{g2} (\kperp) + L_{g3} (\kperp)], \\
L_{g1} (\kperp) = &
-\frac{\alpha_s N_c}{2\pi^2} \Sperp \int \frac{\dd^2\rperp}{2\pi} e^{-i\kperp\cdot\rperp} S^{(2)} (\rperp) S^{(2)} (\rperp) \ln^2 \frac{\kperp^2 \rperp^2}{c_0^2}, \\
L_{g2} (\kperp) = &
\frac{\alpha_s N_c}{\pi} \Sperp \int \dd^2\qa \dd^2 \qb F(\kperp - \qa-\qb) F(\kperp -\qb) F(\qb) \ln^2\frac{\kperp^2}{\qa^2}, \\
L_{g3} (\kperp) = &
- \frac{2\alpha_s N_c}{\pi^2} \Sperp \int \dd^2\qa \dd^2\qb \dd^2\qc F(\qc) F(\kperp - \qb -\qc) F(\kperp-\qa-\qc) \frac{\qa \cdot \qb}{\qa^2\qb^2} \ln \frac{\kperp^2}{\qa^2}.
\end{align}
Using the same procedure laid out in the $q\to q$ channel, we can convert the above cross-section into the expressions in the momentum space as follows
\begin{equation}
\frac{\dd \sigma_{gg}}{\dd y\dd^{2}\pperp}
=
\frac{\dd \sigma^\text{LO}_{gg}}{\dd y\dd^{2}\pperp}
+
\frac{\dd\sigma^\text{NLO}_{gg}}{\dd y\dd^{2}\pperp}
=
\frac{\dd \sigma^\text{LO}_{gg}}{\dd y\dd^{2}\pperp}
+
\sum_{i=1}^6
\frac{\dd\sigma_{gg}^i}{\dd y\dd^{2}\pperp},
\end{equation}
with the LO and NLO cross-sections defined as
\begin{align}
\frac{\dd \sigma^\text{LO}_{gg}}{\dd y\dd^{2}\pperp}
= &
\Sperp  \int_{\tau}^{1}\frac{\dd z}{z ^2} x_{p} g(x_{p},\mu^{2})D_{h/g}(z,\mu^{2})  \int \dd^2 \qa F(\qa) F(\kperp-\qa),
\\
\frac{\dd\sigma_{gg}^1}{\dd y\dd^{2}\pperp}
=
&
\frac{\alpha_s}{2\pi} N_c \Sperp \int_{\tau}^1 \frac{\dd z}{z^2} \int_{\tau/z}^1 d\xi \int \dd^2 \qa x g(x,\mu^2) D_{h/g} (z,\mu^2) \mathcal{P}_{gg} (\xi) \nonumber\\
& \times
\ln\frac{\Lambda^2}{\mu^2} 
\Bigl[
F(\kperp-\qa) + \frac{1}{\xi^2} F(\kperp/\xi-\qa) 
\Bigr]F(\qa),
\\
\frac{\dd\sigma_{gg}^2}{\dd y\dd^{2}\pperp}
=
&
\frac{\alpha_s}{2\pi} N_c \Sperp 
4\beta_0 \int_{\tau}^1 \frac{\dd z}{z^2} \int \dd^2 \qa xg(x,\mu^2) D_{h/g} (z,\mu^2) 
F(\kperp-\qa) F(\qa)
\ln \frac{\kperp^2}{\Lambda^2},
\\
\frac{\dd\sigma_{gg}^3}{\dd y\dd^{2}\pperp}
=
&
\frac{\alpha_s}{\pi^2} N_c \Sperp
\int_\tau^1 \frac{\dd z}{z^2} \int_{\tau/z}^1 \dd \xi \int \dd^2 \qa \dd^2 \qb \dd^2 \qc xg(x,\mu^2) D_{h/g}(z,\mu^2) \frac{[1-\xi(1-\xi)]^2}{\xi(1-\xi)_+} 
\nonumber\\
&
\times \mathcal{T}_{gg}^{(1)} (\xi,\qa,\qb,\qc,\kperp),
\\
\frac{\dd\sigma_{gg}^4}{\dd y\dd^{2}\pperp}
=
& -
2N_f T_R \frac{\alpha_s}{2\pi} \Sperp \int_\tau^1 \frac{\dd z}{z^2} \int_0^1 \dd\xi' \dd^2 \qa x_p g(x_p,\mu^2) D_{h/g}(z,\mu^2) [\xi'^2 + (1-\xi')^2] 
\nonumber\\
&
\times
F(\qa) F(\kperp-\qa) \ln \frac{(\qa-\xi'\kperp)^2}{\kperp^2},
\\
\frac{\dd\sigma_{gg}^5}{\dd y\dd^{2}\pperp}
=
&
- 4N_c \frac{\alpha_s}{2\pi} \Sperp \int_\tau^1 \frac{\dd z}{z^2} \int_0^1 \dd\xi' \dd^2\qa \dd^2\qb x_p g(x_p,\mu^2) D_{h/g}(z,\mu^2) \left[ \frac{\xi'}{(1-\xi')_+} + \frac{1}{2} \xi' (1-\xi') \right] 
\nonumber\\
& 
F(\qa) F(\qb) F(\kperp-\qa) \ln \frac{(\qa+\qb-\xi'\kperp)^2}{\kperp^2}, 
\\
\frac{\dd\sigma_{gg}^6}{\dd y\dd^{2}\pperp}
=& 
\frac{2\alpha_s}{\pi^2} N_c \Sperp 
\int_\tau^1 \frac{\dd z}{z^2} \int \dd^2 \qa \int \dd^2 \qb x_p g(x_p,\mu^2) D_{h/q} (z,\mu^2)
\frac{1}{\qb^2} \ln\frac{\kperp^2}{\qb^2} F(\kperp-\qa) 
\nonumber\\
&
\times [F(\qa+\qb) - \theta(\kperp^2 - \qb^2) F(\qa)] 
\nonumber\\
+& 
\frac{\alpha_s}{\pi} N_c \Sperp
\int_\tau^1 \frac{\dd z}{z^2} \int \dd^2 \qa \int \dd^2 \qb x_p g(x_p,\mu^2) D_{h/q} (z,\mu^2)
F(\qa) F(\qb) F(\kperp-\qb)
\nonumber\\
&
\times \ln^2 \frac{\kperp^2}{(\qa+\qb-\kperp)^2}
\nonumber \\
- & 
\frac{2\alpha_s}{\pi^2} N_c \Sperp 
\int_\tau^1 \frac{\dd z}{z^2} \int \dd^2 \qa \int \dd^2 \qb \int \dd^2 \qc x_p g(x_p,\mu^2) D_{h/q} (z,\mu^2) F(\qa) F(\qb) F(\qc)
\nonumber \\
& \times 
\frac{(\kperp - \qa + \qc) \cdot (\kperp - \qb + \qc)}{(\kperp - \qa + \qc)^2 (\kperp - \qb + \qc)^2}
\ln \frac{\kperp^2}{(\kperp - \qa + \qc)^2},
\end{align}
where $\mathcal{P}_{gg} (\xi) = 2 \left[\frac{\xi}{(1-\xi)_+}+\frac{1-\xi}{\xi}+\xi(1-\xi)\right] + \left(\frac{11}{6} - \frac{2 N_f T_R}{3N_c}\right) \delta(1-\xi)$ and
\begin{align}
\mathcal{T}_{gg}^{(1)} (\xi,\qa,\qb,\qc,\kperp) =
&
\frac{1}{\xi^2} \frac{[(1-\xi)\qa +\qc - \xi\qb]^2}{(\qa+\qc-\kperp)^2 (\qa + \qb - \kperp/\xi)^2} F(\qa) F(\qb) F(\qc)
\nonumber\\
- & 
\frac{1}{(\qa+\qc-\kperp)^2} \frac{\Lambda^2}{\Lambda^2 + (\qa+\qc-\kperp)^2} F(\kperp-\qc) F(\qb) F(\qc) 
\nonumber \\
- &
\frac{1}{\xi^2} \frac{1}{(\qa + \qb - \kperp/\xi)^2} \frac{\Lambda^2}{\Lambda^2 + (\qa + \qb - \kperp/\xi)^2} F(\kperp/\xi-\qb) F(\qb) F(\qc).
\end{align}
Similar to the case for $q\to q$ channel, the $\Lambda$-dependence in $\sigma_{gg}^{1}$, $\sigma_{gg}^{2}$ and $\sigma_{gg}^{3}$ cancels when they are summed. We obtain the $\ln \frac{\Lambda^2}{\mu^2}$ logarithm in $\sigma_{gg}^{1}$ associated with the collinear divergence and the single Sudakov logarithm in terms of $\ln\frac{\kperp^2}{\Lambda^2}$ in $\sigma_{gg}^{2}$. The resummation of these logarithms will be presented in the next section.

Furthermore, we can extract the Sudakov double logarithm from $\sigma_{gg}^{6}$. We have numerically checked that the ratio of $L_{g} \equiv L_{g1} + L_{g2} + L_{g3}$ to $L_{q}\equiv L_{q1} + L_{q2} + L_{q3}$ is about $2$ at large $\kperp$ limit. Therefore, we identify the coefficient of the double logarithm as $\frac{\alpha_s}{2\pi} N_c$, which is twice as large as that in the $q\to q$ channel at the large $N_c$ limit. Similar to the quark case, according to the counting rule, the corresponding coefficient for the $g\to g$ channel double logarithm is $2\times \frac{\alpha_s N_c}{4\pi}$, since both initial state and final state gluon emissions conttribute. Therefore, the obtained coefficient also agrees with the counting rule in Refs.~\cite{Mueller:2013wwa,Sun:2014gfa}. By decomposing the theta function the same way discussed in the last subsection, we have
\begin{align}
\frac{\dd\sigma_{gg}^6}{\dd y\dd^{2}\pperp}
=& 
\frac{\dd\sigma_{gg}^{6a}}{\dd y\dd^{2}\pperp}
+
\frac{\dd\sigma_{gg}^{6b}}{\dd y\dd^{2}\pperp},
\end{align}
where
\begin{align}
\frac{\dd\sigma_{gg}^{6a}}{\dd y\dd^{2}\pperp}
=& 
- \frac{\alpha_s}{2\pi} N_c \Sperp 
\int_\tau^1 \frac{\dd z}{z^2} x_p g(x_p,\mu^2) D_{h/q} (z,\mu^2)
\ln^2 \frac{\kperp^2}{\Lambda^2} \int \dd^2 \qa F(\kperp-\qa)F(\qa),
\\
\frac{\dd\sigma_{gg}^{6b}}{\dd y\dd^{2}\pperp}
=& 
\frac{2\alpha_s}{\pi^2} N_c \Sperp 
\int_\tau^1 \frac{\dd z}{z^2} \int \dd^2 \qa \int \dd^2 \qb x_p g(x_p,\mu^2) D_{h/q} (z,\mu^2)
\frac{1}{\qb^2} \ln\frac{\kperp^2}{\qb^2} F(\kperp-\qa) 
\nonumber\\
&
\times [F(\qa+\qb) - \theta(\Lambda^2 - \qb^2) F(\qa)] 
\nonumber\\
- & \frac{\alpha_s}{2\pi} N_c \Sperp 
\int_\tau^1 \frac{\dd z}{z^2} x_p g(x_p,\mu^2) D_{h/q} (z,\mu^2)
\ln^2 \frac{\kperp^2}{\Lambda^2} \int \dd^2 \qa F(\kperp-\qa)F(\qa) \nonumber\\
+& 
\frac{\alpha_s}{\pi} N_c \Sperp
\int_\tau^1 \frac{\dd z}{z^2} \int \dd^2 \qa \int \dd^2 \qb x_p g(x_p,\mu^2) D_{h/q} (z,\mu^2)
F(\qa) F(\qb) F(\kperp-\qb)
\nonumber\\
&
\times \ln^2 \frac{\kperp^2}{(\qa+\qb-\kperp)^2}
\nonumber \\
- & 
\frac{2\alpha_s}{\pi^2} N_c \Sperp 
\int_\tau^1 \frac{\dd z}{z^2} \int \dd^2 \qa \int \dd^2 \qb \int \dd^2 \qc x_p g(x_p,\mu^2) D_{h/q} (z,\mu^2) F(\qa) F(\qb) F(\qc)
\nonumber \\
& \times 
\frac{(\kperp - \qa + \qc) \cdot (\kperp - \qb + \qc)}{(\kperp - \qa + \qc)^2 (\kperp - \qb + \qc)^2}
\ln \frac{\kperp^2}{(\kperp - \qa + \qc)^2}.
\end{align}
$\sigma_{gg}^{6a}$ will be resummed in the Sudakov resummation and $\sigma_{gg}^{6b}$ contributes to the finite NLO corrections. 

\subsection{$q \to g$ channel}

For the $q \to g$ channel, there is no LO contribution. We have cross-section in the coordinate space as
\begin{align}
\frac{\dd\sigma_{gq}}{\dd y\dd^{2}\pperp}
=
\frac{\dd\sigma^\text{NLO}_{gq}}{\dd y\dd^{2}\pperp}
= 
\sum_{i=a}^c
\frac{\dd\sigma_{gq}^i}{\dd y\dd^{2}\pperp},
\end{align}
where,
\begin{align}
\frac{\dd\sigma_{gq}^a}{\dd y\dd^{2}\pperp}
= &
\frac{\alpha_s}{2\pi} \Sperp C_F \int_\tau^1 \frac{\dd z}{z^2} \int_{\tau/z}^1 \dd\xi xq(x,\mu^2) D_{h/g} (z,\mu^2) 
\int \frac{\dd^2\rperp}{(2\pi)^2} e^{-i\frac{1}{\xi} \kperp\cdot\rperp} S^{(2)} (\rperp) 
\nonumber\\
&\times
\frac{1}{\xi^2} \mathcal{P}_{gq} (\xi) \ln \frac{c_0^2}{\rperp^2\mu^2},
\\
\frac{\dd\sigma_{gq}^b}{\dd y\dd^{2}\pperp}
= &
\frac{\alpha_s}{2\pi} \Sperp C_F \int_\tau^1 \frac{\dd z}{z^2} \int_{\tau/z}^1 \dd\xi xq(x,\mu^2) D_{h/g} (z,\mu^2) 
\int \frac{\dd^2\rperp}{(2\pi)^2} e^{-i\kperp\cdot\rperp} S^{(2)} (\rperp) S^{(2)} (\rperp) 
\nonumber\\
&\times
\mathcal{P}_{gq} (\xi) \ln \frac{c_0^2}{\rperp^2\mu^2},
\\
\frac{\dd\sigma_{gq}^c}{\dd y\dd^{2}\pperp}
= &
8\pi \Sperp C_F \frac{\alpha_s}{2\pi}
\int_\tau^1 \frac{\dd z}{z^2} \int_{\tau/z}^1 \dd\xi xq(x,\mu^2) D_{h/g} (z,\mu^2) 
\int\frac{\dd^2\uperp\dd^2\vperp}{(2\pi)^4} 
e^{-i\frac{1}{\xi}\kperp\cdot(\uperp-\vperp)} e^{-i\kperp\cdot\vperp}
\nonumber\\
&\times
S^{(2)} (\uperp) S^{(2)} (\vperp) \mathcal{P}_{gq} (\xi) \frac{1}{\xi} \frac{(\uperp-\vperp) \cdot \vperp}{(\uperp-\vperp)^2 \vperp^2},
\end{align}
where $\mathcal{P}_{gq} (\xi) = \frac{1}{\xi} [1+(1-\xi)^2]$.

After the Fourier transform, we obtain
\begin{align}
\frac{\dd\sigma_{gq}}{\dd y\dd^{2}\pperp}
=
\frac{\dd\sigma^\text{NLO}_{gq}}{\dd y\dd^{2}\pperp}
= 
\sum_{i=1}^3
\frac{\dd\sigma_{gq}^i}{\dd y\dd^{2}\pperp},
\end{align}
where
\begin{align}
\frac{\dd\sigma_{gq}^1}{\dd y\dd^{2}\pperp} = 
& 
\frac{\alpha_s}{2\pi} C_F \Sperp
\int_\tau^1 \frac{\dd z}{z^2} \int_{\tau/z}^1 \dd \xi \int \dd^2 \qa
x q(x,\mu^2) D_{h/g} (z,\mu^2) \mathcal{P}_{gq} (\xi) \ln \frac{\Lambda^2}{\mu^2}
F(\qa) F(\kperp-\qa),
\\
\frac{\dd\sigma_{gq}^2}{\dd y\dd^{2}\pperp} = 
& 
\frac{\alpha_s}{2\pi} C_F \Sperp
\int_\tau^1 \frac{\dd z}{z^2} \int_{\tau/z}^1 \dd \xi
x q(x,\mu^2) D_{h/g} (z,\mu^2) \mathcal{P}_{gq} (\xi) \ln \frac{\Lambda^2}{\mu^2}
\frac{1}{\xi^2} F(\kperp/\xi),
\\
\frac{\dd\sigma_{gq}^3}{\dd y\dd^{2}\pperp} = 
&
\frac{\alpha_s}{2\pi^2} C_F \Sperp
\int_\tau^1 \frac{\dd z}{z^2} \int_{\tau/z}^1 \dd \xi \int \dd^2 \qa \int \dd^2\qb
x q(x,\mu^2) D_{h/g} (z,\mu^2) \mathcal{P}_{gq} (\xi) 
\mathcal{T}_{gq}^{(1)} (\xi, \qa,\qb,\kperp),
\end{align}
with
\begin{align}
\mathcal{T}_{gq}^{(1)} (\xi, \qa,\qb,\kperp) 
= 
&
\left( \frac{\kperp -\qa - \qb}{(\kperp -\qa - \qb)^2} - \frac{\kperp - \xi \qb}{(\kperp - \xi\qb)^2} \right)^2
F(\qa) F(\qb)
\nonumber\\
&
- \frac{\Lambda^2}{\Lambda^2 + (\kperp -\qa - \qb)^2} \frac{1}{(\kperp -\qa - \qb)^2} 
F(\qb) F(\kperp-\qb)
\nonumber\\
& 
- \frac{\Lambda^2}{\Lambda^2 + (\kperp/\xi-\qb)^2} \frac{1}{(\kperp-\xi\qb)^2} F(\qa) F(\kperp/\xi).
\end{align}
For this channel, we obtain the $\ln\frac{\Lambda^2}{\mu^2}$ logarithm which is associated with the collinear divergence in $\sigma_{gq}^{1}$ and $\sigma_{gq}^{2}$. The Sudakov double logarithm ($\ln^2\frac{\kperp^2}{\Lambda^2}$) and single logarithm ($\ln\frac{\kperp^2}{\Lambda^2}$) do not appear in the off-diagonal channels.

\subsection{$g \to q$ channel}

To complete the calculation for all the channels, we should also compute the $g\to q\bar{q}$ channel. The cross-section of the $g \to q$ channel in the coordinate space is
\begin{align}
\frac{\dd\sigma_{qg}}{\dd y\dd^2 \pperp} = \sum_{i=a}^c \frac{\dd\sigma_{qg}^i}{\dd y\dd^2 \pperp},
\end{align}
where
\begin{align}
\frac{\dd\sigma_{qg}^a}{\dd y\dd^2 \pperp}
= &
\frac{\alpha_s}{2\pi} \Sperp T_R \int_\tau^1 \frac{\dd z}{z^2} \int_{\tau/z}^1 \dd \xi x g(x,\mu^2) D_{h/q} (z,\mu^2) 
\int \frac{\dd^2\rperp}{(2\pi)^2} e^{-i\kperp\cdot\rperp} S^{(2)} (\rperp) 
\nonumber\\
& \times
\mathcal{P}_{qg} (\xi) 
\left[ 
\ln\frac{c_0^2}{\rperp^2\mu^2} - 1
\right],
\\
\frac{\dd\sigma_{qg}^b}{\dd y\dd^2 \pperp}
= &
\frac{\alpha_s}{2\pi} \Sperp T_R \int_\tau^1 \frac{\dd z}{z^2} \int_{\tau/z}^1 \dd \xi x g(x,\mu^2) D_{h/q} (z,\mu^2) 
\int \frac{\dd^2\rperp}{(2\pi)^2} e^{-i\frac{1}{\xi}\kperp\cdot\rperp} S^{(2)} (\rperp) S^{(2)} (\rperp) 
\nonumber\\
& \times
\frac{1}{\xi^2} \mathcal{P}_{qg} (\xi) 
\left[ 
\ln\frac{c_0^2}{\rperp^2\mu^2} - 1
\right],
\\
\frac{\dd\sigma_{qg}^c}{\dd y\dd^2 \pperp}
= &
8\pi \Sperp T_R \frac{\alpha_s}{2\pi}  \int_\tau^1 \frac{\dd z}{z^2} \int_{\tau/z}^1 \dd \xi x g(x,\mu^2) D_{h/q} (z,\mu^2) 
\int \frac{\dd^2\uperp\dd^2\vperp}{(2\pi)^4} e^{-i\kperp\cdot (\uperp-\vperp) -i\frac{1}{\xi} \kperp\cdot\vperp}
\nonumber\\
& \times
\frac{1}{\xi} \mathcal{P}_{qg} (\xi) S^{(2)} (\uperp) S^{(2)} (\vperp) 
\frac{(\uperp-\vperp) \cdot \vperp}{(\uperp-\vperp)^2 \vperp^2}.
\end{align}
Here $\mathcal{P}_{qg} (\xi) = \xi^2 + (1-\xi)^2$. The cross-section in the momentum space reads
\begin{align}
\frac{\dd\sigma_{qg}}{\dd y\dd^{2}\pperp}
= 
\sum_{i=1}^5
\frac{\dd\sigma_{qg}^i}{\dd y\dd^{2}\pperp},
\end{align}
with
\begin{align}
\frac{\dd\sigma_{qg}^1}{\dd y\dd^{2}\pperp}
=
&
\frac{\alpha_s}{4\pi} \Sperp
\int_\tau^1 \frac{\dd z}{z^2} \int_{\tau/z}^1 \dd \xi 
x g(x,\mu^2) D_{h/q}(z,\mu^2) \mathcal{P}_{qg} (\xi) \ln \frac{\Lambda^2}{\mu^2} 
F(\kperp),
\\
\frac{\dd\sigma_{qg}^2}{\dd y\dd^{2}\pperp}
=
&
\frac{\alpha_s}{4\pi} \Sperp
\int_\tau^1 \frac{\dd z}{z^2} \int_{\tau/z}^1 \dd \xi \int \dd^2 \qa
x g(x,\mu^2) D_{h/q}(z,\mu^2) \frac{1}{\xi^2} \mathcal{P}_{qg} (\xi) \ln \frac{\Lambda^2}{\mu^2} 
F(\qa) F(\kperp/\xi-\qa),
\\
\frac{\dd\sigma_{qg}^3}{\dd y\dd^{2}\pperp}
=
&
-\frac{\alpha_s}{4\pi} \Sperp
\int_\tau^1 \frac{\dd z}{z^2} \int_{\tau/z}^1 \dd \xi 
x g(x,\mu^2) D_{h/q}(z,\mu^2) \mathcal{P}_{qg} (\xi) 
F(\kperp),
\\
\frac{\dd\sigma_{qg}^4}{\dd y\dd^{2}\pperp}
=
&
-\frac{\alpha_s}{4\pi} \Sperp
\int_\tau^1 \frac{\dd z}{z^2} \int_{\tau/z}^1 \dd \xi \int \dd^2 \qa
x g(x,\mu^2) D_{h/q}(z,\mu^2) \frac{1}{\xi^2} \mathcal{P}_{qg} (\xi) 
F(\qa) F(\kperp/\xi-\qa),
\\
\frac{\dd\sigma_{qg}^5}{\dd y\dd^{2}\pperp}
= 
&
\frac{\alpha_s}{4\pi^2} \Sperp
\int_\tau^1 \frac{\dd z}{z^2} \int_{\tau/z}^1 \dd \xi \int \dd^2 \qa \int \dd^2 \qb
xg(x,\mu^2) D_{h/q}(z,\mu^2) \mathcal{P}_{qg} (\xi)
\mathcal{T}_{qg}^{(1)} (\xi, \qa,\qb,\kperp),
\end{align}
where 
\begin{align}
\mathcal{T}_{qg}^{(1)} (\xi, \qa,\qb,\kperp)
=
& 
\left( 
\frac{\kperp- \xi\qa -\xi\qb}{(\kperp - \xi\qa -\xi\qb)^2} 
- \frac{\kperp-\qb}{(\kperp-\qb)^2}
\right)^2
F(\qa) F(\qb) 
\nonumber\\
& 
- \frac{1}{(\kperp - \xi\qa -\xi\qb)^2}\frac{\Lambda^2}{\Lambda^2 + (\kperp/\xi - \qa - \qb)^2}
F(\qb) F(\kperp/\xi-\qb)
\nonumber\\
&
- \frac{1}{(\kperp-\qb)^2} \frac{\Lambda^2}{\Lambda^2 + (\kperp-\qb)^2}
F(\qa) F(\kperp).
\end{align}
Again, we obtain the $\ln \frac{\Lambda^2}{\mu^2}$ logarithm in $\sigma_{qg}^{1}$ and $\sigma_{qg}^{2}$ which will be resummed making use of the DGLAP evolution equation. The soft logarithms do not contribute in the off-diagonal channels.

\section{Kinematics near the threshold}
\label{sec:tau}

In this section, we discuss the kinematics of the inclusive hadron production in forward $pA$ collisions and illustrate the kinematic region where the threshold resummation becomes important. The longitudinal momentum fraction of the produced hadron $\tau$, which is defined as $\tau=\frac{p_T}{\sqrt{s}}e^y$, is the key physical quantity. It can also be written as $\tau=xz\xi$, where $x$ is the momentum fraction carried by the projectile parton, $z$ is the momentum fraction carried by the final state hadron and $\xi$ is the momentum fraction in the partonic splitting. We show the plots of $\tau$ as a function of $y$ and $p_T$ at the RHIC and LHC energies in Fig.~\ref{fig:tau-plot}.

\begin{figure}[htb]
\includegraphics[width=0.4\textwidth]{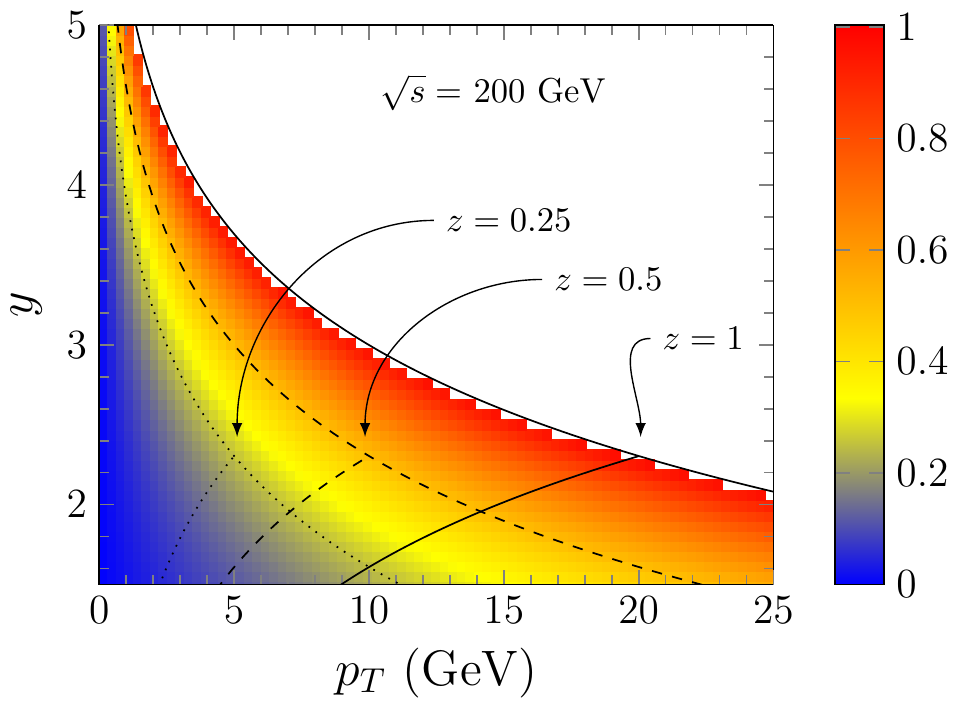}
\includegraphics[width=0.4\textwidth]{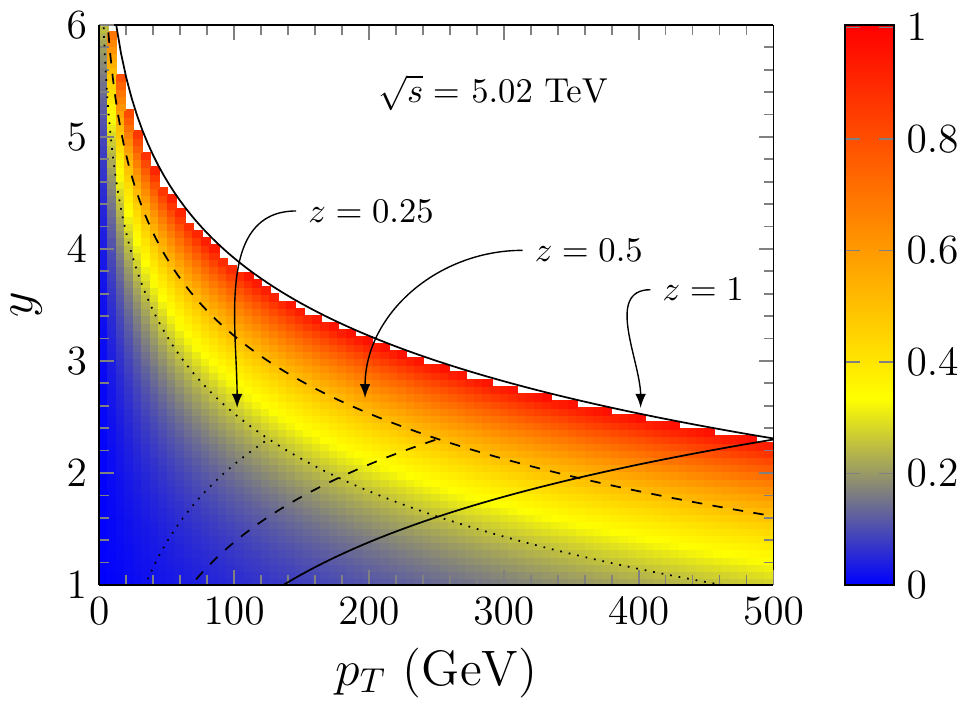}
\caption{The plots of $\tau$ as a function of $y$ and $p_T$ at two different collision energies ($\sqrt{s}$). Various values of $\tau$ are represented by different colors as shown above. The rapidity $y$ is defined in the center of the mass frame of the two colliding hadrons.}
\label{fig:tau-plot}
\end{figure}

These two plots illustrate the allowed ranges of $p_T$ at fixed rapidity $y$ at both RHIC and the LHC energies. The upper-most solid line in Fig.~\ref{fig:tau-plot} represents the boundary determined by the kinematic threshold, which is given by $\tau= \frac{p_T}{\sqrt{s}}e^y=1$ and $z=1$. The region above this boundary is forbidden by kinematics and it implies that the rapidity or $p_T$ of the produced hadron can not be exceedingly large. When $\tau$ is near $1$, $x$, $z$ and $\xi$ are all forced to be near $1$, namely the kinematic threshold. The region in red and yellow in Fig.~\ref{fig:tau-plot} indicates where the threshold resummation is expected to be important. 

Meanwhile, the dilute-dense factorization used in the NLO hadron production calculation assumes that the small-$x$ gluon in the nuclear target is dense enough and the CGC formalism can apply. Since the small-$x$ gluon distribution and the dipole scattering amplitude usually start from the initial condition at $x_g=0.01$ and evolve to smaller $x$ regime, in order to justify the small-$x$ assumptions used in the CGC formalism, we usually need to require $x_g = \frac{p_T}{z\sqrt{s}}e^{-y} \leq 0.01$ with $x_g$ being the momentum fraction of the target gluon. The lower-most solid line is corresponding to the boundary given by $x_g \leq 0.01$ beyond which the dilute-dense factorization no longer applies. It is interesting to note that these two boundaries always intersect at $y_0 =\frac{1}{2}\ln 100 =2.3$. This implies that for rapidities $y\geq y_0$, the gluon inside the target hadron is always located in the small-$x$ regime. Furthermore, in the regime where $0<y <y_0$, the CGC calculation should only be applied to the low $p_T$ region. Additional numerical results are presented in the end and they also indicate that the CGC calculation is expected to break down in the high $p_T$ region when the rapidity $y$ is not forward enough. 

It is worth noting that these two boundaries moves to the left at smaller value of $z$ when the kinematic constraint requires $x_p\equiv \frac{p_T}{z\sqrt{s}}e^y = \xi x \leq 1$. The dashed and dotted lines represent the resulting boundaries at $z=0.5$ and $z=0.25$, respectively. In these cases, $x$ and $\xi$ can approach $1$ with fixed values of $z$. Clearly, due to the initial state soft gluon radiation, both the Sudakov resummation and the collinear threshold resummation with respect to the PDF must be taken into account even when $z$ is not large. Similarly, one can imagine that the Sudakov resummation and the collinear threshold resummation with respect to the FF become important when $\xi$ and $z$ are kept large while $x$ is far away from $1$.

Throughout this paper except the discussion in Sec.~\ref{sec:applicable-range}, we set the small-$x$ gluon distribution to be $0$ when $x_g$ is larger than $0.01$. This prescription, referred as the ``fixed boundary condition'', effectively removes all the contribution from the kinematic regime where dilute-dense factorization does not apply. In Sec.~\ref{sec:applicable-range}, we also present another prescription which freezes ${\cal F}^{x_g} (\kperp)$ at $x_g=0.01$. The numerical results show little difference between these two presciptions at small-$p_T$. While the visible difference between these two prescriptions can be found for the large $p_T$ events measured in the middle rapidity region, we expect that this kinematic region is no longer within the applicable regime of the CGC formalism and the dilute-dense factorization. 

\section{The Resummation of the collinear logarithms}

As mentioned previously in the main text, there are two types of threshold logarithms. The first type is proportional to the logarithm $\alpha_s \ln\Lambda^2/\mu^2 $ and the corresponding partonic splitting function as shown in the above-mentioned NLO corrections (see $\sigma_{qq}^{1}$, $\sigma_{gg}^{1}$, $\sigma_{gq}^{1}$ and $\sigma_{qg}^{1}$), and it is associated with the collinear branching of partons. The second type is proportional to either $\alpha_s \ln k_\perp^2/\Lambda^2 $ or $\alpha_s \ln^2 k_\perp^2/\Lambda^2$, and this type originates from the soft emission of gluons near the kinematic threshold. We first address the issue of the resummation of the collinear part in this section, and then take care of the soft logarithms via the Sudakov factor in the next section. 

\subsection{Resummation of the collinear logarithms via the DGLAP evolution}
\label{sec:dglap-resum}

Following the same idea proposed in Ref.~\cite{Xiao:2018zxf}, we can resum the collinear part of the threshold logarithms~\cite{Becher:2006qw, Becher:2006nr,Becher:2006mr} with the help of the DGLAP evolution, by setting the factorization scale $\mu$ to be $\Lambda$. To deal with the first term of $\sigma_{qq}^{1}$, the first term of $\sigma_{gg}^{1}$, $\sigma_{gq}^{1}$ and $\sigma_{qg}^{1}$, we apply the following replacement
\begin{align}
\left[
\begin{array}{c}
q\left(x_p,\mu \right) \\
g\left(x_p,\mu \right)
\end{array}
\right] 
+ \frac{\alpha_s}{2\pi} \ln\frac{\Lambda^2}{\mu^2} \int_{x_p}^1 \frac{\dd\xi}{\xi}
\left[
\begin{array}{cc}
C_F \mathcal{P}_{qq} (\xi) & T_R \mathcal{P}_{qg} (\xi) \\
C_F \mathcal{P}_{gq} (\xi) & N_C \mathcal{P}_{gg} (\xi)
\end{array}
\right]
\left[
\begin{array}{c}
q\left(x_p/\xi,\mu \right) \\
g\left(x_p/\xi,\mu \right)
\end{array}
\right]
\Rightarrow
\left[
\begin{array}{c}
q\left(x_p,\Lambda \right) \\
g\left(x_p,\Lambda \right)
\end{array}
\right]. 
\end{align}
Upon the above replacement, the threshold logarithm $\ln\frac{\Lambda^2}{\mu^2}$ combined with the corresponding splitting function effectively evolves the factorization scale of the PDFs in the LO cross-section from $\mu^2$ back to $\Lambda^2$. 

The same procedure also can be applied to the FF part. To see this more clearly, we need to rewrite $\sigma_{qq}^{\rm LO}$ and the second term of $\sigma_{qq}^{1}$ as
\begin{align}
\frac{\dd \sigma^{\rm LO}_{qq}}{\dd y\dd^2\pperp} = & 
\Sperp  \int_\tau^1 \frac{\dd z}{z^2} x_p q (x_p,\mu^2)D_{h/q}(z,\mu^2)  F(\kperp) \Bigl|_{x_p=\frac{\tau}{z}, \kperp = \frac{\pperp}{z}},
\\
\frac{\dd\sigma_{qq}^{1'}}{\dd y\dd^{2}\pperp} 
=&
\frac{\alpha_{s}}{2\pi}C_{F}\Sperp
\int \dd \xi \int \frac{\dd z'}{z'^{2}} \theta(z'-\tau) \theta(1-z') \theta(\xi-\frac{\tau}{z'}) \theta(1-\xi) \nonumber\\
&\times x q(x,\mu^{2})D_{h/q}(z',\mu^{2})
\mathcal{P}_{qq}(\xi) 
\ln\frac{\Lambda^{2}}{\mu^{2}}\frac{1}{\xi^{2}}F(\frac{\kperp}{\xi})
\Bigl|_{x=\frac{\tau}{z'\xi}, \kperp =\frac{\pperp}{z'}}
\nonumber\\
=&
\frac{\alpha_{s}}{2\pi}C_{F}\Sperp
\int_{\tau}^{1} \frac{\dd z}{z^{2}}  x q(x,\mu^{2})  F(\frac{\pperp}{z})\ln\frac{\Lambda^{2}}{\mu^{2}} \int_z^1 \frac{\dd \xi}{\xi} D_{h/q}(\frac{z}{\xi},\mu^{2})
\mathcal{P}_{qq}(\xi) 
\Bigl|_{x=\frac{\tau}{z},z=z'\xi}
.\label{eq:sigqq1sec}
\end{align}
The first line of Eq.~(\ref{eq:sigqq1sec}) is exactly the same as the second term of Eq.~(\ref{eq:sigqq1fir}), albeit in a slightly different form. From the first line of Eq.~(\ref{eq:sigqq1sec}) to the second line, we changed the variable $z$ to $z'\xi$. It is then apparent that we can resum the second term of $\sigma_{qq}^{1}$, the second term of $\sigma_{gg}^{1}$, $\sigma_{gq}^{2}$ and $\sigma_{qg}^{2}$ through the DGLAP evolution of the FFs through the following replacement
\begin{align}
\left[
\begin{array}{c}
D_{h/q}\left(z,\mu \right) \\
D_{h/g}\left(z,\mu \right)
\end{array}
\right] 
+ \frac{\alpha_s}{2\pi} \ln\frac{\Lambda^2}{\mu^2} \int_{z}^1 \frac{\dd\xi}{\xi}
\left[
\begin{array}{cc}
C_F \mathcal{P}_{qq} (\xi) & C_F \mathcal{P}_{gq} (\xi) \\
T_R \mathcal{P}_{qg} (\xi) & N_C \mathcal{P}_{gg} (\xi)
\end{array}
\right]
\left[
\begin{array}{c}
D_{h/q}\left(z/\xi,\mu \right) \\
D_{h/g}\left(z/\xi,\mu \right)
\end{array}
\right] 
\Rightarrow
\left[
\begin{array}{c}
D_{h/q}\left(z,\Lambda \right) \\
D_{h/g}\left(z,\Lambda \right)
\end{array}
\right].
\end{align}

To conclude, we have taken care of the resummation of the collinear threshold logarithms in the following NLO correction terms $\sigma_{qq}^1$, $\sigma_{gg}^1$, $\sigma_{gq}^1$, $\sigma_{gq}^2$, $\sigma_{qg}^1$ and $\sigma_{qg}^2$ by setting the factorization scales in $\sigma_{qq}^{\rm LO}$ and $\sigma_{gg}^{\rm LO}$ to be $\Lambda^2$. Since $\Lambda^2$ is usually smaller than $\mu^2$, we refer to this appoach as the reverse-evolution method in this paper.

\subsection{An alternative formulation of the threshold reummation}
\label{sec:rge-resum}

Alternatively, there is another analytical approach to resum the above mentioned collinear logarithms ($\alpha_s\ln \frac{\mu^2}{\Lambda^2}$) in the threshold limit. This approach is similar to the renormalization-group method first developed in Refs.~\cite{Becher:2006nr, Becher:2006mr} for DIS within the SCET framework. Our strategy is laid out as follows. First, we transform the terms which contain large logarithms of $\ln \frac{\mu^2}{\Lambda^2}$ into the Mellin space. Second, we resum the corresponding large logarithms in the Mellin space in the large-$N$ limit. In the end, we perform the inverse Mellin transform back to the momentum space. 

As a matter of fact, the analytical results obtained in this subsection are consistent with those in the SCET approach. Furthermore, we have checked that this alternative approach numerically also agrees well with the resummation method mentioned in the above subsection.

\subsubsection{Mellin Transform}

Let us use the $q\to q$ channel as an example to demonstrate how to perform Mellin transform and carry out resummation in the Mellin space. The derivation for the $g\to g$ channel is similar. Due to the existence of the endpoint singularity in the splitting functions $\mathcal{P}_{qq}$ and $\mathcal{P}_{gg}$ when $\xi \to 1$, the Mellin transform integral is dominated by the endpoint for sufficiently large $N$. In contrast, the off-diagonal splitting functions contain no plus-functions or $\delta$ functions. Therefore, the threshold effects from the off-diagonal channels are much smaller than those in the diagonal channels. 

The Mellin transform and the inverse Mellin transform are usually defined as
\begin{align}
& f(N) =\int_0^1 \dd x x^{N-1} f(x),   \\
& f(x)=\frac{1}{2\pi i}\int_{\mathcal{C}} \dd N x^{-N} f(N),  \label{eq:mellin}
\end{align}
where $\mathcal{C}$ stands for the proper contour which puts all the poles to its left. 

Following the same strategy developed in the last subsection, we resum the collinear logarithms associated with PDFs and FFs seperately. For the first term of $\sigma_{qq}^{1}$, we carry out the Mellin transform as follows
\begin{align}
\int_0^1 \dd x_p x_p^{N-1} \int_{x_p}^1 \frac{\dd\xi}{\xi} q(\frac{x_p}{\xi}) {\cal P}_{qq} (\xi)
= \int_0^1 \dd\xi \xi^{N-1} {\cal P}_{qq} (\xi) \int_0^1 \dd x x^{N-1} q(x)= {\cal P}_{qq}(N) q(N),
\end{align}
where $q(N) \equiv \int_0^1 \dd x x^{N-1} q(x)$ and $\mathcal{P}_{qq}(N) \equiv \int_0^1\dd\xi\xi^{N-1}\mathcal{P}_{qq}(\xi)$. Similarly, for the second term in $\sigma_{qq}^{1}$, we obtain
\begin{align}
\int_0^1 \dd z z^{N-1} \int_z^1 \frac{\dd\xi}{\xi} D_{h/q}(\frac{z}{\xi}) {\cal P}_{qq} (\xi) = {\cal P}_{qq}(N) D_{h/q}(N),
\end{align}
with $D_{h/q}(N) \equiv \int_0^1 \dd z z^{N-1}  D_{h/q}(z)$. Furthermore, we can evaluate ${\cal P}_{qq}(N)$ and find
\begin{align}
\mathcal{P}_{qq}(N) = -2 \gamma_E -2 \psi(N) + \frac{3}{2} - \frac{1}{N} - \frac{1}{N+1} 
= -2 \gamma_E -2 \ln N + \frac{3}{2} + {\cal O}(\frac{1}{N}),
\end{align}
where $\psi(N) = \ln N + {\cal O}(\frac{1}{N})$ is the polygamma function. We have taken the large-$N$ limit in the last step.

In the threshold limit, the resummation of the collinear logarithm in the $\sigma_{qq}^{1}$ term in the Mellin space results in an exponential \cite{Sterman:1986aj,Catani:1989ne, Catani:1996yz, deFlorian:2008wt,Bosch:2004th,Becher:2006qw, Becher:2006nr,Becher:2006mr}. It is worth mentioning that the corresponding contribution from the off-diagonal channels is suppressed in the large-$N$ limit. The resummed quark PDFs and FFs in the Mellin space can be cast into
\begin{align}
& q^{\rm res} (N) = q (N) \exp \left[ -\frac{\alpha_s}{\pi} C_F \ln\frac{\Lambda^2}{\mu^2} (\gamma_E-\frac{3}{4} +\ln N) \right], \\
& D_{h/q}^{\rm res} (N) = D_{h/q} (N)\exp \left[ -\frac{\alpha_s}{\pi} C_F \ln\frac{\Lambda^2}{\mu^2} (\gamma_E-\frac{3}{4} +\ln N) \right].
\end{align} 
Then we perform the inverse Mellin transform with respect to $q^{\rm res} (N)$ and get 
\begin{eqnarray}
q^{\rm res} (x_p,\Lambda ^2, \mu ^2   ) & = &  \int  _{  \mathcal C }  \frac{\dd N}{ 2\pi i}  x_p ^ {-N}  q ( N) \exp \left[ -\frac{\alpha_s}{\pi} C_F \ln\frac{\Lambda^2}{\mu^2} (\gamma_E-\frac{3}{4} +\ln N) \right] \nonumber \\
& =& 
\exp \left[ -\frac{\alpha_s}{\pi} C_F \ln\frac{\Lambda^2}{\mu^2} (\gamma_E-\frac{3}{4}) \right] \int ^1_0 \frac{\dd x  }{x} q(x, \mu ^2) \int _{\mathcal C}   \frac{\dd N}{2\pi i} \left(\frac{x}{x_p}\right)^{N}  \exp \left[ -\frac{\alpha _s}{\pi} C_F \ln\frac{\Lambda^2}{\mu^2} \ln N \right].
\end{eqnarray}
Using the following identity 
\begin{align}
& \int _{\mathcal C}  \frac{\dd N}{2\pi i}    \left(\frac{x}{x_p}\right)^{N} 
  e^{  -\gamma^q_{\Lambda, \mu} \ln N }
  =
  \frac{\theta\left(   x -x_p  \right) }{\Gamma( \gamma^q_{\Lambda, \mu}   )   } 
  \left( \ln \frac{x}{x_p }  \right)  ^{ \gamma^q_{\Lambda, \mu} -1  },   
&& \text{Re}\left[  \gamma^q_{\Lambda, \mu}  \right]  >0 ,  \label{gamma-identity}
\end{align}
with $\gamma^q_{\Lambda, \mu} = \frac{\alpha _s}{\pi} C_F \ln\frac{\Lambda^2}{\mu^2}$,  we reach the resummed expression for the quark distribution
\begin{align}
&   q^{\rm res} (x_p,\Lambda ^2, \mu ^2   ) 
  =  
   \frac{  e^{ - \gamma^q_{\Lambda, \mu} (\gamma_E-\frac{3}{4}) }  }{\Gamma( \gamma^q_{\Lambda, \mu}   )   }   
 \int ^1_{x_p} \frac{\dd x  }{x} q(x, \mu ^2)    
  \left( \ln \frac{x }{x_p}  \right)  ^{ \gamma^q_{\Lambda, \mu} -1  },
&& \text{Re}\left[  \gamma^q_{\Lambda, \mu}  \right]  >0.
\end{align}
Similarly, for the collinear threshold logarithm associated with the quark FF, we have
\begin{align}
&  D_{h/q}^{\rm res} (z,\Lambda ^2, \mu ^2   ) 
  =  
   \frac{  e^{ - \gamma^q_{\Lambda, \mu} (\gamma_E-\frac{3}{4}) }  }{\Gamma( \gamma^q_{\Lambda, \mu}   )   }   
 \int ^1_{z} \frac{\dd z'  }{z'}D_{h/q} (z', \mu ^2)    
  \left( \ln \frac{z'}{z}  \right)  ^{ \gamma^q_{\Lambda, \mu} -1  },
&& \text{Re}\left[  \gamma^q_{\Lambda, \mu}  \right]  >0.
\end{align}
In the running coupling case, the anomalous dimension $\gamma^q_{\Lambda, \mu}$ reads
\begin{equation}
\gamma^q_{\Lambda, \mu} = C_F \int ^{\Lambda^2} _{\mu ^2} \frac{\dd {\mu'}^2}{{\mu'}^2} \frac{\alpha_s ( {\mu'}^2 )}{\pi}.
\end{equation}

For the $g\to g$ channel, the color factor and the splitting function are different from those in the $q\to q$ channel. The Mellin transform of ${\cal P}_{gg} (\xi)$ is given by
\begin{align}
{\cal P}_{gg} (N) \equiv \int_0^1 \dd\xi \xi^{N-1} {\cal P}_{gg} (\xi) = 
- 2 \left[ \gamma_E + \psi(N) - \beta_0 - \frac{2}{N(N^2-1)} + \frac{1}{N+2} \right]
= - 2 \left[ \gamma_E - \beta_0 + \ln N \right] + {\cal O}(\frac{1}{N}),
\end{align}
where $\beta_0 = \frac{11}{12} - \frac{n_f}{6N_c}$ and we have taken large-$N$ limit in the last step. Therefore, for the gluon case, we obtain the following expressions for the resummed gluon PDF and FF
\begin{align}
& g^{\rm res} (x_p, \Lambda^2, \mu^2) = 
\frac{e^{-\gamma^g_{\Lambda,\mu} (\gamma_E-\beta_0)}}{\Gamma(\gamma^g_{\Lambda,\mu})} \int_{x_p}^1 \frac{\dd x}{x} g(x,\mu^2)
\left(\ln \frac{x}{x_p} \right)^{\gamma^g_{\Lambda,\mu} - 1},
&& \text{Re}\left[  \gamma^g_{\Lambda, \mu}  \right]  >0, 
\\
& D_{h/g}^{\rm res} (z, \Lambda^2, \mu^2) = 
\frac{e^{-\gamma^g_{\Lambda,\mu} (\gamma_E-\beta_0)}}{\Gamma(\gamma^g_{\Lambda,\mu})} \int_z^1 \frac{\dd z'}{z'} D_{h/g} (z', \mu^2)
\left( \ln \frac{z'}{z} \right)^{\gamma^g_{\Lambda,\mu} - 1},
&& \text{Re}\left[  \gamma^g_{\Lambda, \mu}  \right]  >0, 
\end{align}
where the gluon channel anomalous dimension reads
\begin{align}
\gamma^g_{\Lambda,\mu}=N_c \int_{\mu^2}^{\Lambda^2} \frac{\dd \mu'^2}{\mu'^2} \frac{\alpha_s (\mu'^2)}{\pi}.
\end{align}

\subsubsection{The forward threshold jet function}

Analogous to the jet function defined in Refs.~\cite{Becher:2006nr, Becher:2006mr}, we can also define the so-called forward threshold jet functions $\Delta^q (\Lambda^2,\mu^2,\omega)$ and $\Delta^g (\Lambda^2,\mu^2,\omega)$ in the quark and gluon channels, respectively. These two functions can be written as
\begin{align}
& \Delta^q (\Lambda^2,\mu^2,\omega) = \frac{e^{-\gamma^q_{\Lambda,\mu} (\gamma_E - \frac{3}{4})}}{\Gamma(\gamma^q_{\Lambda,\mu})} \omega^{\gamma^q_{\Lambda,\mu}-1},
&&
{\rm Re}\left[\gamma^q_{\Lambda,\mu}\right] >0,
\label{eq:jet-function-q}
\\
& \Delta^g (\Lambda^2,\mu^2,\omega) = \frac{e^{-\gamma^g_{\Lambda,\mu} (\gamma_E - \beta_0)}}{\Gamma(\gamma^g_{\Lambda,\mu})} \omega^{\gamma^g_{\Lambda,\mu}-1},
&&
{\rm Re}\left[\gamma^g_{\Lambda,\mu}\right] >0,
\label{eq:jet-function-g}
\end{align}
with $\omega \equiv\ln\frac{1}{\xi}$. Here the splitting fraction of the longitudinal momentum $\xi$ is $\frac{x_p}{x}$ for the initial state gluon emission and it should be identified as $\frac{z}{z'}$ in the case of final state gluon emission. The resummed PDFs and FFs derived in the last section can then be written as
\begin{align}
& q^{\rm res} (x_p, \Lambda^2, \mu^2) = \int_{x_p}^1 \frac{\dd x}{x} q(x,\mu^2) \Delta^q (\Lambda^2, \mu^2, \omega=\ln \frac{x}{x_p}),
\\
& g^{\rm res} (x_p, \Lambda^2, \mu^2) = \int_{x_p}^1 \frac{\dd x}{x} g(x,\mu^2) \Delta^g (\Lambda^2, \mu^2, \omega=\ln \frac{x}{x_p}),
\\
& D_{h/q}^{\rm res} (z,\Lambda ^2,\mu^2) = \int^1_{z} \frac{\dd z'}{z'}D_{h/q} (z',\mu^2) \Delta^q (\Lambda^2, \mu^2, \omega=\ln \frac{z}{z'}),
\\
& D_{h/g}^{\rm res} (z,\Lambda ^2,\mu^2) = \int^1_{z} \frac{\dd z'}{z'}D_{h/g} (z',\mu^2) \Delta^qg(\Lambda^2, \mu^2, \omega=\ln \frac{z}{z'}).
\end{align}

To connect and compare with the renormalization group approach in Refs.~\cite{Becher:2006nr, Becher:2006mr}, we differentiate Eqs.~(\ref{eq:jet-function-q}-\ref{eq:jet-function-g}) with respect to $\ln\mu^2$ and find 
\begin{align}
& 
\frac{\dd\Delta^q (\Lambda^2,\mu^2,\omega)}{\dd\ln\mu^2}=-\frac{\alpha_s C_F}{\pi}\left[-\gamma_E+\frac{3}{4}-\psi(\gamma^q_{\Lambda,\mu})+\ln\omega \right]\Delta^q(\Lambda^2,\mu^2,\omega),
\\
&
\frac{\dd\Delta^g (\Lambda^2,\mu^2,\omega)}{\dd\ln\mu^2}=-\frac{\alpha_s N_c}{\pi}\left[-\gamma_E+\beta_0-\psi(\gamma^g_{\Lambda,\mu})+\ln\omega \right]\Delta^g(\Lambda^2,\mu^2,\omega).
\end{align}
Due to the scale dependence in the anomalous dimensions, the flow directions of the renormalization group equation for the $\mu$ and $\Lambda$ scales are opposite to each other. Employing the identity of the digamma function $\psi(\gamma)=-\gamma_E+\int_0^1\dd u\frac{1-u^{\gamma-1}}{1-u}$, we can show that the collinear jet threshold functions $\Delta^q$ and $\Delta^g$ defined above satisfy the following integro-differential equations
\begin{align}
&
\frac{\dd\Delta^q (\Lambda^2,\mu^2,\omega)}{\dd\ln\mu^2} = -\frac{\alpha_s C_F}{\pi}\left[\ln\omega+\frac{3}{4}\right]
\Delta^q (\Lambda^2,\mu^2,\omega) + \frac{\alpha_s C_F}{\pi}\int_0^\omega \dd \omega' \frac{\Delta^q (\Lambda^2,\mu^2,\omega)-\Delta^q(\Lambda^2,\mu^2,\omega')}{\omega-\omega'}, \label{eq:rge-quark}
\\
&
\frac{\dd\Delta^g (\Lambda^2,\mu^2,\omega)}{\dd\ln\mu^2} = -\frac{\alpha_s N_c}{\pi}\left[\ln\omega+ \beta_0 \right]
\Delta^g (\Lambda^2,\mu^2,\omega) + \frac{\alpha_s N_c}{\pi}\int_0^\omega \dd \omega' \frac{\Delta^g (\Lambda^2,\mu^2,\omega)-\Delta^g(\Lambda^2,\mu^2,\omega')}{\omega-\omega'}, \label{eq:rge-gluon}
\end{align}
respectively. In deriving the above result, we have used $\Delta^{q/g} (\Lambda^2, \mu^2, \omega) \propto \omega^{\gamma_{\Lambda,\mu}^{q/g} - 1}$ together with the identity 
\begin{align}
\int_0^1 \dd u u^{\gamma_{\Lambda,\mu}^{q/g} - 1} \Delta^{q/g} (\Lambda^2, \mu^2, \omega) = \int_0^\omega \frac{\dd \omega'}{\omega} \Delta^{q/g} (\Lambda^2, \mu^2, \omega').
\end{align}

In the threshold limit which gives rise to the approximation $\ln\frac{1}{\xi}|_{\xi\to 1} \approx 1-\xi$, the above evolution equation for the jet function $\Delta^q (\Lambda^2, \mu^2, \omega)$ looks rather similar to that developed in Refs.~\cite{Becher:2006nr, Becher:2006mr} within the SCET framework. The only difference lies in the absence of the Sudakov double logarithm ($-\frac{\alpha_s C_F}{2\pi} L^2$) and the single logarithm ($\frac{3}{2} \frac{\alpha_s C_F}{\pi} L$) with $L=\ln k_\perp^2/\Lambda^2$. To simplify the theoretical derivations presented in this paper, we choose to first resum the collinear logarithms using the renormalization group equations given by Eqs.~(\ref{eq:rge-quark}-\ref{eq:rge-gluon}). Then, we deal with the resummation of the single and double soft logarithms separately through the Sudakov factor in Sec.~\ref{sec:sudakov}. The bottom line is that the threshold implemented in our calculation is consistent with the systematic renormalization group equation approach discussed in Refs.~\cite{Becher:2006nr, Becher:2006mr}. 

\subsubsection{Analytic Continuation}

The resummed results obtained above are well-defined in the ${\rm Re}[\gamma^{q/g}_{\Lambda,\mu}]>0$ region. However, they become singular at $\xi=1$ (or equivlently speaking, $x=x_p$ or $z=z'$) when ${\rm Re}[\gamma^{q/g}_{\Lambda,\mu}]\leq 0$. This is simply due to the fact that Eq.~(\ref{gamma-identity}) requires ${\rm Re}[\gamma^{q/g}_{\Lambda,\mu}]>0$ in order to close the integral contour to the left. Similar to the analytic continuation of the gamma function\footnote{To understand this point better, we briefly recall the analytic continuation of the gamma function $\Gamma(z)$. Conventionally, $\Gamma(z)$ is defined via the integral,
$\Gamma(z) = \int_0^\infty dx x^{z-1} e^{-x}$,
when ${\rm Re}[z] >0 $. This integral is divergent at ${\rm Re}[z]\leq 0$ and therefore $\Gamma(z)$ is not properly defined with this expression in this region. The relation $\Gamma(z)=\frac{1}{z}\Gamma(z+1)$ can be employed to uniquely extend the gamma function to the $-1<{\rm Re}[z]\leq 0$ region. Furthermore, using the above relation repeatedly, we can further extend $\Gamma(z)$ to the entire negative half plane except zero and negative integers.}, the identity shown in Eq.~(\ref{gamma-identity}) can be extended to the entire complex plane. For example, we can analytically continue the resummed results to the region where $-1< {\rm Re}[\gamma^{q/g}_{\Lambda,\mu}]\leq 0$ in the complex plane by reconsidering the following inverse Mellin transform
\begin{align}
&
\int_0^1 \frac{\dd x}{x} q(x, \mu^2) \int_{\mathcal C} \frac{\dd N}{2\pi i} \left(\frac{x}{x_p}\right)^N e^{-\gamma_{\Lambda,\mu}^q \ln N} 
\nonumber \\
=
& 
\int_0^1 \frac{\dd x}{x} [q(x, \mu^2) - q(x_p, \mu^2)]\int_{\mathcal C} \frac{\dd N}{2\pi i} \left(\frac{x}{x_p}\right)^N e^{-\gamma_{\Lambda,\mu}^q \ln N}
+
\int_0^1 \frac{\dd x}{x} q(x_p, \mu^2) \int_{\mathcal C} \frac{\dd N}{2\pi i} \left(\frac{x}{x_p}\right)^N e^{-\gamma_{\Lambda,\mu}^q \ln N} 
\nonumber \\
= & 
\int_{x_p}^1 \frac{\dd x}{x} [q(x,\mu^2) - q(x_p, \mu^2)] \frac{1}{\Gamma(\gamma^q_{\Lambda,\mu})} \left( \ln \frac{x}{x_p} \right)^{\gamma^q_{\Lambda,\mu} - 1} 
+
q(x_p,\mu^2)
\frac{1}{\Gamma(\gamma^q_{\Lambda,\mu}+1)} \left( \ln \frac{1}{x_p} \right)^{\gamma^q_{\Lambda,\mu}}.
\end{align}
It is then clear that the applicable range of the resummed results is extended and now it includes the $-1< {\rm Re}[\gamma^{q/g}_{\Lambda,\mu}]\leq 0$ region. To extend to the smaller ${\rm Re}[\gamma_{\Lambda,\mu}^q]$ region, we can repeat the same strategy and consider the following subtractions
\begin{align}
&
\int_0^1 \frac{\dd x}{x} q(x, \mu^2) \int_{\mathcal C} \frac{\dd N}{2\pi i} \left(\frac{x}{x_p}\right)^N e^{-\gamma_{\Lambda,\mu}^q \ln N} 
\nonumber \\
=
& 
\int_0^1 \frac{\dd x}{x} \left[q(x, \mu^2) - q(x_p, \mu^2) - x_p q'(x_p) \ln\frac{x}{x_p} \right]\int_{\mathcal C} \frac{\dd N}{2\pi i} \left(\frac{x}{x_p}\right)^N e^{-\gamma_{\Lambda,\mu}^q \ln N}
\nonumber\\
&
+
\int_0^1 \frac{\dd x}{x} \left[q(x_p, \mu^2) + x_p q'(x_p) \ln\frac{x}{x_p} \right] \int_{\mathcal C} \frac{\dd N}{2\pi i} \left(\frac{x}{x_p}\right)^N e^{-\gamma_{\Lambda,\mu}^q \ln N} 
\nonumber\\
= & 
\int_{x_p}^1 \frac{\dd x}{x} \left[q(x,\mu^2) - q(x_p, \mu^2) - x_p q'(x_p, \mu^2) \ln\frac{x}{x_p} \right] \frac{1}{\Gamma(\gamma^q_{\Lambda,\mu})} \left( \ln \frac{x}{x_p} \right)^{\gamma^q_{\Lambda,\mu} - 1}  
\nonumber\\
& 
+ q(x_p,\mu^2) \frac{1}{\Gamma(\gamma^q_{\Lambda,\mu}+1)} \left( \ln \frac{1}{x_p} \right)^{\gamma^q_{\Lambda,\mu}}
+ x_p q'(x_p,\mu^2) \frac{\gamma^q_{\Lambda,\mu}}{\Gamma(\gamma^q_{\Lambda,\mu} + 2)}
\left( \ln\frac{1}{x_p}\right)^{\gamma^q_{\Lambda,\mu} + 1},
\end{align}
where $q'(x_p, \mu^2) \equiv \frac{\dd q(x,\mu^2)}{\dd x}|_{x=x_p}$. The above subtraction method extends the validation range of our results to ${\rm Re}[\gamma^q_{\Lambda,\mu}]>-2$. In principle, we can repeat the same trick and cover the entire complex plane. In practice, it is quite challenging to numerically evaluate higher order derivatives of PDFs or FFs for the regime with ${\rm Re} [\gamma^q_{\Lambda,\mu}]<-2$. Conventionally, one can introduce the star distribution~\cite{Becher:2006nr,Becher:2006mr,Bosch:2004th} and rewrite the resummed PDFs and FFs as
\begin{align}
&   q^{\rm res} (x_p,\Lambda ^2, \mu ^2) 
  =  
   \frac{e^{ - \gamma^q_{\Lambda, \mu} (\gamma_E-\frac{3}{4})}}{\Gamma(\gamma^q_{\Lambda, \mu})}   
 \int ^1_{x_p} \frac{\dd x}{x} q(x, \mu ^2)    
  \left( \ln \frac{x }{x_p}  \right)_*^{ \gamma^q_{\Lambda, \mu} -1  },
\\
& g^{\rm res} (x_p, \Lambda^2, \mu^2) = 
\frac{e^{-\gamma^g_{\Lambda,\mu} (\gamma_E-\beta_0)}}{\Gamma(\gamma^g_{\Lambda,\mu})} \int_{x_p}^1 \frac{\dd x}{x} g(x,\mu^2)
\left(\ln \frac{x}{x_p} \right)_*^{\gamma^g_{\Lambda,\mu} - 1},
\\
& D_{h/q}^{\rm res} (z,\Lambda ^2, \mu ^2) 
  =  
   \frac{  e^{ - \gamma^q_{\Lambda, \mu} (\gamma_E-\frac{3}{4}) }  }{\Gamma( \gamma^q_{\Lambda, \mu}   )   }   
 \int ^1_{z} \frac{\dd z'  }{z'}D_{h/q} (z', \mu ^2)    
  \left( \ln \frac{z'}{z}  \right)_*^{ \gamma^q_{\Lambda, \mu} -1  },
\\
& D_{h/g}^{\rm res} (z, \Lambda^2, \mu^2) = 
\frac{e^{-\gamma^g_{\Lambda,\mu} (\gamma_E-\beta_0)}}{\Gamma(\gamma^g_{\Lambda,\mu})} \int_z^1 \frac{\dd z'}{z'} D_{h/g} (z', \mu^2)
\left( \ln \frac{z'}{z} \right)_*^{\gamma^g_{\Lambda,\mu} - 1}, 
\end{align}
where the star distribution represents the analytical expression obtained from the analytic continuation and it is properly defined in the full $\gamma^{q/g}_{\Lambda,\mu}$ complex plane. 

\begin{figure}[hbt]
\includegraphics[width=0.4\textwidth]{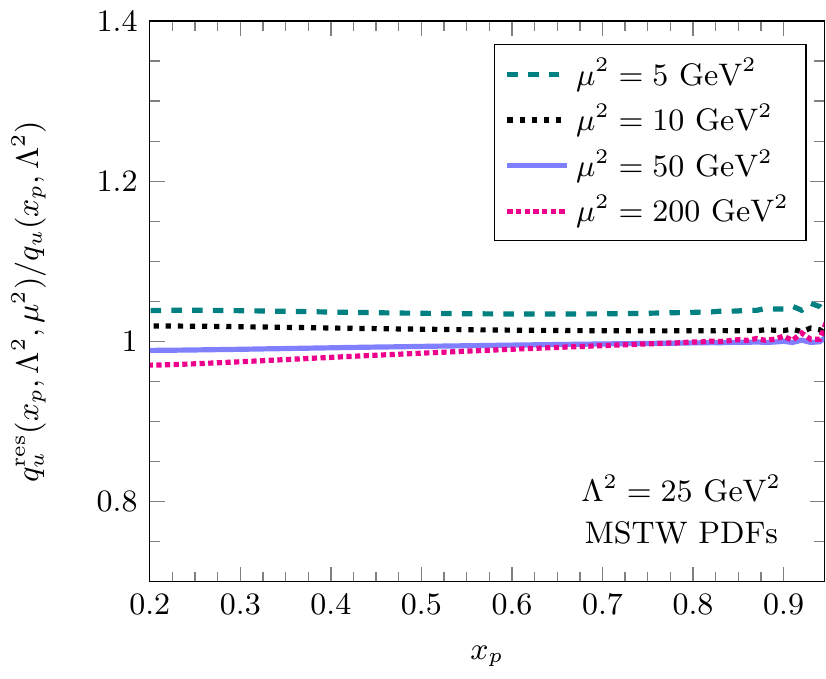}
\caption{The ratios of $q_{u}^{\text{res}}(x_p,\Lambda^2, \mu^2)$ to $q_u(x_p, \Lambda ^2)$ as a function of $x_p$ with various values of $\mu^2$.}
\label{fig:RGE_ratio}
\end{figure}

With the above analytic continuation technique, we can finally compute the resummed PDFs and FFs numerically for a broad range of $\mu^2$. To compare this renormalization group equation approach with the reverse-evolution approach developed in the last subsection, we show the ratio of the up quark distributions in Fig.~\ref{fig:RGE_ratio}. Here the up quark distributions $q_{u}^{\rm res} (x_p,\Lambda^2, \mu^2)$ and $q_u (x_p, \Lambda^2)$ are computed from the renormalization group approach and the reverse DGLAP approach, respectively. As shown in Fig.~\ref{fig:RGE_ratio}, the ratio is close to $1$ in the intermediate and large $x_p$ regions and it indicates that the numerical difference between these two approaches is small.

The reverse-evolution approach developed in Sec.~\ref{sec:dglap-resum} does not rely on the large-$N$ and $\xi\to 1$ approximations, which are vital in the renormalization group equation approach discussed in this subsection. Furthermore, it automatically takes care of the off-diagonal channels. Therefore, we employ the first approach in our numerical evaluations. Nonetheless, the fact that the ratio of $q_{u}^{\rm res} (x_p,\Lambda^2, \mu^2)$ and $q_u (x_p, \Lambda^2)$ is rather close to $1$ for various values of $\mu^2$ suggests these two approaches are numerically equivalent.

\section{The Resummation of the soft logarithms} 
\label{sec:sudakov}

Let us discuss the resummation of the soft part of the threshold logarithms via the Sudakov factor in this section. Soft logarithms only appear in the $q\to q$ and $g\to g$ channels. Conventionally, these double and single logarithms can be resummed through the Sudakov factors. For the $q\to q$ channel, we have the following logarithms from the $\sigma_{qq}^{5a}$ and $\sigma_{qq}^2$ terms,
\begin{align}
-\frac{\alpha_s}{2\pi} C_F \ln^2 \frac{\kperp^2}{\Lambda^2} + 3 \frac{\alpha_s}{2\pi} C_F \ln \frac{\kperp^2}{\Lambda^2}.
\end{align}
The first term is the double logarithm term and the second one is the single logarithm term derived with the fixed strong coupling. As the common practice, we need to convert the above expression from the fixed coupling one to the running coupling one in phenomenology. Therefore, in the threshold resummation, we employ the following Sudakov factor 
\begin{align}
S_{\rm Sud}^{qq} 
= C_F \int_{\Lambda^2}^{\kperp^2} \frac{\dd \mu^2}{\mu^2} \frac{\alpha_s(\mu^2)}{\pi} \ln \frac{\kperp^2}{\mu^2}
- 3 C_F \int_{\Lambda^2}^{\kperp^2} \frac{\dd \mu^2}{\mu^2} \frac{\alpha_s(\mu^2)}{2\pi}.
\end{align}
The resummation of soft logarithms becomes an exponential of the Sudakov factor. The extraction of the double logarithms is quite challenging in the running coupling case. Alternatively, we first extract those soft logarithms with the fixed coupling and then compute the mismatch term to take into account the difference. The difference between the running coupling Sudakov factor and the fixed coupling NLO correction is cast into the following matching term,
\begin{align}
S_{\rm Sud}^{qq} - C_F \frac{\alpha_s}{2\pi} \left( \ln^2 \frac{\kperp^2}{\Lambda^2} - 3 \ln\frac{\kperp^2}{\Lambda^2} \right).
\end{align}
The discussion for the $g\to g$ channel also follows suit. At the end of the day, the resummed formula reads
\begin{align}
\frac{\dd\sigma_{\rm resummed}}{\dd y\dd^2\pperp}
= & 
\Sperp  \int_\tau^1 \frac{\dd z}{z ^2} x_p q(x_p,\Lambda^2)D_{h/q}(z,\Lambda^2) F(\kperp) e^{-S_{\rm Sud}^{qq}}
\nonumber\\
 &+ 
\Sperp  \int_\tau^1 \frac{\dd z}{z ^2} x_p g(x_p,\Lambda^2)D_{h/g}(z,\Lambda^2) \int \dd^2 \qa F(\qa) F(\kperp-\qa) e^{-S_{\rm Sud}^{gg}}, \label{eq:resed}
\end{align}
where $S_{\text{Sud}}^{qq}$ and $S_{\text{Sud}}^{gg}$ are Sudakov factors for $q\to q$ and $g\to g$ channels which are given by
\begin{align}
& S_{\rm Sud}^{qq} 
= C_F \int_{\Lambda^2}^{\kperp^2} \frac{\dd \mu^2}{\mu^2} \frac{\alpha_s(\mu^2)}{\pi} \ln \frac{\kperp^2}{\mu^2}
- 3 C_F \int_{\Lambda^2}^{\kperp^2} \frac{\dd \mu^2}{\mu^2} \frac{\alpha_s(\mu^2)}{2\pi},
\\
& S_{\rm Sud}^{gg}
= N_c \int_{\Lambda^2}^{\kperp^2} \frac{\dd \mu^2}{\mu^2} \frac{\alpha_s(\mu^2)}{\pi} \ln \frac{\kperp^2}{\mu^2}
- \frac{11N_c-2n_f}{3}\int_{\Lambda^2}^{\kperp^2} \frac{\dd \mu^2}{\mu^2} \frac{\alpha_s(\mu^2)}{2\pi}.
\end{align}
The Sudakov factor follows the counting rule which is given in Refs.~\cite{Mueller:2013wwa,Sun:2014gfa}. As discussed in the last section, the factorization scale in Eq.~(\ref{eq:resed}) is set to be $\Lambda$ as a result of the resummation of collinear logarithms.

Thus, the Sudakov matching term, which is treated as part of the NLO correction, is given by
\begin{align}
\frac{\dd\sigma_{\rm Sud~matching}}{\dd y\dd^2\pperp}
= & 
\Sperp  \int_\tau^1 \frac{\dd z}{z ^2} x_p q(x_p,\mu^2)D_{h/q}(z,\mu^2) F(\kperp) \left\{S_{\rm Sud}^{qq} - \left[ C_F\frac{\alpha_s}{2\pi}\left(\ln^2 \frac{\kperp^2}{\Lambda^2}-3\ln \frac{\kperp^2}{\Lambda^2}\right)\right]\right\}
\nonumber\\
 &+ 
\Sperp  \int_\tau^1 \frac{\dd z}{z ^2} x_p g(x_p,\mu^2)D_{h/g}(z,\mu^2) \int \dd^2 \qa F(\qa) F(\kperp-\qa) 
\nonumber\\
& \phantom{xx} 
\times
\left\{ S_{\rm Sud}^{gg} - \left[ N_c\frac{\alpha_s}{2\pi}\left(\ln^2 \frac{\kperp^2}{\Lambda^2}-4\beta_0\ln \frac{\kperp^2}{\Lambda^2}\right) \right] \right\}. \label{eq:sudakov-mismatch}
\end{align}

\section{Summary of the Full Threshold Resummed Results}
\label{sec:resummed-full-expression}

To make the resummed results more accessible to the interested readers, we provide a thorough summary of the full NLO cross-section after the threshold resummation in the large $N_c$ limit, which have been numerically evaluated and referred to as the ``Resummed'' results in the plots throughout the paper. 

First, the resummation of the collinear logarithms in $\sigma_{qq}^1$, $\sigma_{gg}^1$, $\sigma_{gq}^1$, $\sigma_{gq}^2$, $\sigma_{qg}^1$ and $\sigma_{qg}^2$ terms sets the factorization scales in $\sigma_{qq}^{\rm LO}$ and $\sigma_{gg}^{\rm LO}$ to be $\Lambda^2$. Second, the resummation of the soft logarithms in $\sigma_{qq}^{5a}$, $\sigma_{qq}^2$, $\sigma_{gg}^{6a}$ and $\sigma_{gg}^2$ terms yields the exponential expression of the Sudakov factor. The rest of the NLO corrections together with the matching terms do not contain apparent large logarithms and they are numerically small. Therefore, we treat them as the new NLO hard factors after the subtraction of logarithms. The resummation improved NLO cross-section is then given by
\begin{align}
\frac{\dd\sigma}{\dd y\dd^2\pperp} =
\frac{\dd\sigma_{\rm resummed}}{\dd y\dd^2\pperp} + \frac{\dd\sigma_{\rm NLO~matching}}{\dd y\dd^2\pperp}
+ \frac{\dd\sigma_{\rm Sud~matching}}{\dd y\dd^2\pperp},
\label{eq:final-prescription}
\end{align}
where 
\begin{align}
\frac{\dd\sigma_{\rm resummed}}{\dd y\dd^2\pperp}
= & 
\Sperp  \int_\tau^1 \frac{\dd z}{z ^2} x_p q(x_p,\Lambda^2)D_{h/q}(z,\Lambda^2) F(\kperp) e^{-S_{\rm Sud}^{qq}}
\nonumber\\
 &+ 
\Sperp  \int_\tau^1 \frac{\dd z}{z ^2} x_p g(x_p,\Lambda^2)D_{h/g}(z,\Lambda^2) \int \dd^2 \qa F(\qa) F(\kperp-\qa) e^{-S_{\rm Sud}^{gg}},
\\
\frac{\dd\sigma_{\rm Sud~matching}}{\dd y\dd^2\pperp}
= & 
\Sperp  \int_\tau^1 \frac{\dd z}{z ^2} x_p q(x_p,\mu^2)D_{h/q}(z,\mu^2) F(\kperp) \left\{S_{\rm Sud}^{qq} - \left[ C_F\frac{\alpha_s}{2\pi}\left(\ln^2 \frac{\kperp^2}{\Lambda^2}-3\ln \frac{\kperp^2}{\Lambda^2}\right)\right]\right\}
\nonumber\\
 &+ 
\Sperp  \int_\tau^1 \frac{\dd z}{z ^2} x_p g(x_p,\mu^2)D_{h/g}(z,\mu^2) \int \dd^2 \qa F(\qa) F(\kperp-\qa) 
\nonumber\\
& \phantom{xx} 
\times
\left\{ S_{\rm Sud}^{gg} - \left[ N_c\frac{\alpha_s}{2\pi}\left(\ln^2 \frac{\kperp^2}{\Lambda^2}-4\beta_0\ln \frac{\kperp^2}{\Lambda^2}\right) \right] \right\},
\\
\frac{\dd\sigma_{\rm NLO~matching}}{\dd y\dd^2\pperp}
= &
\sum_{i=3}^4 \frac{\dd\sigma_{qq}^i}{\dd y\dd^2\pperp} + \frac{\dd\sigma_{qq}^{5b}}{\dd y\dd^2\pperp}
+ \sum_{i=3}^5 \frac{\dd\sigma_{gg}^i}{\dd y\dd^2\pperp} + \frac{\dd\sigma_{gg}^{6b}}{\dd y\dd^2\pperp}
+ \frac{\dd\sigma_{gq}^3}{\dd y\dd^2\pperp}
+ \frac{\dd\sigma_{qg}^2}{\dd y\dd^2\pperp}
+ \sum_{i=4}^5\frac{\dd\sigma_{qg}^i}{\dd y\dd^2\pperp}.
\end{align}
The Sudakov factors are 
\begin{align}
& S_{\rm Sud}^{qq} 
= C_F \int_{\Lambda^2}^{\kperp^2} \frac{\dd \mu^2}{\mu^2} \frac{\alpha_s(\mu^2)}{\pi} \ln \frac{\kperp^2}{\mu^2}
- 3 C_F \int_{\Lambda^2}^{\kperp^2} \frac{\dd \mu^2}{\mu^2} \frac{\alpha_s(\mu^2)}{2\pi},
\\
& S_{\rm Sud}^{gg}
= N_c \int_{\Lambda^2}^{\kperp^2} \frac{\dd \mu^2}{\mu^2} \frac{\alpha_s(\mu^2)}{\pi} \ln \frac{\kperp^2}{\mu^2}
- \frac{11N_c-2n_f}{3}\int_{\Lambda^2}^{\kperp^2} \frac{\dd \mu^2}{\mu^2} \frac{\alpha_s(\mu^2)}{2\pi}.
\end{align}
For the reader's convenience, we list all the updated NLO matching terms in the following
\begin{align}
\frac{\dd\sigma_{qq}^3}{\dd y\dd^{2}\pperp}
=& \frac{\alpha_{s}}{2\pi^2} C_F \Sperp \int_{\tau}^{1} \frac{\dd z}{z^{2}}\int_{\tau/z}^{1}\dd\xi\int\dd^{2}\qa\dd^{2}\qb
 x q(x,\mu^{2})D_{h/q}(z,\mu^{2})
\frac{1+\xi^2}{(1-\xi)_+} \mathcal{T}_{qq}^{(1)}(\xi,\qa,\qb,\kperp), 
\\
\frac{\dd\sigma_{qq}^4}{\dd y\dd^{2}\pperp}
=& 
- \frac{\alpha_{s}}{\pi} C_F \Sperp
 \int_{\tau}^{1} \frac{\dd z}{z^{2}}\int_{0}^{1}\dd\xi'\int\dd^{2}\qa 
 x_p q(x_p,\mu^{2})D_{h/q}(z,\mu^{2})
\frac{1+{\xi'}^{2}}{\left(1-\xi'\right)_+}\ln\frac{(\qa-\xi'\kperp)^{2}}{\kperp^{2}}
F(\qa)F(\kperp),  
\\
\frac{\dd\sigma_{qq}^{5b}}{\dd y\dd^{2}\pperp}
=& 
\frac{2\alpha_s}{\pi^2} C_F \Sperp 
\int_\tau^1 \frac{\dd z}{z^2} \int \dd^2 \qa x_p q(x_p,\mu^2) D_{h/q} (z,\mu^2) 
\frac{1}{\qa^2} \ln \frac{\kperp^2}{\qa^2} [F(\kperp-\qa) -  \theta(\Lambda^2-\qa^2) F(\kperp)] 
\nonumber\\
- &
\frac{\alpha_s}{2\pi} C_F \Sperp 
\int_\tau^1 \frac{\dd z}{z^2} x_p q(x_p,\mu^2) D_{h/q} (z,\mu^2) F(\kperp)
\ln^2 \frac{\kperp^2}{\Lambda^2}  \nonumber\\
+ & 
\frac{\alpha_s}{\pi} C_F \Sperp 
\int_\tau^1 \frac{\dd z}{z^2} \int \dd^2 \qa x_p q(x_p,\mu^2) D_{h/q} (z,\mu^2) 
F(\qa) F(\kperp) \ln^2 \frac{\kperp^2}{(\kperp - \qa)^2} \nonumber \\
- & \frac{2\alpha_s}{\pi^2} C_F \Sperp 
\int_\tau^1 \frac{\dd z}{z^2} \int \dd^2 \qa \int \dd^2 \qb x_p q(x_p,\mu^2) D_{h/q} (z,\mu^2) 
F(\qa) F(\qb) \ln \frac{\kperp^2}{(\kperp-\qa)^2} \nonumber\\
& \times \frac{(\kperp-\qa)\cdot (\kperp-\qb)}{(\kperp-\qa)^2 (\kperp-\qb)^2} ,
\\
\frac{\dd\sigma_{gg}^3}{\dd y\dd^{2}\pperp}
=
&
\frac{\alpha_s}{\pi^2} N_c \Sperp
\int_\tau^1 \frac{\dd z}{z^2} \int_{\tau/z}^1 \dd \xi \int \dd^2 \qa \dd^2 \qb \dd^2 \qc xg(x,\mu^2) D_{h/g}(z,\mu^2) \frac{[1-\xi(1-\xi)]^2}{\xi(1-\xi)_+} 
\nonumber\\
&
\times \mathcal{T}_{gg}^{(1)} (\xi,\qa,\qb,\qc,\kperp),
\\
\frac{\dd\sigma_{gg}^4}{\dd y\dd^{2}\pperp}
=
& -
2N_f T_R \frac{\alpha_s}{2\pi} \Sperp \int_\tau^1 \frac{\dd z}{z^2} \int_0^1 \dd\xi' \dd^2 \qa x_p g(x_p,\mu^2) D_{h/g}(z,\mu^2) [\xi'^2 + (1-\xi')^2] 
\nonumber\\
&
\times
F(\qa) F(\kperp-\qa) \ln \frac{(\qa-\xi'\kperp)^2}{\kperp^2},
\\
\frac{\dd\sigma_{gg}^5}{\dd y\dd^{2}\pperp}
=
&
- 4N_c \frac{\alpha_s}{2\pi} \Sperp \int_\tau^1 \frac{\dd z}{z^2} \int_0^1 \dd\xi' \dd^2\qa \dd^2\qb x_p g(x_p,\mu^2) D_{h/g}(z,\mu^2) \left[ \frac{\xi'}{(1-\xi')_+} + \frac{1}{2} \xi' (1-\xi') \right] 
\nonumber\\
& 
F(\qa) F(\qb) F(\kperp-\qa) \ln \frac{(\qa+\qb-\xi'\kperp)^2}{\kperp^2}, 
\\
\frac{\dd\sigma_{gg}^{6b}}{\dd y\dd^{2}\pperp}
=& 
\frac{2\alpha_s}{\pi^2} N_c \Sperp 
\int_\tau^1 \frac{\dd z}{z^2} \int \dd^2 \qa \int \dd^2 \qb x_p g(x_p,\mu^2) D_{h/q} (z,\mu^2)
\frac{1}{\qb^2} \ln\frac{\kperp^2}{\qb^2} F(\kperp-\qa) 
\nonumber\\
&
\times [F(\qa+\qb) - \theta(\Lambda^2 - \qb^2) F(\qa)] 
\nonumber\\
- & \frac{\alpha_s}{2\pi} N_c \Sperp 
\int_\tau^1 \frac{\dd z}{z^2} x_p g(x_p,\mu^2) D_{h/q} (z,\mu^2)
\ln^2 \frac{\kperp^2}{\Lambda^2} \int \dd^2 \qa F(\kperp-\qa)F(\qa) \nonumber\\
+& 
\frac{\alpha_s}{\pi} N_c \Sperp
\int_\tau^1 \frac{\dd z}{z^2} \int \dd^2 \qa \int \dd^2 \qb x_p g(x_p,\mu^2) D_{h/q} (z,\mu^2)
F(\qa) F(\qb) F(\kperp-\qb)
\nonumber\\
&
\times \ln^2 \frac{\kperp^2}{(\qa+\qb-\kperp)^2}
\nonumber \\
- & 
\frac{2\alpha_s}{\pi^2} N_c \Sperp 
\int_\tau^1 \frac{\dd z}{z^2} \int \dd^2 \qa \int \dd^2 \qb \int \dd^2 \qc x_p g(x_p,\mu^2) D_{h/q} (z,\mu^2) F(\qa) F(\qb) F(\qc)
\nonumber \\
& \times 
\frac{(\kperp - \qa + \qc) \cdot (\kperp - \qb + \qc)}{(\kperp - \qa + \qc)^2 (\kperp - \qb + \qc)^2}
\ln \frac{\kperp^2}{(\kperp - \qa + \qc)^2},
\\
\frac{\dd\sigma_{gq}^3}{\dd y\dd^{2}\pperp} = 
&
\frac{\alpha_s}{2\pi^2} C_F \Sperp
\int_\tau^1 \frac{\dd z}{z^2} \int_{\tau/z}^1 \dd \xi \int \dd^2 \qa \int \dd^2\qb
x q(x,\mu^2) D_{h/g} (z,\mu^2) \mathcal{P}_{gq} (\xi) 
\mathcal{T}_{gq}^{(1)} (\xi, \qa,\qb,\kperp),
\\
\frac{\dd\sigma_{qg}^2}{\dd y\dd^{2}\pperp}
=
&
-\frac{\alpha_s}{4\pi} \Sperp
\int_\tau^1 \frac{\dd z}{z^2} \int_{\tau/z}^1 \dd \xi \int \dd^2 \qa
x g(x,\mu^2) D_{h/q}(z,\mu^2) \frac{1}{\xi^2} \mathcal{P}_{qg} (\xi) 
F(\qa) F(\kperp/\xi-\qa),
\\
\frac{\dd\sigma_{qg}^4}{\dd y\dd^{2}\pperp}
=
&
-\frac{\alpha_s}{4\pi} \Sperp
\int_\tau^1 \frac{\dd z}{z^2} \int_{\tau/z}^1 \dd \xi 
x g(x,\mu^2) D_{h/q}(z,\mu^2) \mathcal{P}_{qg} (\xi) 
F(\kperp),
\\
\frac{\dd\sigma_{qg}^5}{\dd y\dd^{2}\pperp}
= 
&
\frac{\alpha_s}{4\pi^2} \Sperp
\int_\tau^1 \frac{\dd z}{z^2} \int_{\tau/z}^1 \dd \xi \int \dd^2 \qa \int \dd^2 \qb
xg(x,\mu^2) D_{h/q}(z,\mu^2) \mathcal{P}_{qg} (\xi)
\mathcal{T}_{qg}^{(1)} (\xi, \qa,\qb,\kperp),
\end{align}
where we have defined 
\begin{align}
\mathcal{T}_{qq}^{(1)}(\xi,\qa,\qb,\kperp)= 
& \frac{(\qb-\qa/\xi)^{2}}{(\kperp+\qa)^2 (\kperp/\xi+\qb)^2} F(\qa)F(\qb)
\nonumber\\
& - \frac{1}{(\kperp+\qa)^2} \frac{\Lambda^{2}}{\Lambda^{2}+(\kperp+\qa)^2}
F(\qb)F(\kperp)
\nonumber \\
& 
- \frac{1}{(\kperp +\xi\qb)^2 } \frac{\Lambda^{2}}{\Lambda^{2}+(\kperp/\xi+\qb)^2}
F(\kperp/\xi)F(\qa),
\\
\mathcal{T}_{gg}^{(1)} (\xi,\qa,\qb,\qc,\kperp) =
&
\frac{1}{\xi^2} \frac{[(1-\xi)\qa +\qc - \xi\qb]^2}{(\qa+\qc-\kperp)^2 (\qa + \qb - \kperp/\xi)^2} F(\qa) F(\qb) F(\qc)
\nonumber\\
- & 
\frac{1}{(\qa+\qc-\kperp)^2} \frac{\Lambda^2}{\Lambda^2 + (\qa+\qc-\kperp)^2} F(\kperp-\qc) F(\qb) F(\qc) 
\nonumber \\
- &
\frac{1}{\xi^2} \frac{1}{(\qa + \qb - \kperp/\xi)^2} \frac{\Lambda^2}{\Lambda^2 + (\qa + \qb - \kperp/\xi)^2} F(\kperp/\xi-\qb) F(\qb) F(\qc),
\\
\mathcal{T}_{gq}^{(1)} (\xi, \qa,\qb,\kperp) 
= 
&
\left( \frac{\kperp -\qa - \qb}{(\kperp -\qa - \qb)^2} - \frac{\kperp - \xi \qb}{(\kperp - \xi\qb)^2} \right)^2
F(\qa) F(\qb)
\nonumber\\
&
- \frac{\Lambda^2}{\Lambda^2 + (\kperp -\qa - \qb)^2} \frac{1}{(\kperp -\qa - \qb)^2} 
F(\qb) F(\kperp-\qb)
\nonumber\\
& 
- \frac{\Lambda^2}{\Lambda^2 + (\kperp/\xi-\qb)^2} \frac{1}{(\kperp-\xi\qb)^2} F(\qa) F(\kperp/\xi),
\\
\mathcal{T}_{qg}^{(1)} (\xi, \qa,\qb,\kperp)
=
& 
\left( 
\frac{\kperp- \xi\qa -\xi\qb}{(\kperp - \xi\qa -\xi\qb)^2} 
- \frac{\kperp-\qb}{(\kperp-\qb)^2}
\right)^2
F(\qa) F(\qb) 
\nonumber\\
& 
- \frac{1}{(\kperp - \xi\qa -\xi\qb)^2}\frac{\Lambda^2}{\Lambda^2 + (\kperp/\xi - \qa - \qb)^2}
F(\qb) F(\kperp/\xi-\qb)
\nonumber\\
&
- \frac{1}{(\kperp-\qb)^2} \frac{\Lambda^2}{\Lambda^2 + (\kperp-\qb)^2}
F(\qa) F(\kperp).
\end{align}

It is important to note that the factorization scale in $\sigma_{\rm resummed}$ now becomes $\Lambda$ due to the resummation of the threshold collinear logarithms, while that in $\sigma_{\rm NLO~matching}$ remains as $\mu$. This replacement of the factorization scale is akin to the common practice (setting $\mu$ to $\mu_b$) in the transverse momentum-dependent distribution factorization\cite{Collins:1984kg}. 

\section{Natural Choices of the Auxiliary Scale}
\label{kin-choice}

Here we illustrate how to determine the proper value of semi-hard scale $\Lambda$ in our numerical calculations, since this scale plays an important role in numerical results. In the case of fixed coupling, we first use an intuitive method and find $\Lambda^2 \sim (1-\xi) k_\perp^2$ when the threshold logarithms become important. In addition, by using the saddle point approximation, we find that the semi-hard scale $\Lambda$ can be determined by dominant scale determined by the saddle point of the resummed formula in both the fixed and running couple scenarios. 

\subsection{An Intuitive Derivation with the Fixed coupling}

\begin{figure}[htp]
\centering
\includegraphics[width=0.4\textwidth]{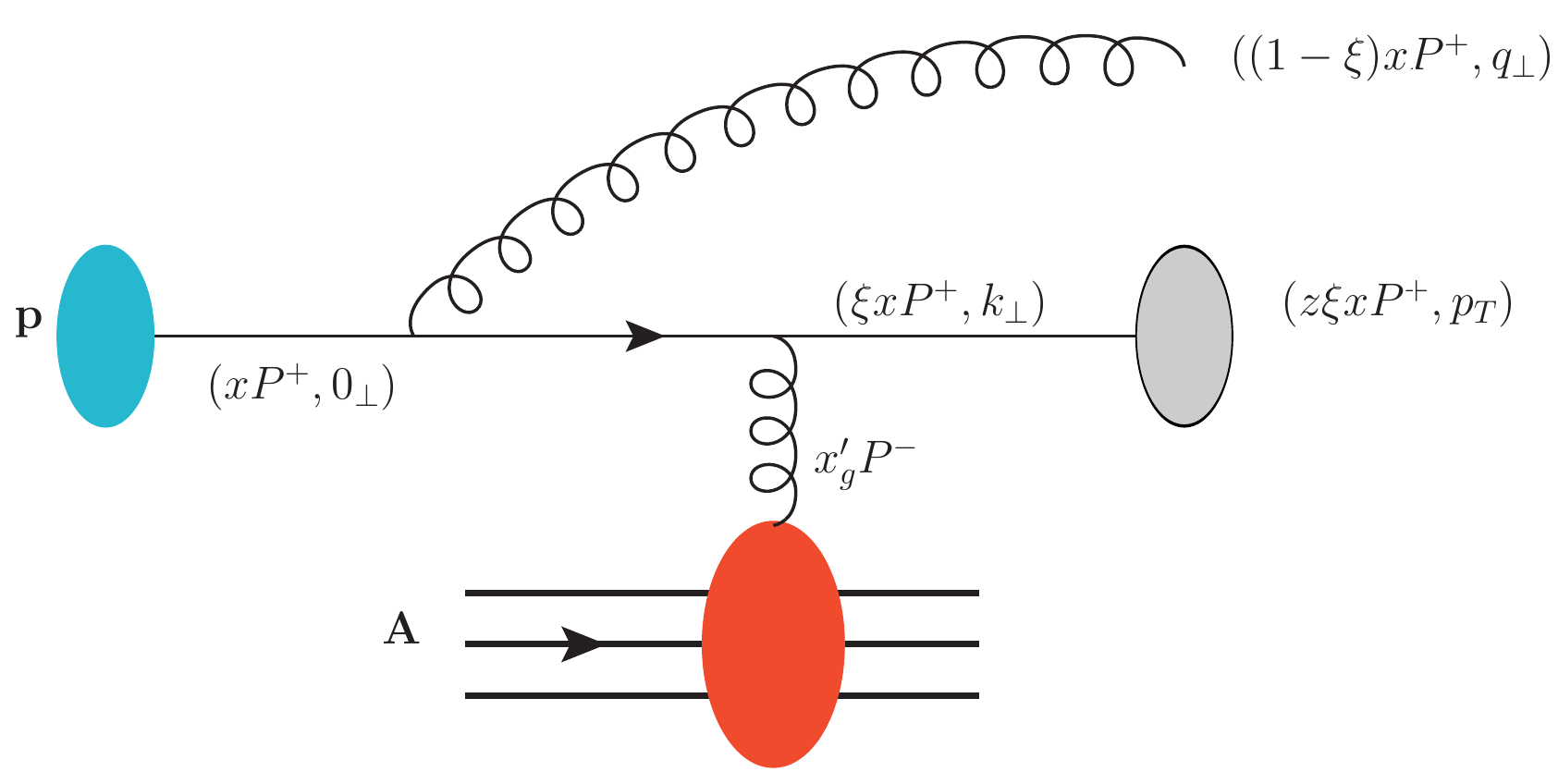}
\caption{The kinematics of the real gluon emission.}
\label{real-emission}
\end{figure}
To illustrate the physical interpretation of the semi-hard scale $\Lambda^2$, it is instructive to consider the real emission of gluons as shown in Fig.~\ref{real-emission}. Following the discussion outlined in Ref.~\cite{Watanabe:2015tja}, we use the light-cone perturbation theory and define $p^+=\frac{p^0+p^3}{\sqrt{2}}$ and $p^-=\frac{p^0-p^3}{\sqrt{2}}$. According to the momentum conservation before and after the splitting, we get the following kinematic constraint for the radiated gluon in the $\xi \to 1$ limit
\begin{equation}
q^-=\frac{\qperp^2}{2(1-\xi)x_pP^+}\leq P^- \quad \Rightarrow \quad \xi \leq 1-\frac{x_g\qperp^2}{\kperp^2}. 
\end{equation}
Then the upper limit of the divergent integral of $\dd\xi$ should be modified as following
\begin{align}
\int_0^{1-\frac{x_g\qperp^2}{\kperp^2}}\frac{\dd\xi}{1-\xi}=\ln\frac{\kperp^2}{\qperp^2}+\ln\frac{1}{x_g}.
\end{align}
It is important to note that these two logarithms arise two physical regions. First, in the region $0<\xi<1-\frac{\qperp^2}{\kperp^2}$ with finite longitudinal momentum $q^-$, one gets Sudakov logarithm $\ln\frac{\kperp^2}{\qperp^2}$ corresponding to real gluon emission. On the other hand, $q^-\to \infty$ in the region $1-\frac{\qperp^2}{\kperp^2}<\xi<1-\frac{\qperp^2}{\kperp^2}x_g$, then one gets $\ln\frac{1}{x_g}$ which corresponds to part of the small-$x$ evolution. For virtual gluon, there is no such requirement. 

If we introduce the semi-hard scale $\Lambda^2$ to represent the typical transverse momentum associated with the Sudakov real gluon emission, then we can find that the real and virtual contributions would cancel each other in the region $\qperp^2\leq \Lambda^2$. When the saturation momentum is not large, this scale is estimated to be $(1-\xi)\kperp^2 \sim (1-\tau)k_\perp^2$ since the real gluon emission requires $\xi<1-\frac{\qperp^2}{\kperp^2}$. On the other hand, when the saturation effect is dominant, we expect that the real and virtual contributions cancel up to the saturation momentum $Q_s^2$. With these overall considerations, we choose this semi-hard scale $\Lambda^2$ to be
\begin{equation}
\Lambda^2_{\rm fixed} \simeq \max\left[(1-\tau)\kperp^2,Q_s^2 \right] \gg \Lambda^2_{\rm QCD},
\end{equation}
where the subscript ``fixed'' indicates that this expression for the auxiliary scale is derived with fixed coupling. The auxiliary scale $\Lambda^2$ in our calculation is similar to the intermediate scale $\mu_i^2$ in SCET~\cite{Becher:2006nr,Becher:2006mr}. In the region $\Lambda^2<\qperp^2 <k_\perp^2$, the remaining virtual contribution is found to be
\begin{align}
-\int_{\Lambda^2}^{\kperp^2}\frac{\dd\qperp^2}{\qperp^2}\ln\frac{\kperp^2}{\qperp^2}\Rightarrow -\frac{1}{2}\ln^2\frac{\kperp^2}{\Lambda^2},
\end{align}
which can be identified as the Sudakov double logarithmic contribution. The above intuitive discussion of the scale choice is based on the separation of the kinematic region, and a more rigorous derivation using the saddle point approximation is provided in the next subsection. It is important to note that the $\Lambda^2$ scale can also be determined from the scale $\mu_r^2 \equiv c_0^2/\rperp^2$ with $\rperp$ being the typical scale in the coordinate space. There are two competing mechanisms: the threshold soft gluon emission and the saturation effects when we try to locate the region where the dominant contribution arises. In addition, for convenience, we use a fixed estimated value of $\Lambda^2$ in the numerical evaluation for a given kinematic region.

\subsection{Saddle Point Approximation}
\label{sec:saddle-point}

In addition to the above intuitive derivation, we can analytically study the choice of $\Lambda^2$ via the saddle point approximation. The saddle point approximation method, also known as the method of steepest descent, allows one to locate the region where the most important contribution arises in the resummed results and therefore identify the natural choice of $\Lambda^2$. Similar ideas have also been used in Refs.~\cite{Parisi:1979se,Collins:1984kg,Qiu:2000ga}. 

To determine the value of the auxiliary scale $\Lambda^2$ in the resumed expression in Eq.~(\ref{eq:final-prescription}), let us first consider the corresponding results for the $q\to q$ channel in the coordinate space
\begin{align}
\frac{\dd\sigma_{\text{resummed}}^{qq}}{\dd y \dd^2p_T} 
= & S_\perp\int_\tau^1\frac{\dd z}{z^2}\int\frac{\dd^2\rperp}{(2\pi)^2} e^{-i \kperp \cdot \rperp} S^{(2)}(\rperp) 
e^{-S_{\rm Sud}^{qq}} 
\int_{x_p}^1\frac{\dd x}{x}q(x,\mu)\frac{e^{\left(3/4-\gamma_E\right)\gamma^q_{\mu_r,\mu}}}{\Gamma(\gamma^q_{\mu_r,\mu})}\left[\ln\frac{x}{x_p}\right]_*^{\gamma^q_{\mu_r,\mu}-1} 
\nonumber\\
& \times
\int_{z}^1\frac{\dd z'}{z'}D_{h/q}(z')\frac{e^{\left(3/4-\gamma_E\right)\gamma^q_{\mu_r,\mu}}}{\Gamma(\gamma^q_{\mu_r,\mu})}\left[\ln\frac{z'}{z}\right]_*^{\gamma^q_{\mu_r,\mu}-1}, 
\label{eq:qq-rperp}
\end{align}
with $\mu_r = c_0/\rperp $, $c_0 = 2e^{-\gamma_E}$ and $\gamma_{\mu_r,\mu}^q=C_F\int_{\mu^2}^{\mu_r^2}\frac{\dd\mu^{\prime 2}}{\mu^{\prime 2}}\frac{\alpha_s(\mu^{\prime 2})}{\pi}$. The Sudakov factor $S_{\rm Sud}^{qq}$ in the coordinate space is
\begin{align}
S_{\rm Sud}^{qq} = C_F  \int_{c_0^2/\rperp^2}^{\kperp^2} \frac{\dd \mu^2}{\mu^2} \frac{\alpha_s (\mu^2)}{\pi} \ln \frac{\kperp^2}{\mu^2} - 3 C_F \int_{c_0^2/\rperp^2}^{\kperp^2} \frac{\dd \mu^2}{\mu^2} \frac{\alpha_s (\mu^2)}{2\pi}.
\end{align}
 
The saddle point of the above $r_\perp$ integral depends on both the strength of the saturation effect (given by the dipole amplitude $S^{(2)}(\rperp)$) and the threshold resummation. To identify the corresponding saddle point of each contribution, it is convenient to rewrite above formula in terms of the convolution of the dipole gluon distribution and the threshold Sudakov factor in the momentum space as follows
\begin{align}
\frac{\dd\sigma_{\text{resummed}}^{qq}}{\dd y \dd^2p_T} 
= S_\perp\int_\tau^1\frac{\dd z}{z^2}\int_{x_p}^1\frac{\dd x}{x}q(x,\mu)\int_{z}^1\frac{\dd z'}{z'}D_{h/q}(z') \, \int d^2 q_\perp F(k_\perp - q_\perp) G_{\text{th}}(q_\perp), \label{convolution}
\end{align}
where $F(k_\perp - q_\perp)$ is the Fourier transform of $S^{(2)}(\rperp)$ and  
\begin{align}
G_{\text{th}} (q_\perp)\equiv \int \frac{\dd^2 r_\perp}{(2\pi)^2} e^{-iq_\perp \cdot r_\perp} e^{-S_{\rm Sud}^{qq}} \frac{e^{\left(3/4-\gamma_E\right)\gamma^q_{\mu_r,\mu}}}{\Gamma(\gamma^q_{\mu_r,\mu})}\left[\ln\frac{x}{x_p}\right]_*^{\gamma^q_{\mu_r,\mu}-1}  \frac{e^{\left(3/4-\gamma_E\right)\gamma^q_{\mu_r,\mu}}}{\Gamma(\gamma^q_{\mu_r,\mu})}\left[\ln\frac{z'}{z}\right]_*^{\gamma^q_{\mu_r,\mu}-1} . \label{th-function}
\end{align}
When $k_\perp$ is large, the dominant contribution of Eq.~(\ref{convolution}) comes from the regions near the vicinity of $q_\perp \sim 0$ or $q_\perp \sim k_\perp$. To proceed, let us first consider the saddle point of the $r_\perp$ integral in Eq.~(\ref{th-function}) at $q_\perp\sim 0$ as follows
\begin{align}
\int_0^\infty \dd \rperp^2 
\exp\left[- S_{\rm Sud}^{qq} \right] 
\frac{e^{2\left(3/4-\gamma_E\right)\gamma^q_{\mu_r,\mu}}}{\Gamma(\gamma^q_{\mu_r,\mu})\Gamma(\gamma^q_{\mu_r,\mu})}\left[\ln \frac{1}{\xi}\ln \frac{1}{\xi'} \right]_*^{\gamma^q_{\mu_r,\mu}-1},
\end{align}
where $1/\xi = x/x_p$ and $1/\xi' = z'/z$. In order to apply the saddle point approximation, we first switch the integration variable from $\rperp^2$ to $\ln (\rperp^2 \Lambda_{\rm QCD}^2)$ so that the integration range becomes $[-\infty, +\infty]$. Second, the phase factor $e^{-iq_\perp \cdot \rperp}$ or the Bessel function $J_{0} (|q_\perp||\rperp|)$ has been removed at the $q_\perp \sim 0$ limit. Then the above integral can be cast into the following integral
\begin{align}
\int_{-\infty}^{+\infty} \dd \ln (\rperp^2 \Lambda_{\rm QCD}^2) \exp [E (\rperp)],
\end{align}
where the exponent is defined as
\begin{align}
E (\rperp) = \ln (\rperp^2 \Lambda_{\rm QCD}^2) - S_{\rm Sud}^{qq} 
+ 2 \left( \frac{3}{4} - \gamma_E \right) \gamma^q_{\mu_r,\mu} -2 \ln \Gamma(\gamma^q_{\mu_r,\mu})
+ \gamma^q_{\mu_r,\mu} \ln [(1-\xi)(1-\xi')].
\end{align}
Here, we take the limit that $\ln\frac{1}{\xi} \approx 1-\xi$ and $\ln\frac{1}{\xi'} \approx 1-\xi'$ when $\xi$ and $\xi'$ are both close to 1. This means that both the initial state and the final state radiations are near threshold. The phase space for such scenario is quite limited. In most cases, only either the initial state radiation or the final state radiation is near threshold. Therefore, only one of $\xi$ and $\xi'$ is close to 1 and the other is far away from $1$. Then, the term $\ln[(1-\xi)(1-\xi')]$ in the above equation should be replaced by $\ln(1-\xi)$ or $\ln(1-\xi')$ accordingly. Also, we approximately write $\ln \Gamma(x) \approx - \gamma_E x$ when $x$ is of the order of $1$. In the case of the fixed strong coupling, the saddle point $r_{\rm sp}$ is determined by 
\begin{align}
\frac{\dd E(\rperp)}{\dd \rperp} \Big{|}_{\rperp = r_{\rm sp}}^{\text{fixed}\, \alpha_s} = 0 \quad \Rightarrow \quad 1 - \frac{\alpha_s C_F}{\pi} \ln \frac{\kperp^2 (1-\xi) r_{\rm sp}^2}{c_0^2} = 0.
\end{align}
Solving the above equation yields the position of the saddle point, which reads
\begin{align}
\mu_{\rm sp}^2 \equiv \frac{c_0^2}{r_{\rm sp}^2} = \kperp^2 (1-\xi) \exp\left[ - \frac{\pi}{\alpha_s C_F} \right].
\end{align}
This indicates that the nature choice of $\Lambda^2$ is roughly $\kperp^2 (1-\xi)$. This result parametrically agrees with the choice of the semi-hard scale $\Lambda^2$ discussed in our previous intuitive derivation and the intermediate scale $\mu_i$ suggested in Refs.~\cite{Becher:2006nr,Becher:2006mr}.

Similarly, we consider the case for the running coupling and write down the $E (\rperp)$ function explicitly as follows,
\begin{align}
& E (\rperp) = \ln (\rperp^2 \Lambda_{\rm QCD}^2) + \frac{C_F}{N_c\beta_0} \ln \frac{\kperp^2}{\Lambda_{\rm QCD}^2} \ln\ln\frac{c_0^2}{\rperp^2 \Lambda_{\rm QCD}^2} + \frac{C_F}{N_c \beta_0} \ln\frac{\kperp^2 \rperp^2}{c_0^2} + \frac{C_F}{N_c \beta_0} \left[ \ln\ln \frac{c_0^2}{\rperp^2\Lambda_{\rm QCD}^2} \right] \ln (1-\xi) + \cdots,
\end{align}
where $\beta_0 = \frac{11}{12}-\frac{n_f}{6N_c}$. Here, we have only kept the $\rperp$-dependent terms. Then we find the saddle point as
\begin{align}
\mu_{\rm sp}^2 \equiv \frac{c_0^2}{r_{\rm sp}^2} = \Lambda_{\rm QCD}^2 \left[ \frac{\kperp^2 (1-\xi)}{\Lambda^2_{\rm QCD}} \right]^{\frac{C_F}{C_F+N_c \beta_0}}. \label{th-saddle}
\end{align}
This result is in line with the saddle point found in Ref.~\cite{Collins:1984kg} in the so-called CSS formalism. In particular, if we replace $\kperp^2 (1-\xi)(1-\xi')$ with $Q^2$ and set $C_F=4/3$, $N_c=3$ and $n_f=5$, we can exactly reproduce the result in Ref.~\cite{Collins:1984kg} and find $\frac{1}{\mu_{\rm sp}} = \frac{1}{\Lambda_{\rm QCD}} \left[ \frac{Q}{\Lambda_{\rm QCD}} \right]^{-0.41}$.

Next, let us study $F(k_\perp - q_\perp)$ in the second region where $q_\perp \sim \kperp$, and find the saddle point in the $\rperp$ integral involving the dipole gluon distribution. In this case with $q_\perp \sim \kperp$, the phase factor $e^{-i(\kperp-q_\perp)\cdot \rperp}$ or the Bessel function $J_0 (|\kperp-q_\perp||\rperp|)$ can set to $1$. The integral of interest becomes
\begin{align}
\int_0^\infty \dd \rperp^2 S^{(2)} (\rperp) = \int_0^\infty \dd \rperp^2 \exp\left[ - \frac{1}{4} Q_s^2 \rperp^2 \right]
= \int_{-\infty}^{+\infty} \dd \ln (\rperp^2 \Lambda_{\rm QCD}^2) \exp\left[ \ln(\rperp^2 \Lambda_{\rm QCD}^2) - \frac{1}{4} Q_s^2 \rperp^2 \right],
\end{align}
where $S^{(2)} (\rperp)$ is approximately equal to $\exp\left[ - \frac{1}{4} Q_s^2 \rperp^2 \right]$ as suggested in the GBW model for the dipole scattering amplitude with $Q_s$ the saturation momentum. It is clear that the saddle point of this integral locates at $\frac{1}{r_{\rm sp}^2} = \frac{Q_s^2}{4}$. In the low momentum region where $q_\perp \sim \kperp$, the GBW model usually provides a good description of the transverse momentum distribution for $F(k_\perp - q_\perp)$, while it does not have the power law tail in the high $q_\perp$ region.  

To summarize, the semi-hard auxiliary scale $\Lambda^2$ in Eq.~(\ref{eq:final-prescription}) is determined by the dominant region of $\rperp$-integral in Eq.~(\ref{eq:qq-rperp}). Physically speaking, there are two competing mechanisms which controls the $r_\perp$ integral in Eq.~(\ref{eq:qq-rperp}). When the final state jet transverse momentum $\kperp$ mainly comes from the dipole gluon distribution, we find that the semi-hard scale is given by Eq.~(\ref{th-saddle}). On the other hand, when the saturation effect is strong, we can see that the typical semi-hard scale should be of the order $Q_s^2$. Near the kinematic threshold when $k_\perp$ is large, we need to minimize the typical $r_\perp$ in Eq.~(\ref{eq:qq-rperp}) to avoid strong cancellation caused by the oscillation phase factor. Thus, the dominant contribution to the whole integral in Eq.~(\ref{eq:qq-rperp}) comes from the larger scale of these two, or equivalently speaking, the smaller $r_{\rm sp}$. Therefore, we arrive at the following quantitative prescription for the choice of $\Lambda^2$ 
\begin{align}
\Lambda^2 \approx \max \left\{ \Lambda_{\rm QCD}^2 \left[ \frac{\kperp^2 (1-\xi)}{\Lambda^2_{\rm QCD}} \right]^{\frac{C_F}{C_F+N_c\beta_0}}, Q_s^2\right\}. \label{lambdavalue2}
\end{align}
To get the result for the $g\to g$ channel, we only need to replace the color factor $C_F$ with $N_c$.

\begin{figure}[htb]
\includegraphics[width=0.4\textwidth]{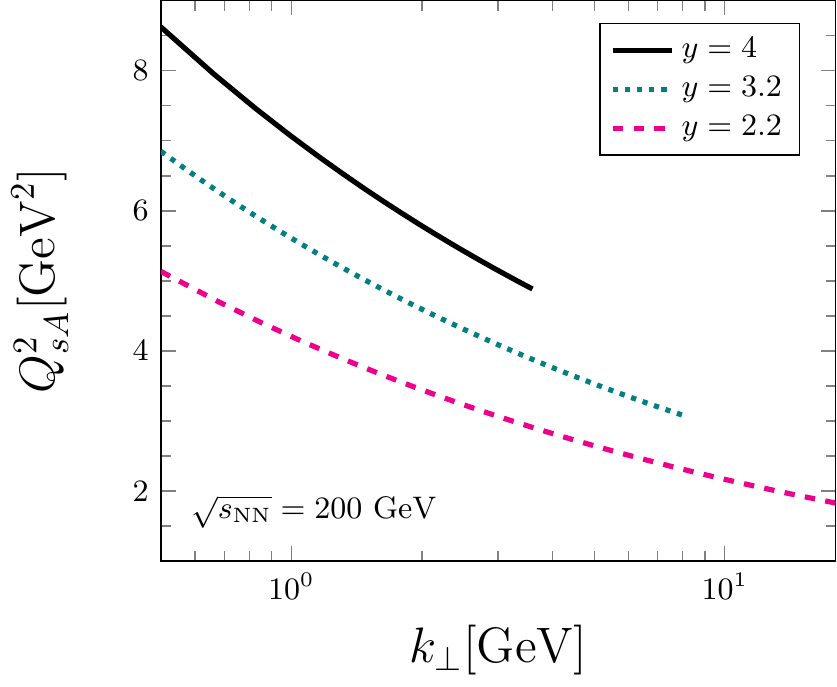}
\caption{Saturation momentum square $Q_{sA}^2=5 \, Q^2_{sp} (x_g)$ as a function of $k_{\perp}$ at different rapidities in the GBW model.}
\label{fig:Qs2}
\end{figure}

It is time to use Eq.~(\ref{lambdavalue2}) to estimate the natural choices of the auxiliary scale $\Lambda^2$ in various kinematic regions. At the RHIC energy, one can take the typical $\kperp \sim 10$ GeV. Assuming $1-\xi \sim 0.5$, the saddle point of the first region is $\mu_{\rm sp}^2 \sim 0.7$ GeV$^2$ for the $q\to q$ channel. To obtain the above numbers, we have used $n_f=4$ and $\Lambda_{QCD}=0.15$ GeV. For the $g\to g$ channel, the value becomes $2$ GeV$^2$, which is not large either. In the large-$x$ region, the cross-section is dominated by the quark channel since the quark density is much larger than the gluon density. Therefore, the scale $\Lambda^2$ at the RHIC energy is mainly determined by the saturation momentum in $dAu$ collisions. In our numerical evaluation, we employ the GBW model to estimate the saturation momentum, which is given by\cite{Golec-Biernat:1998zce}
\begin{equation}
Q^2_{sp} (x_g) = Q^2_0 \left(\frac{x_0}{x_g}\right)^{\lambda},
\end{equation}
where $x_g = \frac{\kperp}{\sqrt{s_{\rm NN}}} e^{-y} $, $x_0 = 3.04\times 10^{-4}$, $\lambda = 0.288$ and $Q_0 ^2 =1$ GeV$^2$. For the gold nucleus target, we use $Q^2_{sA} (x_g) =5 Q^2_{sp}(x_g)$ and show the saturation momentum as a function of $k_{\perp}$ at different rapidities in Fig.~\ref{fig:Qs2}. The corresponding saturation momentum for the proton target in $pp$ collisions is only $1/5$ of the values shown in Fig.~\ref{fig:Qs2}. The semi-hard scale, $\Lambda^2$, needs to be much larger than $\Lambda_{\rm QCD}^2$. Therefore, we choose it to be a few times smaller than that in $dAu$ collisions but still larger than 1 GeV$^2$. We show the values of $\Lambda^2$ used to compute the cross-sections at the RHIC energy in Table~\ref{tab:value-lambda}.

\begin{table}[htb]\centering
\caption{Values of $\Lambda^2$ in calculating the cross-sections at $\sqrt{s_{\rm NN}} = 200$ GeV at RHIC.} 
\label{tab:value-lambda}
\begin{tabular}{c|c|c|c|c|c|c}
\hline
\hline
Rapidity   & \multicolumn{2}{c|}{$y=2.2$} & \multicolumn{2}{c|}{$y=3.2$} & \multicolumn{2}{c}{$y=4$} \\
\hline
Collisional systerms & ~~~~ dAu ~~~~  & ~~~~ pp ~~~~ & ~~~~ dAu ~~~~ & ~~~~ pp ~~~~ & ~~~~ dAu ~~~~ & ~~~~ pp ~~~~ \\
\hline
$\Lambda^2$ (GeV$^2$) & $2\sim 4$ & $\sim 1$ & $3\sim 7$ & $\sim 2$ & $5 \sim 9$ & $\sim 3$ \\
\hline
\hline
\end{tabular}
\end{table}

At the LHC energy, the typical $\kperp$ is usually very large. Taking $\kperp \sim 100$ GeV and $1-\xi \sim 0.5$, we find $\Lambda^2 \sim 4$ GeV$^2$ for $q\to q$ channel and $\Lambda^2 \sim 30$ GeV$^2$ for $g\to g$ channel. Since the $g\to g$ channel becomes more important at the LHC energy, we choose $\Lambda^2$ to be $20$ GeV$^2$ at $\sqrt{s}=5.02$ TeV and set $\Lambda^2$ to be $40$ GeV$^2$ at $\sqrt{s}=13$ TeV.

In principle, we expect that the dependence on the auxiliary scale is mild except for case in the threshold region where the resummed result gets enhanced by the threshold resummation. In Fig.~\ref{fig:lambda-dependence}, the $\Lambda^2$-dependence of the resummed cross-sections at different collisional energies are presented. In general, the $\Lambda^2$-dependence is strong when the threshold resummation is important (i.e., when $p_T$ is closer to the kinematic threshold). Furthermore, at given $p_T$, the resummed cross-section increases with decreasing $\Lambda^2$ due to increased threshold logarithms. These effects are manifest at the RHIC energy as shown in the left panel of Fig.~\ref{fig:lambda-dependence}. At the RHIC energy, we see clearly that the $\Lambda^2$-dependence becomes stronger at larger $p_T$. Taking into account that the maximum $p_T$ allowed by the kinematics is just around $8$ GeV at $y=3.2$, the threshold resummation is actually vital at large $p_T$. At low $p_T$, the resummed cross-section is less sensitive to the value of $\Lambda^2$. At the LHC energy, the $p_T$ regions that we are interested in are actually still far away from the kinematic threshold. Therefore, as shown in the right panel of Fig.~\ref{fig:lambda-dependence}, the $\Lambda^2$ dependence in the resummed cross-section is rather mild. 

\begin{figure}[h!]
\centering
\includegraphics[width=0.4\textwidth]{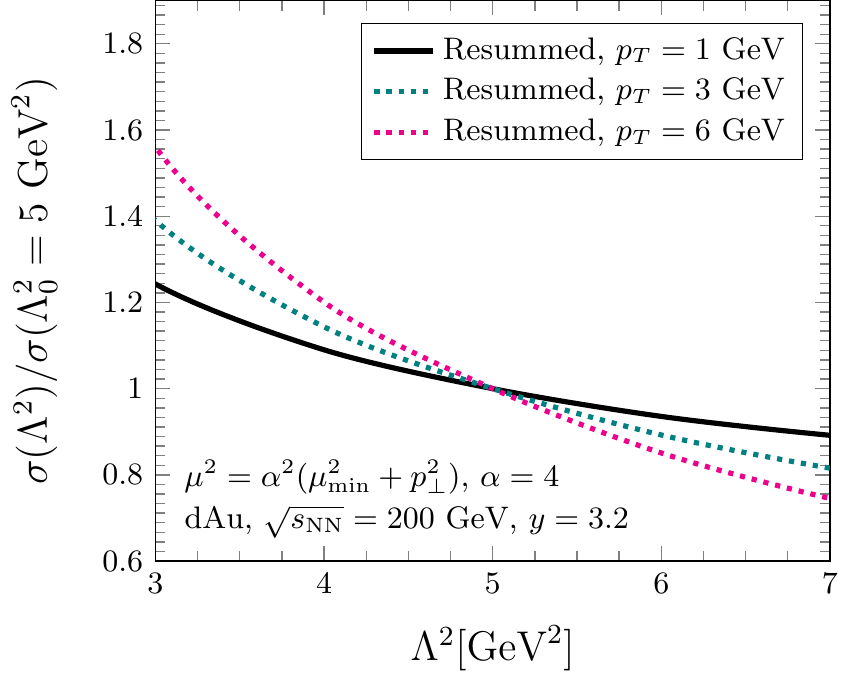}
\includegraphics[width=0.4\textwidth]{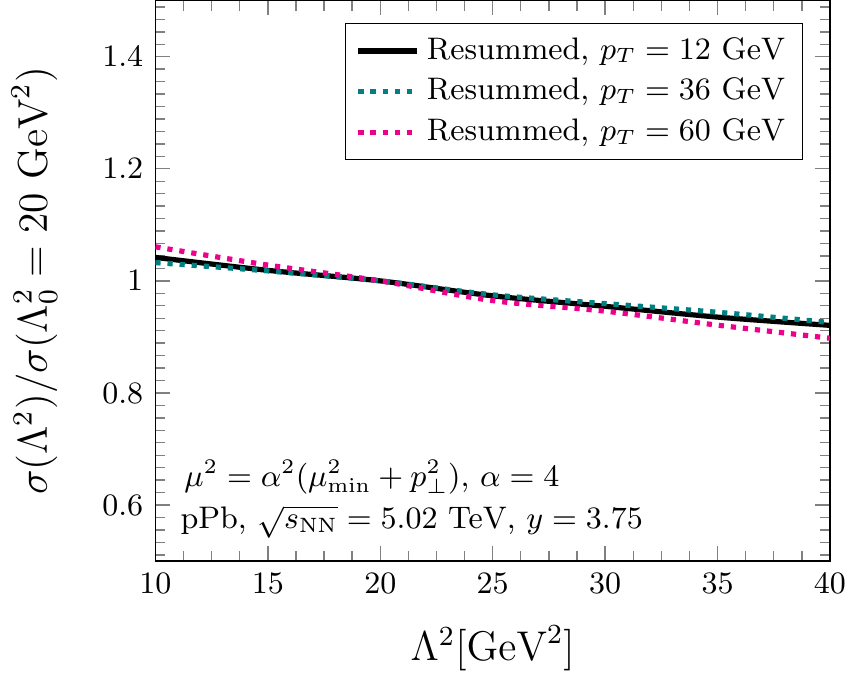}
\caption{The ratios of the resummed cross-section to the reference cross-section at certain value of $\Lambda^2_0$ as a function of $\Lambda ^2$ at $\sqrt{s_\text{NN}} = 200$ GeV (the left plot) and $\sqrt{s_\text{NN}} = 5.02$ TeV (the right plot). At the RHIC energy, we choose $\Lambda^2_0 =5\text{GeV}^2$, while we set $\Lambda^2_0 =20\text{GeV}^2$ at the LHC.}
\label{fig:lambda-dependence}
\end{figure}

\section{Dipole scattering amplitude and small-$x$ gluon distribution}
\label{sec:rcbk}

\begin{figure}[b]
\includegraphics[width=0.4\textwidth]{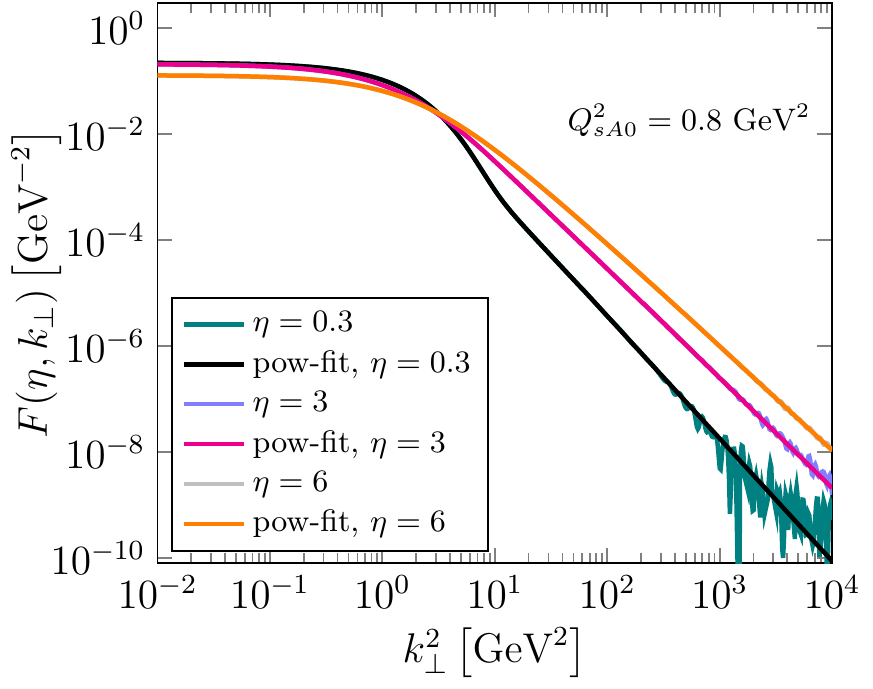}
\caption{
The numerical Fourier transform of $S^{(2)}_{x_g} (\rperp)$ compared with the power-law fit in the high $k_\perp$ region. The oscillating curves indicate that the direct numerical Fourier transform is inaccurate at high $k_\perp$ and they are the artifacts of limited sampling. The power-law fit method is used to capture the key feature of $F(\eta,\kperp)$ in the high $\kperp$ region and remove the non-physical oscillation.
}
\label{fig:rcbk}
\end{figure}

In the above derivation, the dipole gluon distribution $F(\eta,\kperp)$ is defined as the Fourier transform of the dipole scattering amplitude $S^{(2)}_{x_g}( \rperp)$ in the fundamental representation, where $\eta=\ln\frac{x_0}{x_g}$ is the rapidity range of the small-$x$ evolution with $x_0=0.01$. Conventionally, we evolve the dipole scattering amplitude with initial condition starting from $x_g=x_0=0.01$ in the CGC formalism. This means that we usually assume that the CGC formalism starts to apply when $x_g<x_0$. The dimension of $F(\eta,\kperp)$ is GeV$^{-2}$. It is related to $\fcal^{x_g}(\kperp)$ by $F(\eta,\kperp)=\fcal^{x_g}(\kperp)/\Sperp$. In principle, $F(\eta,\kperp)$ and $S^{(2)}_{x_g}(\rperp)$ implicitly depend on $x_g$. In the previous sections, we have suppressed the $x_g$ or $\eta$ dependence for simplicity. In our numerical implementation, we define $S^{(2)}_{x_g}( \rperp)  = 1 - N(\eta, \rperp)$. Here $N(\eta, \rperp)$ is the solution to the rcBK evolution equation~\cite{Golec-Biernat:2001dqn,Balitsky:2006wa, Kovchegov:2006vj, Gardi:2006rp, Balitsky:2007feb,Albacete:2007yr, Albacete:2010sy,Berger:2010sh}, which is given by
\begin{align}
\frac{\dd N (\eta, \rperp)}{\dd \eta} = 
&
\int \dd^2 \rbperp
{\cal K}_{\rm BK} (\rperp,\rbperp,\rcperp)
[N (\eta, \rbperp) + N(\eta, \rcperp) - N(\eta, \rperp) - N(\eta, \rbperp) N(\eta, \rcperp)],
\end{align}
where 
$\rcperp \equiv \rperp - \rbperp$. We employ the approach proposed in Ref.~\cite{Balitsky:2006wa} to take into account the running coupling corrections. The BK evolution kernel, ${\cal K}_{\rm BK}$, is then given by
\begin{align}
{\cal K}_{\rm BK} (\rperp,\rbperp,\rcperp) = \frac{\alpha_s (\rperp) N_c}{2\pi^2} 
\left[ 
\frac{\rperp^2}{\rbperp^2 \rcperp^2} 
+ \frac{\alpha_s(\rbperp) - \alpha_s(\rcperp)}{\alpha_s(\rcperp)} \frac{1}{\rbperp^2}
+ \frac{\alpha_s(\rcperp) - \alpha_s(\rbperp)}{\alpha_s(\rbperp)} \frac{1}{\rcperp^2}
\right],
\end{align} 
where 
\begin{align}
\alpha_s (\rperp) = \frac{4\pi}{(11-\frac{2}{3}n_f) \ln \frac{4}{a_0^2 r_*^2}},
\end{align}
with $a_0 = 0.2$ GeV, $n_f=3$, $r_*^2 = \frac{\rperp^2}{1+\rperp^2/r_{\rm max}^2}$ and $r_{\rm max} = 5$ GeV$^{-1}$. This method is similar to the $b_*$-prescription in the well-known Collins-Soper-Sterman (CSS) formalism and it has also been implemented in Ref.~\cite{Iancu:2020jch}. It ensures a smooth transition from the perturbative region to the infrared regime and freezes the strong coupling constant in the long distance limit. The numerical solution of the rcBK equation with this prescription for $\alpha_s (\rperp)$ is quite close to that with the conventional method where $\alpha_s$ is frozen at a certain value. We set $r_{\rm max}$ to be 5 GeV$^{-1}$, so that $\alpha_s (\rperp \to \infty) = 1$, which is in line with Ref.~\cite{Albacete:2010sy}. 

The above rcBK evolution equation can be solved numerically with a given initial condition. In our calculation, we employ the initial condition given by the modified McLerran-Venugopalan model~\cite{Albacete:2010sy,Fujii:2013gxa}
\begin{align}
N (\eta = 0, \rperp) = 1 - \exp\left[ - \frac{1}{4} (\rperp^2 Q_{s0}^2)^\gamma \ln(e+ \frac{1}{|\rperp| \Lambda_{\rm BK}}) \right],
\end{align}
where $\gamma = 1.118$ and $\Lambda_{\rm BK} = 0.24$ GeV. Here $\Lambda_{\rm BK}$ is used to regulate the infrared physics, and it is usually interpreted as the QCD scale $\Lambda_{\rm QCD}$. As one of the two parameters in the modified MV model for the initial condition, we use the value\cite{Fujii:2013gxa} $\Lambda_{\rm BK} = 0.24$ GeV close to $\Lambda_{\rm QCD}$ which determines the one-loop running coupling in calculating the NLO hard factors. This initial condition is identical to the one used in Ref.~\cite{Albacete:2010sy} and it is extracted from the global analysis to the HERA data. Following Ref.~\cite{Albacete:2010sy}, we use $Q_{sp0}^2 = 0.16$ GeV$^2$ for a proton target and set $Q_{sA0}^2 = 5 Q_{sp0}^2$ for a gold/lead nucleus target for the initial condition, and then solve the above rcBK evolution equation numerically. 

The Fourier transform of the sample solutions are shown in Fig.~\ref{fig:rcbk}, and they agree with previous numerical results well. In the high $k_\perp$ regime, it is quite challenging to obtain a smooth curve for $F(\eta, \kperp)$. This is due to the rapid oscillation of the Bessel function when $k_\perp$ is large and insufficient numerical accuracies. In general, this kind of oscillatory behavior can be mitigated with more computing resources. Nevertheless, in fact, we know that the high $k_\perp$ tail of $F(\eta, k_\perp)$ behaves like a power law. Therefore, in practice, we perform the power-law fit to the numerical result of $F(\eta, k_\perp)$ in the high $\kperp$ region and smoothly connect to the numerical solution at low $\kperp$. The use of the gluon distribution $F(\eta,k_\perp)$ with the power-law fit can significantly improve the numerical accuracy with limited computing resources.

\section{Estimating theoretical uncertainties}
\label{sec:theo-uncertainty}

It is straightforward to find that the $\mu^2$ and $\Lambda^2$ dependences in our calculation exactly vanish at the one-loop order. The residual scale dependences come in terms of higher order corrections in our calculation are due to the truncation of the perturbative expansion of the hard factor. The exact all order results, if they are available, are expected to be independent of the factorization scale and the auxiliary scale. Since we only manage to explicitly obtain the NLO hard factor and we truncate the perturbative expansion of the hard factor at the $\alpha_s$ order, there are still some residual dependence on these two scales in our resummed formalism. Nevertheless, it is the common practice in QCD calculations to use the residual dependence to estimate the unknown higher order corrections. In this paper, we estimate the uncertainty in the theoretical calculation by varying the factorization scale $\mu^2$ from $4(\mu_{\rm min}^2 +\pperp^2)$ to $16(\mu_{\rm min}^2 + \pperp^2)$ and changing $\Lambda^2$ within estimated ranges. Although the uncertainties of PDFs and FFs from the global analysis may also contribute, they are not included in our analysis.  

While estimating the theoretical uncertainties for differential cross-sections is straightforward, the case for the nuclear modification factor $R_{pA}$ is a bit tricky. $R_{pA}$ is defined as  $R_{pA} = \frac{1}{A} \sigma_{pA}/\sigma_{pp}$, which involves the ratio of the $pA$ and $pp$ cross-sections. Experimentally, the measurements of these two cross-sections are independent. On the other hand, in our calculation, they are computed from the same NLO cross-section in the CGC formalism with different initial conditions for the target $k_\perp$ dependent small-$x$ gluon distribution. Therefore, there is some certain level of correlation between theoretical uncertainties of $\sigma_{pA}$ and $\sigma_{pp}$ when we vary scales. 

We use $\delta\sigma_{pA}$ and $\delta\sigma_{pp}$ to represent the theoretical uncertainties of $\sigma_{pA}$ and $\sigma_{pp}$ correspondingly. To estimate the theoretical uncertainty of $R_{pA}$, which is denoted as $\delta {R}$, we need to introduce the correlation between $\sigma_{pA}$ and $\sigma_{pp}$ which is denoted as ${\cal C}(\sigma_{pA},\sigma_{pp})$. The uncertainty of $R_{pA}$ is then given by
\begin{align}
\frac{\delta R}{R_{pA}} = \sqrt{\left(\frac{\delta\sigma_{pA}}{\sigma_{pA}}\right)^2 + \left(\frac{\delta \sigma_{pp}}{\sigma_{pp}}\right)^2 -  2{\cal C}(\sigma_{pA},\sigma_{pp}) \frac{\delta\sigma_{pA}}{\sigma_{pA}} \frac{\delta\sigma_{pp}}{\sigma_{pp}}}. 
\end{align}

\begin{figure}[htb]
\centering
\includegraphics[width=0.4\textwidth]{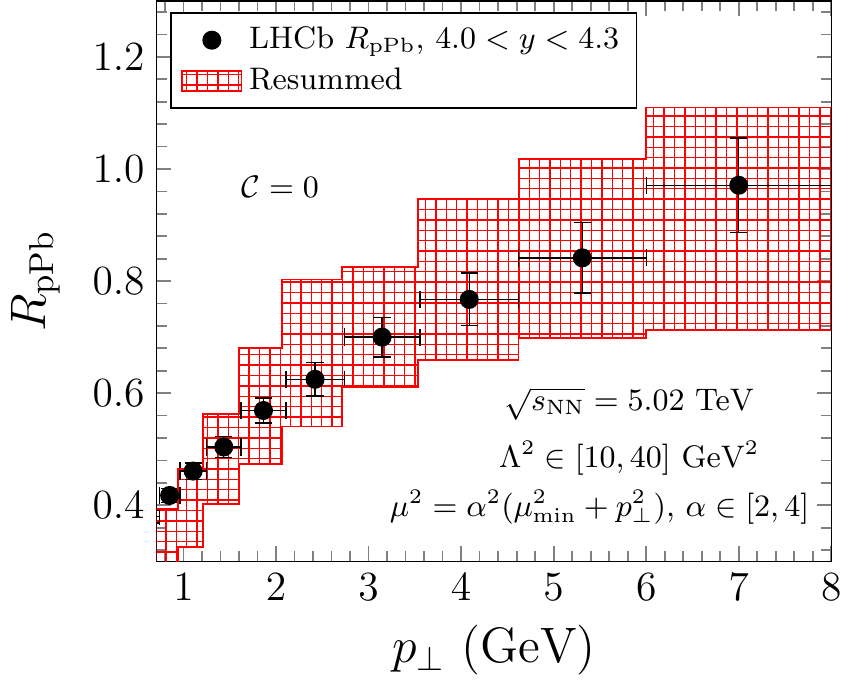}
\includegraphics[width=0.4\textwidth]{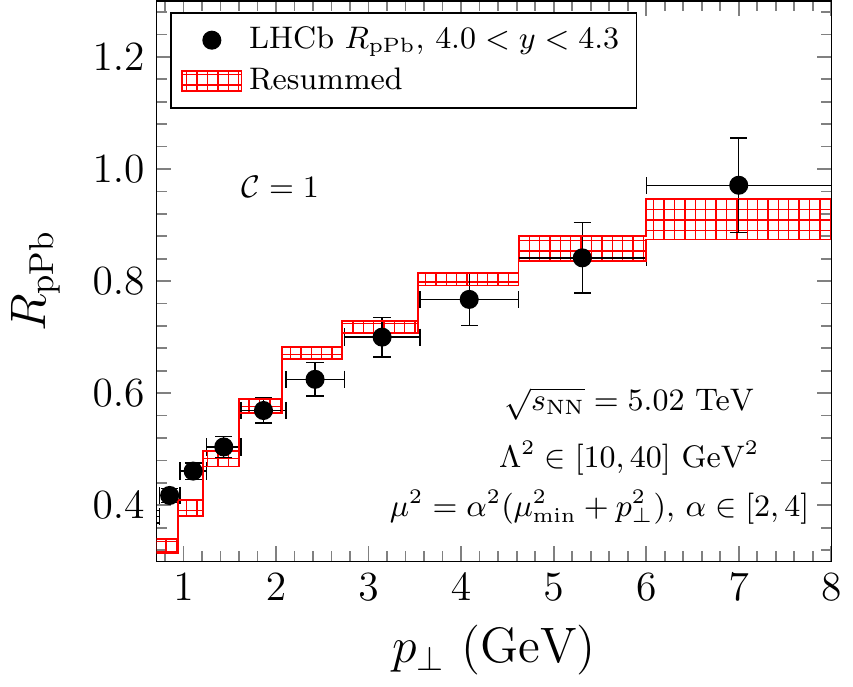}
\caption{Estimates of the theoretical uncertainty for $R_{pPb}$ in two extreme scenarios compared with the LHCb data~\cite{LHCb:2021vww}.}
\label{fig:uncertainties}
\end{figure}

Let us consider two extreme scenarios: (1) $\sigma_{pA}$ and $\sigma_{pp}$ are uncorrelated; (2) $\sigma_{pA}$ and $\sigma_{pp}$ are completely correlated. ${\cal C}(\sigma_{pA},\sigma_{pp}) = 0$ in the first scenario while ${\cal C}(\sigma_{pA},\sigma_{pp})=1$ in the latter. We present our estimates of the theoretical uncertainty for $R_{pPb}$ in these two extreme scenarios compared with the LHCb data in Fig.~\ref{fig:uncertainties}. Clearly, if $\sigma_{pA}$ and $\sigma_{pp}$ are uncorrelated, the uncertainties in these two accumulates. This leads to sizable uncertainties in the final results, as shown in the left panel of Fig.~\ref{fig:uncertainties}. However, if $\sigma_{pA}$ and $\sigma_{pp}$ are completely correlated, the uncertainties largely cancel. Therefore, the uncertainty band in the right panel of Fig.~\ref{fig:uncertainties} is excessively narrow. The actual scenario should be somewhere in-between. In our numerical evaluation, the theoretical uncertainties of $R_{\rm pPb}$ are estimated from the ratios of upper bands of $\sigma_{pA}$ and $\sigma_{pp}$ and those of the lower bands. This simple prescription gives error bands close to the second scenario with the perfect correlation (${\cal C}(\sigma_{pA},\sigma_{pp})=1$). This result can be viewed as the sign that there exists a strong correlation between $\sigma_{pA}$ and $\sigma_{pp}$ in our numerical calculations.

\section{Relating the numerical results to the experimental data}
\label{sec:relating}

In this section, we discuss in detail how to relate our numerical results to experimental measurements, and provide a systematical description of all relevant data with the chosen initial condition and a uniform set of parameters.

Let us first briefly summarize the numerical implementation of threshold resummation for the inclusive hadron production in forward $pp/pA$ collisions. We perform the numerical calculations in the momentum space, since the evaluation in the coordinate space is quite demanding on computation resources and it may result in sizable uncertainties. The LO and one-loop cross-sections are given in Sec.~\ref{sec:full-one-loop}, and the resummed cross-section is given in Sec.~\ref{sec:resummed-full-expression}. We employ the one-loop running coupling for the strong coupling $\alpha_s$, which is given by 
\begin{align}
\alpha_s (\mu^2) = \frac{4\pi}{(11-\frac{2n_f}{3}) \ln \frac{\mu^2}{\Lambda_{\rm QCD}^2}},
\end{align}
where the number of active flavor $n_f=4$ and $\Lambda_{\rm QCD} = 0.15$ GeV. With this set of parameters, we set the one-loop running coupling at the $Z$ pole to be $\alpha_s(\mu=M_Z=91.2 {\rm GeV}) =0.118$. The dipole scattering amplitude, $S^{(2)}(\rperp) $, is obtained by solving the rcBK evolution equation with the initial condition provided by Ref.~\cite{Albacete:2010sy}. The dipole gluon distribution $F(k_\perp)$ is the Fourier transform of $S^{(2)}(\rperp)$. For more details, we refer the reader to Sec.~\ref{sec:rcbk}. Additionally, We utilise the NLO MSTW PDFs \cite{Martin:2009iq} and NLO DEHSS FFs \cite{deFlorian:2014xna} in our numerical evaluation.

On the theoretical calculation side, we execute the numerical implementation mentioned above and numerically calculate the differential cross-section of $\pi^0$ production divided by the overlapping transverse area $\Sperp$ of the collision, which reads
\begin{equation}
\frac{1}{\Sperp} \frac{1}{2\pi \pperp} \frac{\dd^2 \sigma^{\text{pA} \rightarrow \pi \text{X}}}{\dd y \dd \pperp }.
\end{equation} 

On the experimental side, the BRAHMS~\cite{Arsene:2004ux}, ALICE~\cite{ALICE:2012mj} and ATLAS~\cite{ATLAS:2016xpn} collaborations measure the hadron yield, which is related to the differential cross-section as follows
\begin{equation}
\frac{1}{2\pi \pperp} \frac{\dd^2 N ^{\text{pA} \rightarrow h \text{X}}}{\dd y \dd \pperp}
=
\frac{1}{\sigma_{\text{inel}}} \frac{1}{2\pi \pperp} \frac{\dd^2 \sigma^{\text{pA} \rightarrow h\text{X}}}{\dd y \dd \pperp},
\end{equation}
where $h$ represents various kinds of hadrons measured in different experiments and $\sigma_{\text{inel}}$ represents the total inelastic cross-section. Therefore, our results need to be multiplied by a factor of  $\frac{\sigma^h}{\sigma^{\pi}} \frac{\Sperp}{\sigma_{\text{inel}}}$ to compare with these data, where $\sigma^h/\sigma^\pi$ is the rescale factor converting the $\pi$ cross-section calculated in our code to the measured hadron $h$ yields in experiments. We use $\sigma_{\text{inel}} = 2400$ mb in $dAu$ collisions at $\sqrt{s_{\rm NN}} = 200$ GeV~\cite{Arsene:2004ux} and $\sigma_{\text{inel}}  =  2100$ mb in $pPb$ collisions which is measured with at least one charged hadron locating in the range of $2.5 \leq y \leq 4.5$ at the LHC energy~\cite{LHCb:2021vww}. In the $pA/dA$ collisions, we approximately set $\Sperp$ as the transverse area of the target nucleus, since the size of the nucleus is much larger than that of the proton/deuteron. Therefore, we utilize $\Sperp  = 1770$ mb in $dAu$ collisions~\cite{Watanabe:2015tja} and $\Sperp = 1830$ mb in $pPb$ collisions. For the ratio $\sigma^h/\sigma^\pi$, we summarize the parameters used in our numerical evaluations in Table~\ref{table:1}. 

\begin{table}[h]\centering
\caption{List of input values of parameters used in our numerical evaluations.} 
\label{table:1}
\begin{tabular}{c|c|c|c|c}
\hline
\hline
Experiment    & ~~ BRAHMS ~~ & ~~ STAR ~~ & ~~ ATLAS/ALICE ~~ &  LHCb  \\
\hline
Hadron $h$ &  $h^-$  &  $\pi^0$ & $h^{\pm}$  & ~~ prompt $h^\pm$ ~~ \\
\hline
$\sigma^h / \sigma^{\pi}$     & 1.3  &  1 &  2.2 &  2.2 $\times$ 0.85\\ 
\hline
\hline
\end{tabular}
\end{table}

Meanwhile, the STAR~\cite{Adams:2006uz} and LHCb~\cite{LHCb:2021vww,LHCb:2021abm} collaborations directly publish the differential cross-section in the following form
\begin{equation}
 \frac{1}{2\pi \pperp}   \frac{  \dd^2 \sigma ^{\text{pA} \rightarrow   h \text{X} } }{\dd y \dd \pperp }.
\end{equation}
Therefore, only a factor of $\frac{\sigma^h}{\sigma^{\pi}} \Sperp$ is multiplied to our numerical result before it is compared with these experimental data. To account for the fraction of the prompt charged particles measured by the LHCb collaboration\cite{LHCb:2021abm}, we multiply a factor of $0.85$ to the total cross section for charged hadrons as given in the end of Table~\ref{table:1}.

The connection between theoretical calculations and experimental measurements that we adopt remains the same for $pp$ collisions. However, the overlapping transverse area $S_\perp$ needs to be recalibrated in $pp$ collisions, since the approximation which assumes that the size of the target (proton) is much larger than that of the proton projectile is no longer valid. In our calculation, the transverse area $S^{pp}_{\perp}$ becomes a parameter fitted from the data. Typically, its value is expected to vary from $\pi R_p ^2$ to  $4\pi R_p ^2$ with $R_p$ being the proton radius. By using $S^{pp}_{\perp} = 2 \pi (0.9~{\rm fm})^2 = 51$ mb in our numerical evaluation, we find that we can describe the experimental data measured in $pp$ collisions. Furthermore, this value of $S^{pp}_{\perp}$ is close to $\sigma_{\text{inel}} = 41$ mb measured by the BRAHMS collaboration in $pp$ collisions~\cite{Arsene:2004ux}.

\section{Additional Numerical Results}
\label{sec:additional-plots}

In the end, we numerically evaluate the ``LO", ``one-loop", and the ``resummed" NLO cross-sections in the large $N_c$ limit and compare these theoretical curves with RHIC and the LHC data in the plots presented in the main text and additional ones shown below. We set $C_F=N_c/2=3/2$ in our numerical calculation due to the large $N_c$ limit. It is worth-noting that the large $N_c$ approximation is essential since it allows one to significantly simplify both the analytic and numerical calculations. 

As shown above, the threshold resummation helps us systematically take all order threshold logarithms into account and  thus restore predictive power for the one-loop CGC calculations. After including the threshold resummation with the proper choice of the initial condition and scales, the NLO CGC calculation can describe the data from RHIC and the LHC across a wide range of $p_T$ regions and collisional energies as shown above in the main text. In contrast to the LO and one-loop results, the resummed NLO result allows us to make more robust and reliable predictions. 

At the last section of the supplemental material, we provide a lot of additional numerical results and compare the LO, one-loop, and resummed NLO CGC calculations with the rest of all the available data measured in $pA$ ($dAu$) and $pp$ collisions at RHIC and the LHC. As the closing remark, we discuss the applicable regions of the CGC formalism and point out that the small-$x$ calculation may be only valid in the low-$p_T$ region if the rapidity $y$ of the measured hadron is not sufficiently large. 

In our numerical evaluation, we always choose the so-called ``fixed boundary condition'' which sets ${\cal F}^{x_g}(\kperp)$ to $0$ when $x_g>0.01$ and restricts our calculation in the dilute-dense framework. As a comparison in Sec.~\ref{sec:applicable-range}, we also present the ``frozen boundary condition'' which freezes  ${\cal F}^{x_g}(\kperp)$ at $x_g=0.01$ and demonstrate that these two prescriptions lead to little difference numerically in the small-$p_T$ region. One expect that the results from these two prescriptions become distinguishable at high $p_T$ for hadron productions with $y<2.3$. This can be another sign that our NLO CGC should no longer apply.

\subsection{Supplementary Numerical Results for Forward Hadron Productions in $pA$ Collisions}

\begin{figure}[htb]
\includegraphics[width=0.32\textwidth]{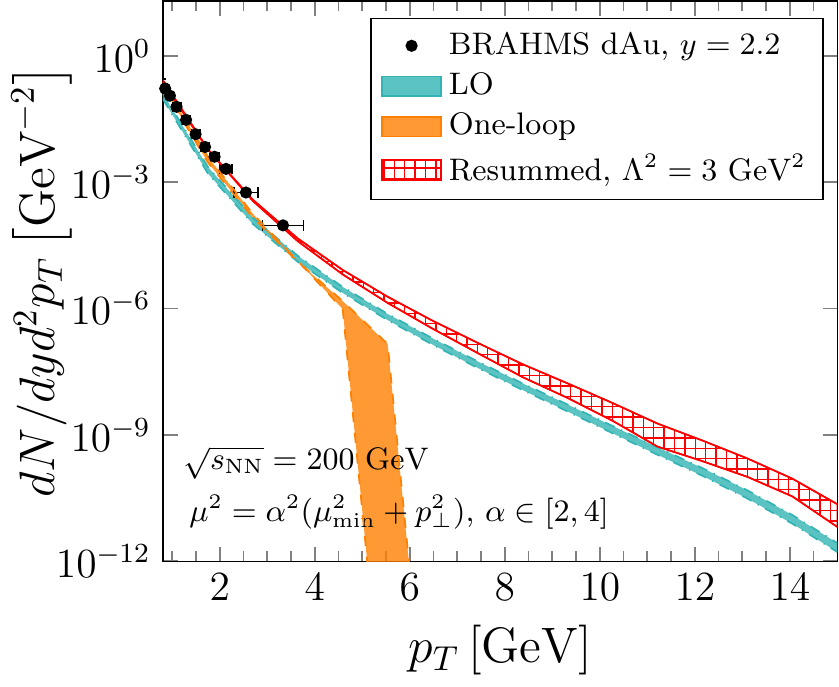}
\includegraphics[width=0.32\textwidth]{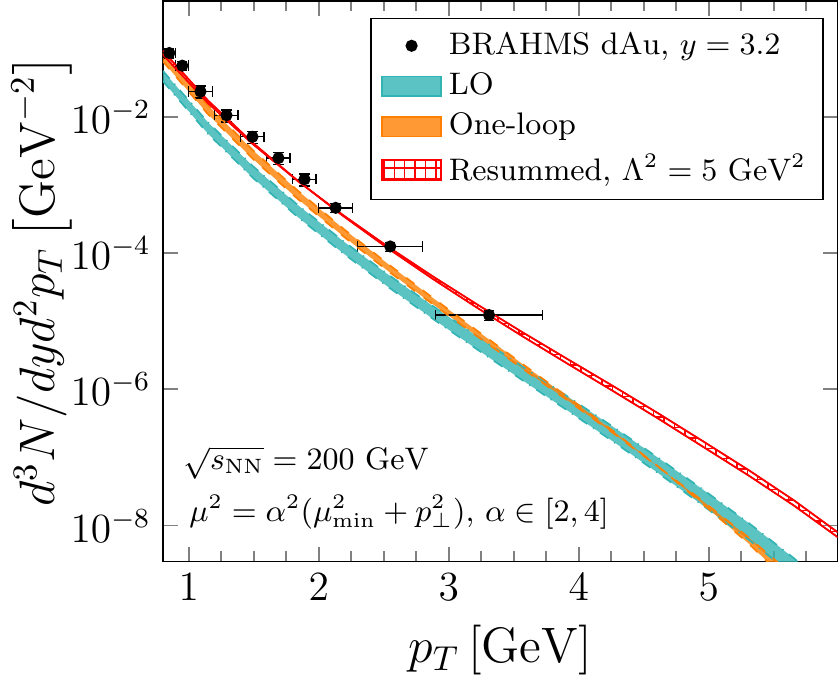}
\includegraphics[width=0.32\textwidth]{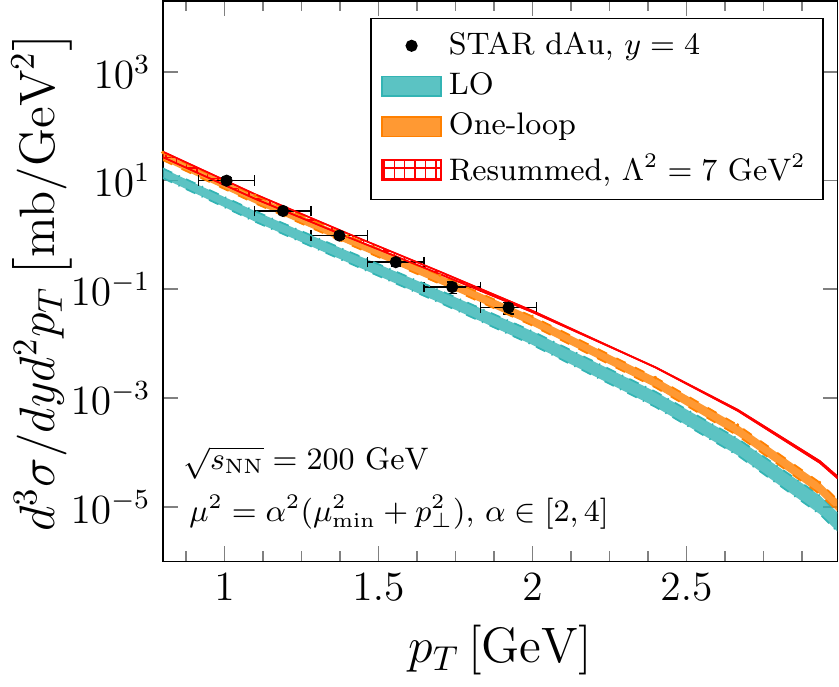}
\caption{Comparisons of the RHIC $dAu$ data \cite{Arsene:2004ux, Adams:2006uz} measured by the BRAHMS and STAR collaborations with the CGC calculations at fixed values of the auxiliary scale $\Lambda$. The error bands are obtained by varying the factorization scale $\mu^2$. }
\label{fig:rhic-dAu-3plots}
\end{figure}

\begin{figure}[htb]
\includegraphics[width=0.32\textwidth]{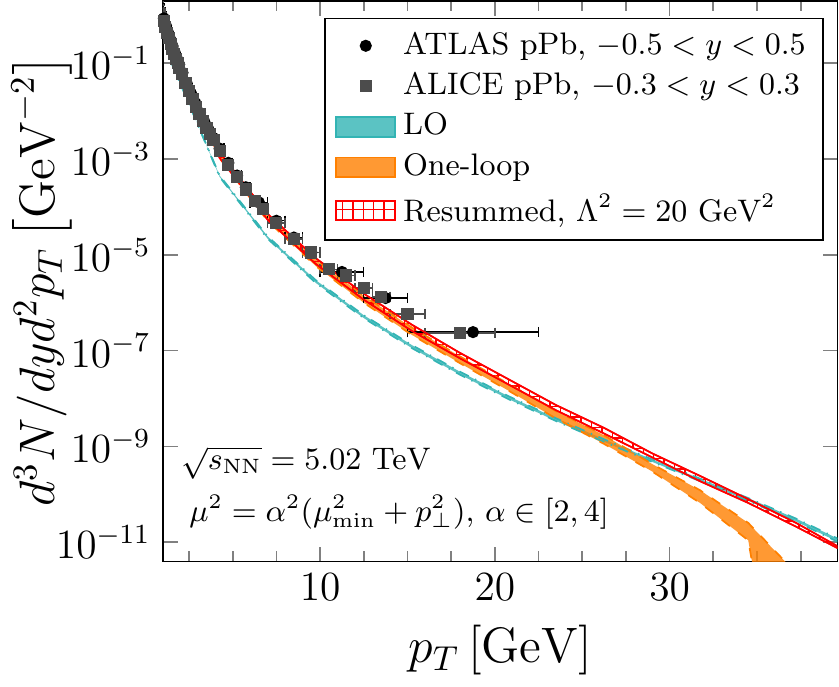}
\includegraphics[width=0.32\textwidth]{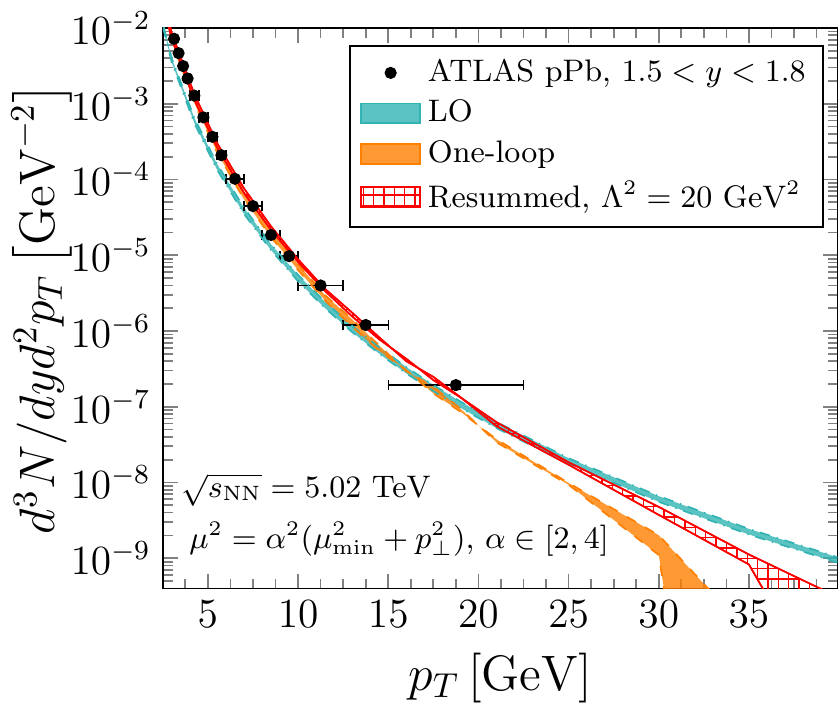}
\includegraphics[width=0.32\textwidth]{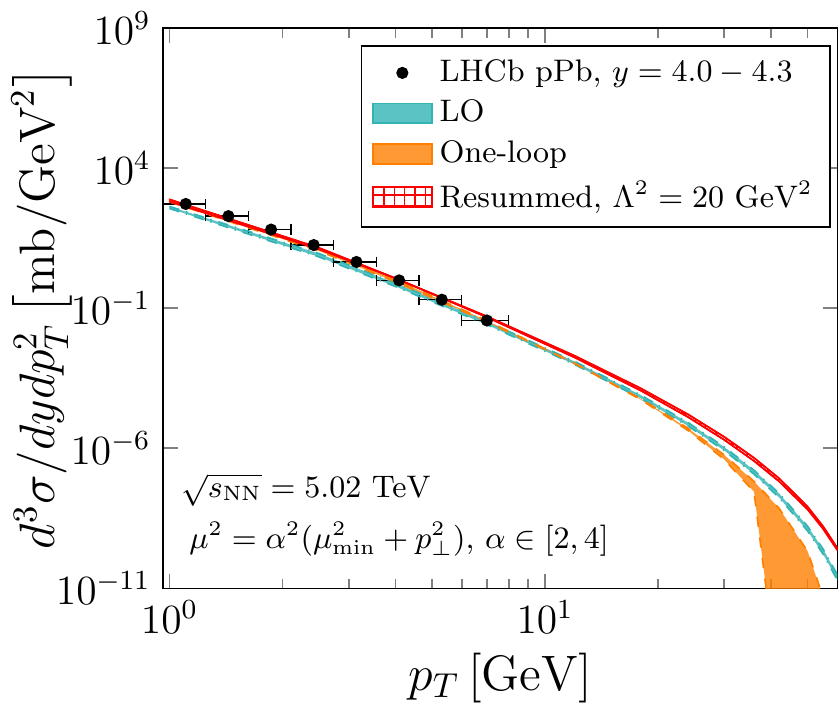}
\caption{Comparisons of the LHC $pPb$ data \cite{ALICE:2012mj, ATLAS:2016xpn,LHCb:2021vww} measured by the ALICE, ATLAS and LHCb collaborations with the CGC calculations with fixed $\Lambda$ scales.}
\label{fig:lhc_2all}
\end{figure}

As discussed briefly earlier in the main text, the inclusive hadron production in forward $pA$ collisions is an ideal process to probe the onset of the gluon saturation phenomenon. In the dilute-dense factorization, the LO calculation is simple and intuitive. However, it contains a rather strong dependence on the choice of the factorization scale $\mu$, and thus it alone can not reliably describe all the experimental data. The comparison between the LO results (represented by teal bands) and experimental data are shown in Fig.~\ref{fig:rhic-dAu-3plots} and Fig~\ref{fig:lhc_2all}. 

The NLO corrections significantly reduces the $\mu$ dependence. The numerical calculations at the one-loop order have already been carried out in the previous works~\cite{Stasto:2013cha,Watanabe:2015tja}. The major issue of the one-loop cross-section is that it turns negative at high-$p_T$ near the threshold region. This negative cross-section issue, as illustrated by the orange bands in the left plot of Fig.~\ref{fig:rhic-dAu-3plots} for RHIC energy and plots in Fig~\ref{fig:lhc_2all} for the LHC energy at $5\text{TeV}$, has attracted a lot of attentions in the community. It becomes manifest that the one-loop cross section consistently turns negative at sufficiently large $p_T$ in the forward rapidity region near the threshold. According to our numerical results, the threshold logarithmic terms are negligible at low $p_T$, whereas they become the dominant contribution in the high $p_T$ region with $p_T\gg Q_s$. At RHIC, since the saturation momentum increases and the kinematic limit of $p_T$ decreases with increasing rapidity $y$, the issue of negative one-loop results becomes less severe in the more forward region. For the rapidity bin around $y = 4$, the negativity does not appear due to the lack of phase space for $p_T$. As laid out above, one can systematically resolve this issue through the implementation of the threshold resummation.

In Fig.~\ref{fig:rhic-dAu-3plots}, we show the comparison between our numerical results and experimental data measured by the BRAHMS and STAR collaborations for $dAu$ collisions at RHIC in three rapidity bins around $y=2.2, 3.2$ and $4$. The resummed calculation has two parameters: the factorization scale $\mu$ and the semi-hard auxiliary scale $\Lambda$. The proper and natural choice of the $\Lambda$ scale is discussed in Sec.~\ref{sec:saddle-point}, and the numerical values in different kinematic regions are shown in Table~\ref{tab:value-lambda}. The central values in Table~\ref{tab:value-lambda} are used in the numerical evaluation. To estimate the theoretical uncertainties at NLO order, we vary the factorization scale $\mu^2$ from $4(\mu_{\rm min}^2+p_T^2)$ to $16(\mu_{\rm min}^2 + p_T^2)$ with $\mu_{\rm min} = 2$ GeV. Remarkably, the resummed calculation not only fixes the negative problem but also improves the quality of the description of the experimental data.

In Fig.~\ref{fig:lhc_2all}, we present the numerical results for $pPb$ collisions at the LHC measured by the ALICE, ATLAS and LHCb collaborations in three rapidity bins near $y=0, \,1.65$ and $4.15$. In the first two middle rapidity regions, our framework can only be applied in the small-$p_T$ region. At high-$p_T$, our numerical results start to deviate from the experimental data since the so-called dilute-dense factorization framework breaks down. More detailed discussions of the applicable windows of our calculation are provided in Sec.~\ref{sec:applicable-range}. Nonetheless, our numerical results yield robust predictions and agree with the experimental data well in the middle rapidity and low-$p_T$ region and in the forward rapidity regime for the entire $p_T$ range.

\subsection{Numerical Results for Forward Hadron Productions in $pp$ Collisions}

\begin{figure}[!h]
\includegraphics[width=0.32\textwidth]{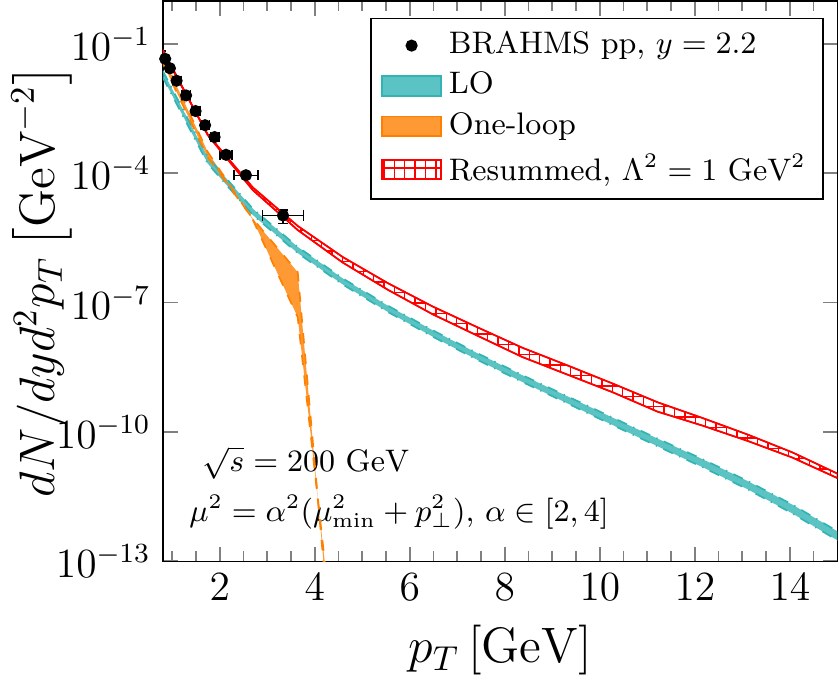}
\includegraphics[width=0.32\textwidth]{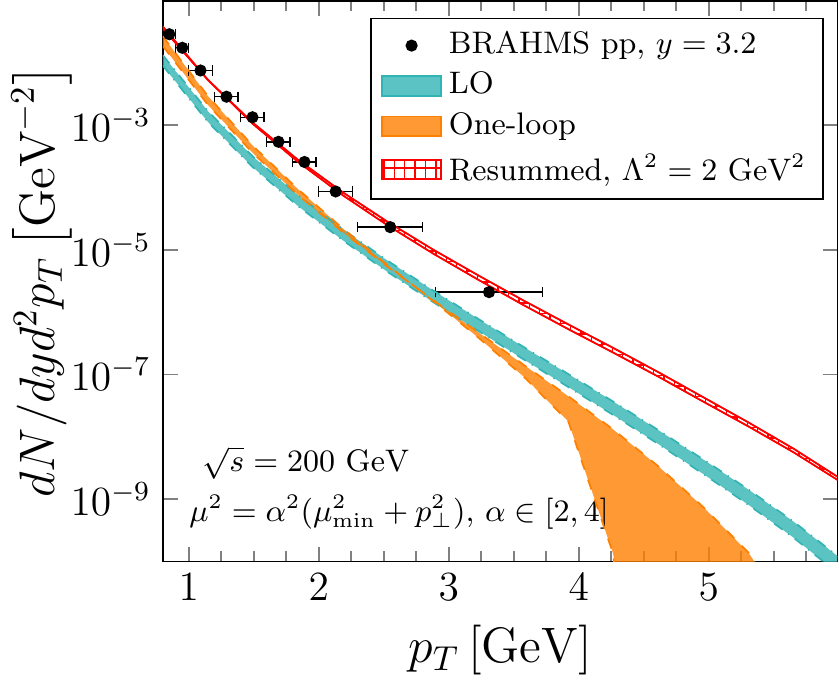}
\includegraphics[width=0.32\textwidth]{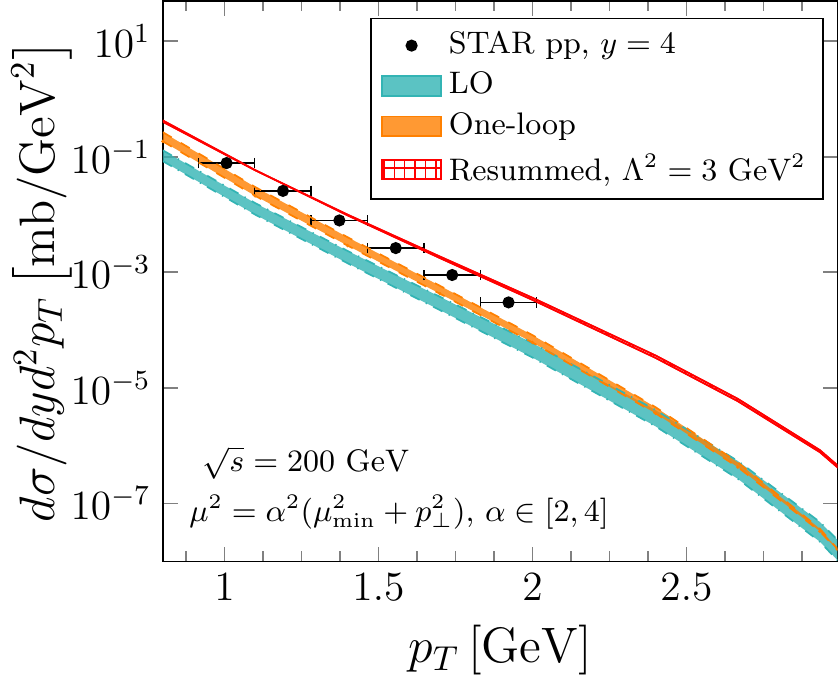}
\caption{Comparisons of the RHIC $pp$ data \cite{Arsene:2004ux, Adams:2006uz} from the BRAHMS and STAR collaborations with our CGC calculations.}
\label{fig:rhic_pp}
\end{figure}

\begin{figure}[!h]
\includegraphics[width=0.32\textwidth]{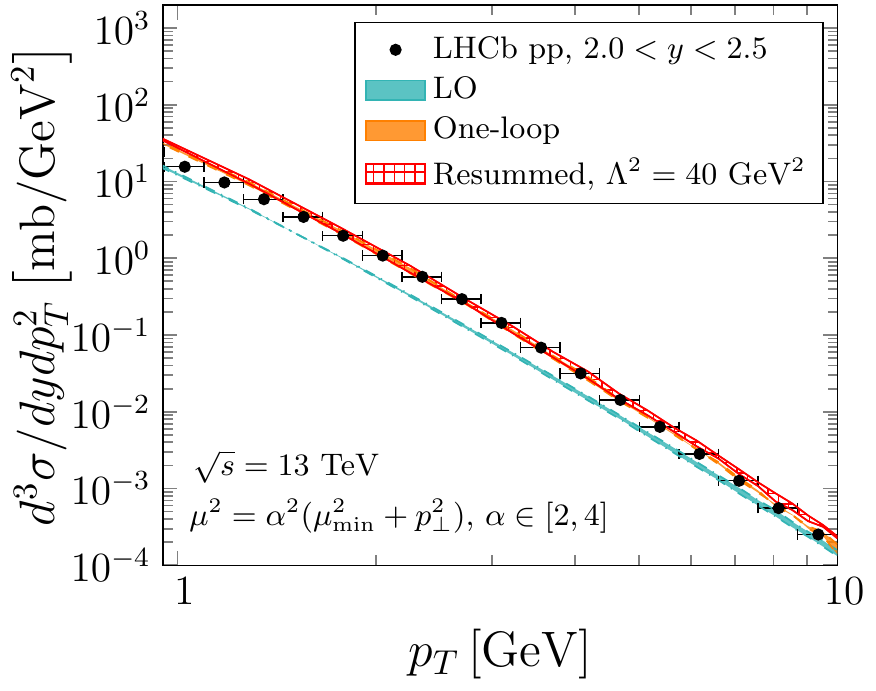}
\includegraphics[width=0.32\textwidth]{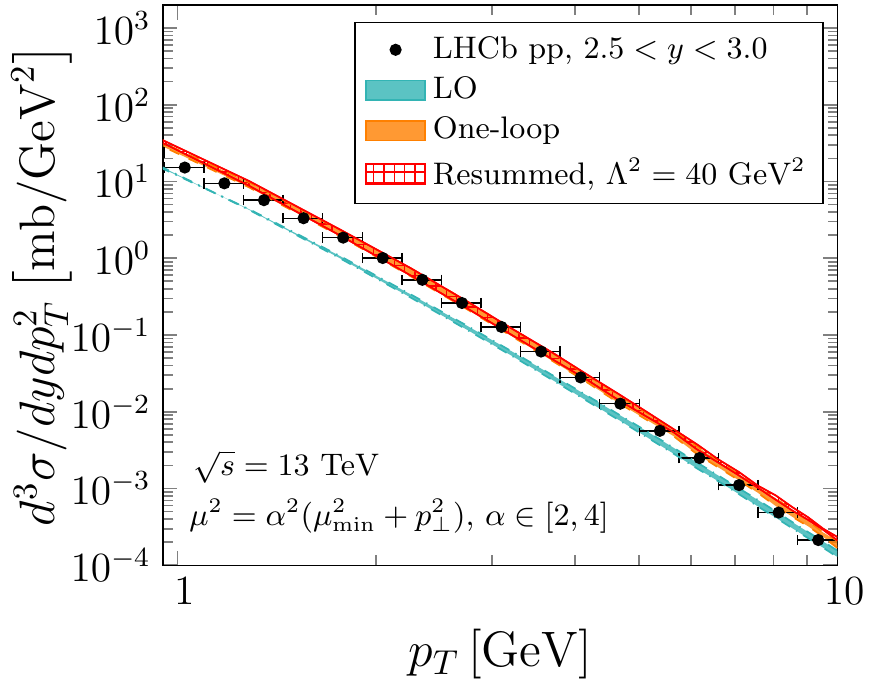}
\includegraphics[width=0.32\textwidth]{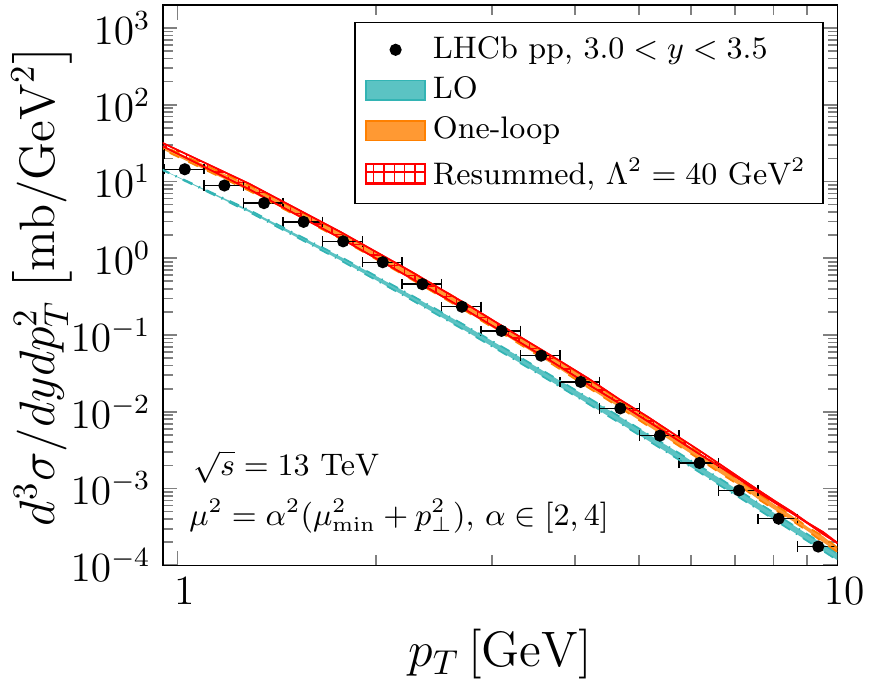}\\
\includegraphics[width=0.32\textwidth]{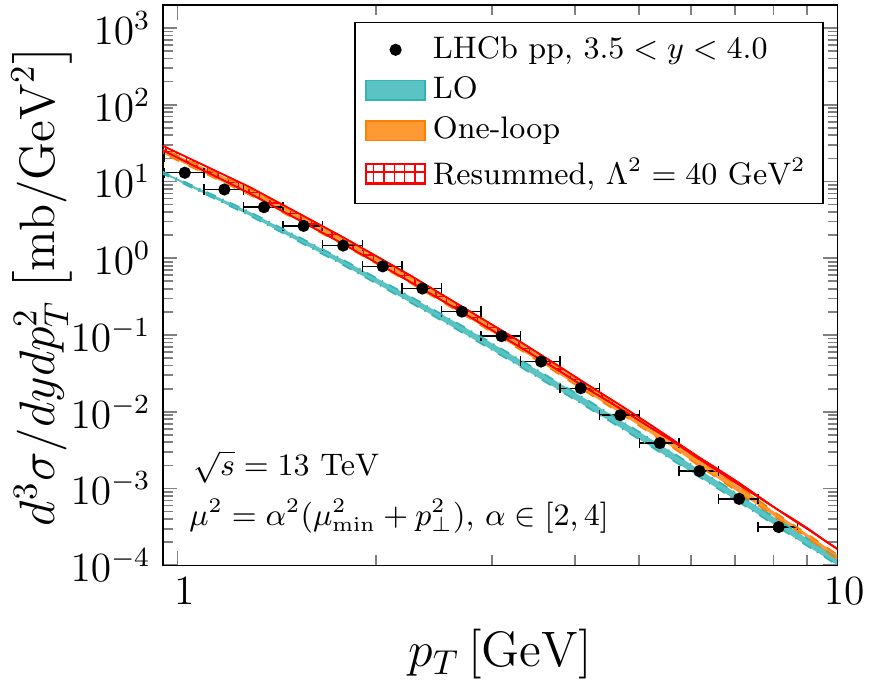}
\includegraphics[width=0.32\textwidth]{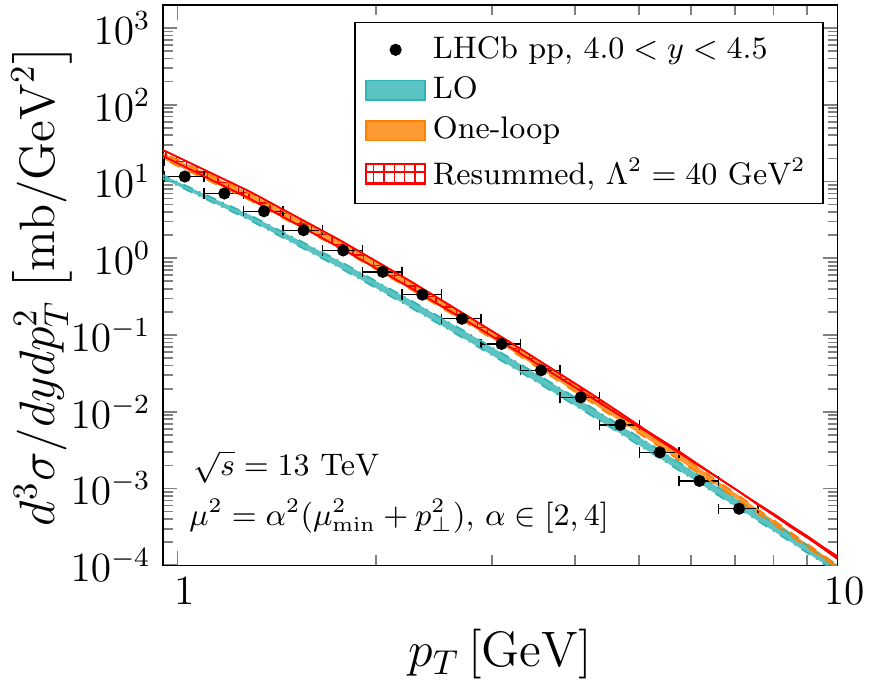}
\includegraphics[width=0.32\textwidth]{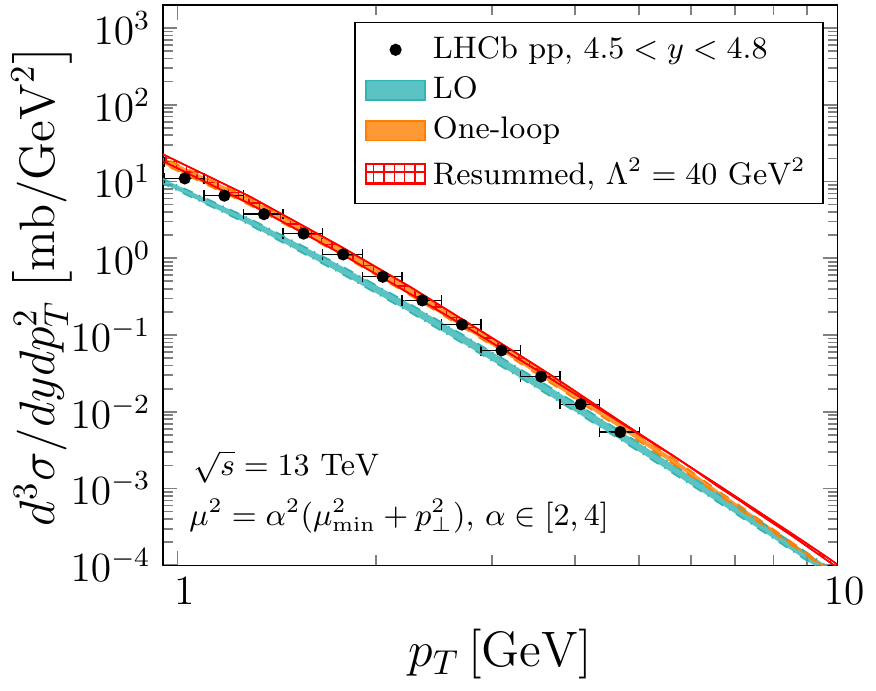}
\caption{Comparisons of the LHCb $pp$ data \cite{LHCb:2021abm} with our CGC calculations at $\sqrt{s}=13$ TeV.   }
\label{fig:lhcb_13tev}
\end{figure}

In principle, the dilute-dense factorization employed in this paper only requires that the gluon density in the target hadron is much higher than the parton density in the projectile. For $pp$ collisions, as long as the rapidity $y$ is sufficiently large (roughly $2$), this requirement can be met. However, as briefly mentioned above, our calculation may not be directly applied to the forward hadron productions $pp$ collisions since we have also assumed that the size of the target hadron (nucleus) is much larger than that of the projectile proton. This simplification allows us to neglect the impact parameter ($b_\perp$) dependence in the dipole scattering amplitude $S^{(2)}(r_\perp)$ and integrate over the impact parameter $b_\perp$ freely. This integral results in an overall normalization $S_\perp$. In $pA$ collisions, $S_\perp$ is approximately the transverse area of the target nucleus. Nevertheless, in $pp$ collisions, $S_\perp$ is supposed to be the overlapping transverse area in which the inelastic $pp$ collision occurs and it is close to the total inelastic cross-section, which is estimated to be a couple times of the target proton transverse area $\pi R_p^2$. Therefore, this overall normalization is less constrained in $pp$ collisions. 

After setting $S_\perp^{pp}=2\pi R_p^2$, we find that our resummed numerical results, which are shown in Figs.~\ref{fig:rhic_pp} and \ref{fig:lhcb_13tev}, can consistently describe the $pp$ data from RHIC at $\sqrt{s_{\text{NN}}}=200 \text{GeV}$ and the recent data measured by LHCb collaboration at $\sqrt{s_{\text{NN}}}=13 \text{TeV}$ for all of the forward rapidity bins. 

Furthermore, due to the rather small saturation momenta in the target proton, the issue of the negative cross-section sets in earlier for the one-loop results in the $p_T \gg Q_s$ region as compared to the case in $dAu$ collisions at RHIC. This can be seen in the left and middle plots of Fig.~\ref{fig:rhic_pp}. Nevertheless, with proper choice of the auxiliary scale $\Lambda$, the resummed results can resolve this issue and restore the predictive power of the CGC NLO calculation in the $p_T \gg Q_s$ region. In comparison, due to the much larger phase space in high energy collisions at the LHC, the negativity issue is much less severe in the observed $p_T$ regions shown in Fig.~\ref{fig:lhcb_13tev}. At the LHC energy, one needs to proceed to a much higher $p_T$ in order to approach the threshold limit, where the negative contributions in the one-loop result start to become dominant.

\subsection{The applicable range of the CGC phenomenology}
\label{sec:applicable-range}

\begin{figure}[hbt]
\begin{center}
\includegraphics[width=0.4\textwidth]{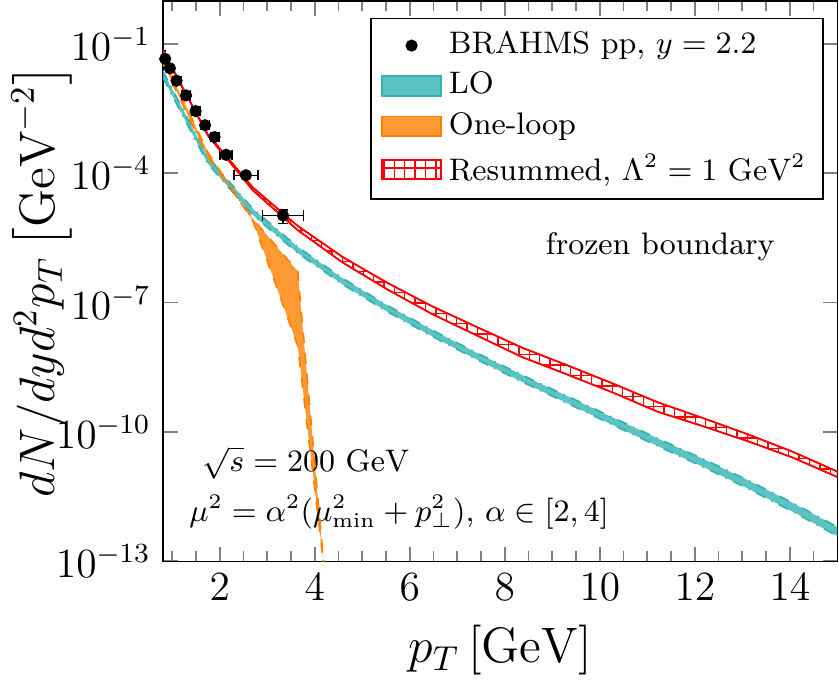}
\includegraphics[width=0.4\textwidth]{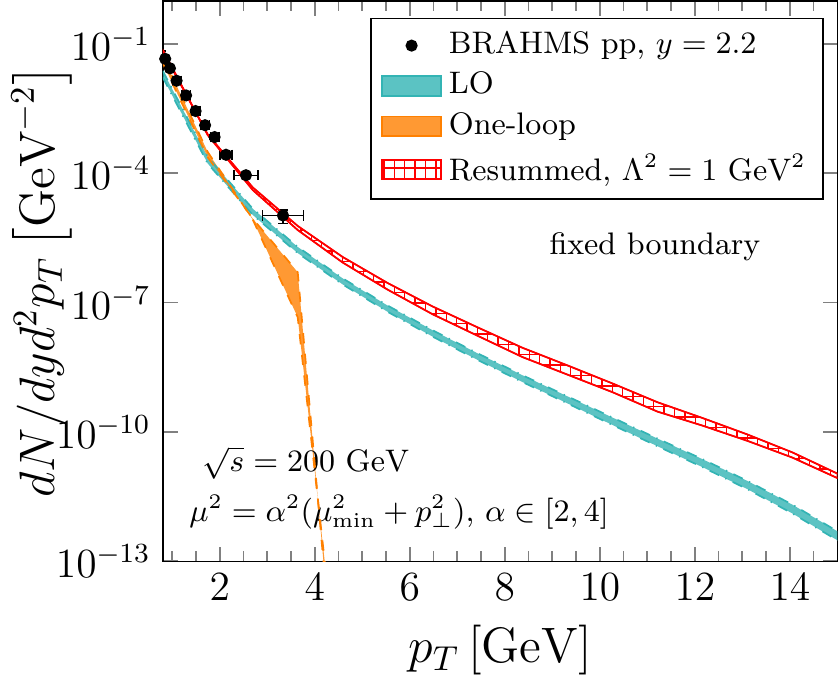}
\end{center}
\caption{Comparisons of numerical results with two different boundary conditions at $\sqrt{s}=200$ GeV in $pp$ collisions\cite{Arsene:2004ux}.}
\label{fig:freeze1}
\end{figure}

\begin{figure}[!h]
\begin{center}
\includegraphics[width=0.4\textwidth]{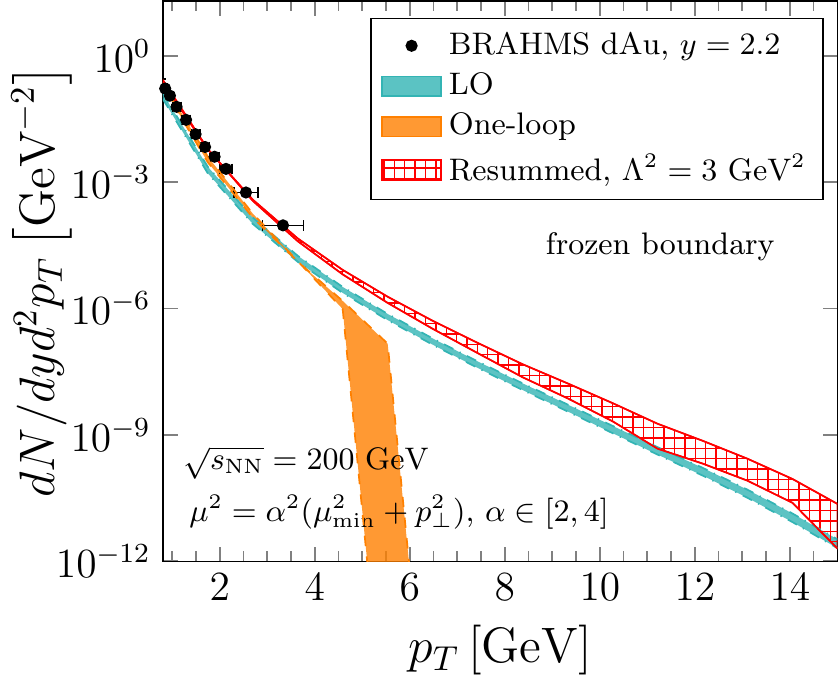}
\includegraphics[width=0.4\textwidth]{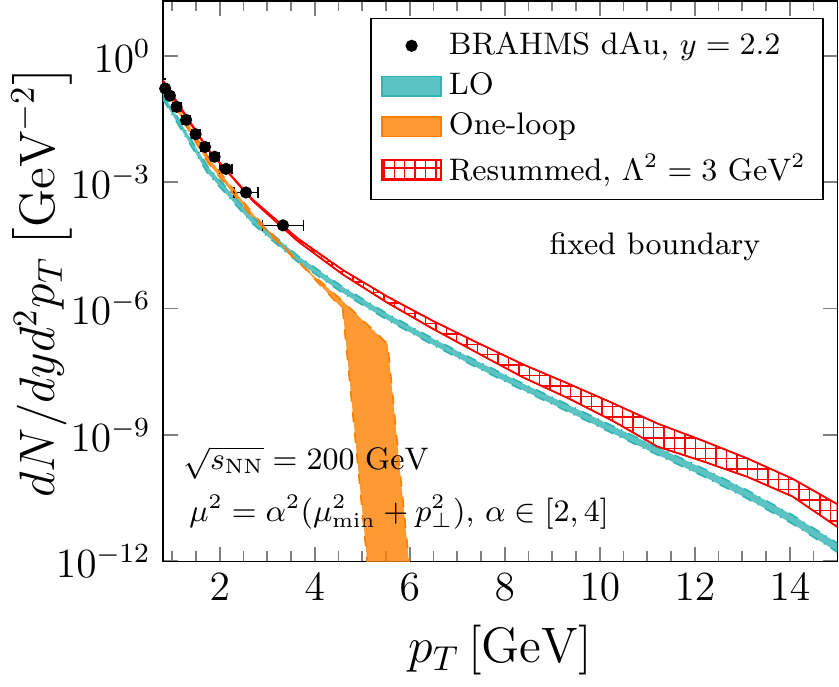}
\end{center}
\caption{Comparisons of numerical results with two different boundary conditions at $\sqrt{s_{\text{NN}}}=200$ GeV in $dAu$ collisions
\cite{Arsene:2004ux}.}
\label{fig:freeze2}
\end{figure}

\begin{figure}[!h]
\begin{center}
\includegraphics[width=0.4\textwidth]{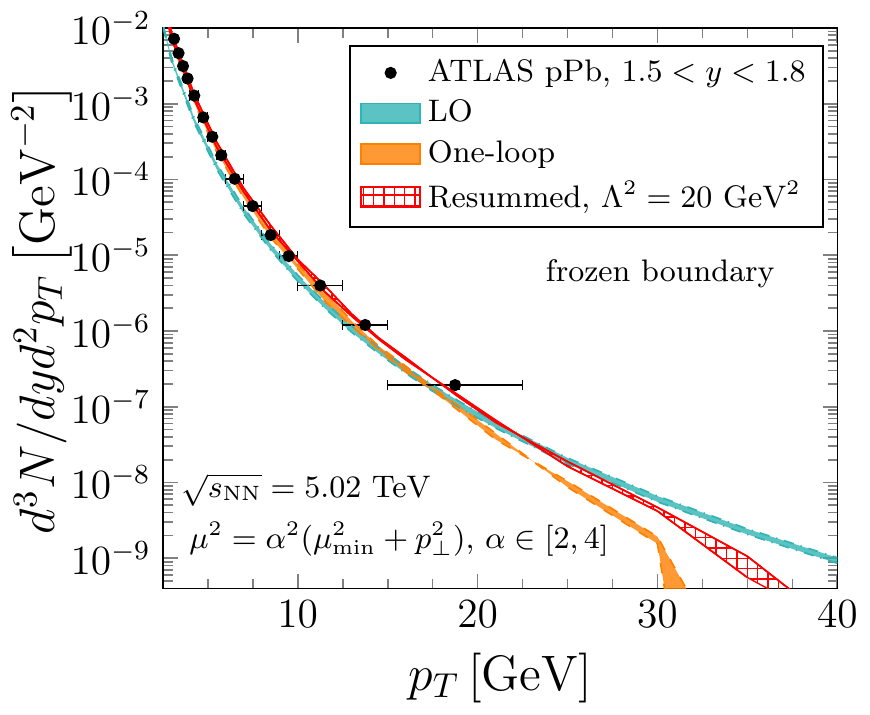}
\includegraphics[width=0.4\textwidth]{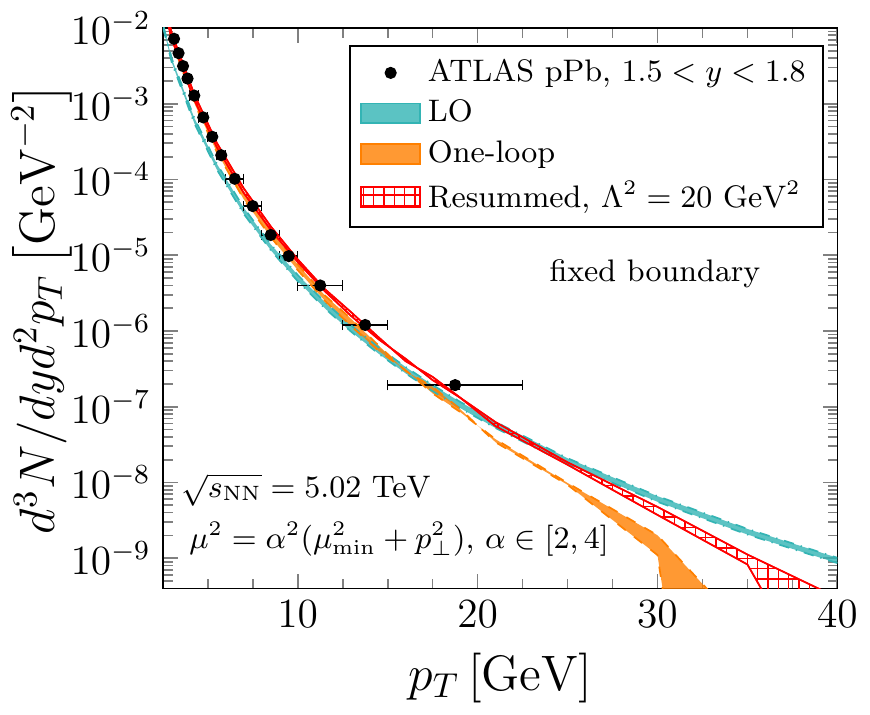}
\end{center}
\caption{Comparisons of the theoretical calculations for hadron yields with two boundary conditions at $\sqrt{s_{\text{NN}}}=5.02$ TeV in $pPb$ collisions\cite{ATLAS:2016xpn}.}
\label{fig:freeze3}
\end{figure}

\begin{figure}[!h]
\begin{center}
\includegraphics[width=0.4\textwidth]{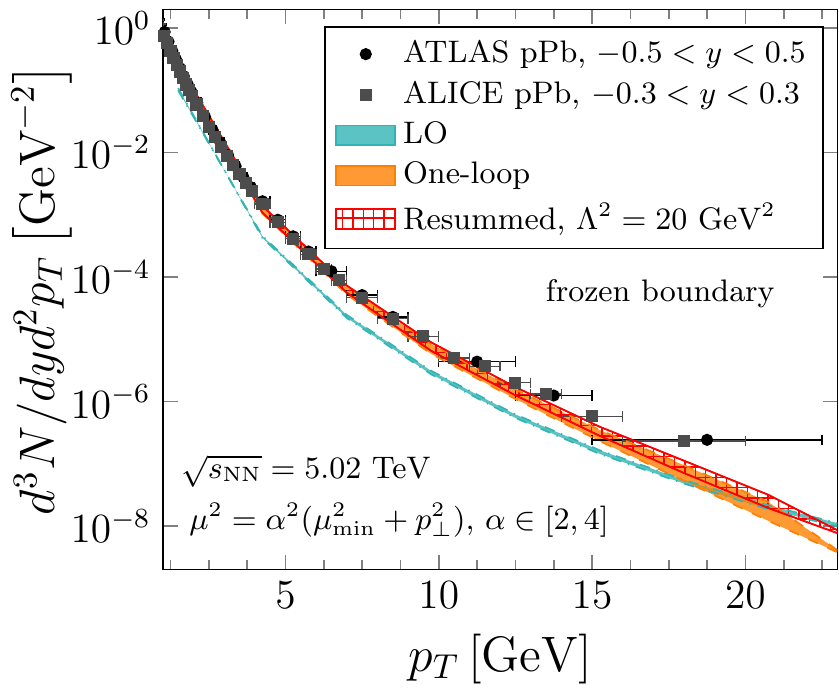}
\includegraphics[width=0.4\textwidth]{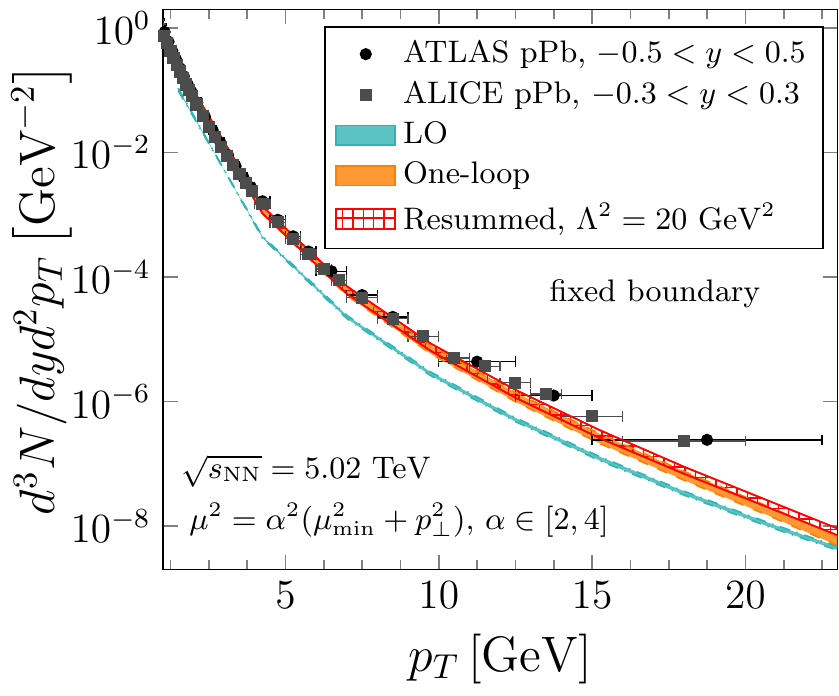}
\end{center}
\caption{Comparisons of the theoretical calculations for hadron yields with two boundary conditions at $\sqrt{s_{\text{NN}}}=5.02$ TeV in $pPb$ collisions\cite{ALICE:2012mj, ATLAS:2016xpn}.}
\label{fig:freeze4}
\end{figure}

This subsection is devoted to the discussion on the applicable region of the CGC calculation in forward hadron productions. More specifically, the dilute-dense factorization used in this work requires that the parton density in the nucleus target is much denser than that in the proton projectile. The deep inelastic scattering experimental data at HERA has revealed an intriguing feature of geometrical scaling at $x_{\text{B}}\equiv Q^2/s<0.01$ \cite{Golec-Biernat:1998zce,Stasto:2000er}, which has been reckoned as a compelling evidence for saturation physics. Therefore, we often set the initial condition for the small-$x$ dipole scattering amplitude at $x_g=0.01$ and evolve it towards lower values of $x_g$. 

For the forward hadron productions in $pA$ collisions, the LO order kinematics give $x_p = \frac{k_\perp}{\sqrt{s}}e^{y}$ and $x_g = \frac{k_\perp}{\sqrt{s}}e^{-y}$. In our NLO calculation and scheme choice, we subtract exactly the amount of the contribution proportional to $\alpha_s \ln 1/x_g$ into the BK equation and resum this logarithm via the numerical solution of the rcBK equation. This indicates that the dipole scattering amplitude $S^{(2)}(r_\perp)$ is evolved to the same value of $x_g$ when they are used as the small-$x$ input in the LO and NLO cross-sections.  

To apply the CGC formalism to the forward hadron production, one needs to require $x_g = x_p e^{-2y}< 0.01$ which implies $y>1/2\ln (100x_p)$. Since $x_p <1$, we find that we can guarantee that $x_g$ is always less than $0.01$ for the entire $k_\perp$ ($p_T$) region if $y\geq \ln10 \simeq 2.3$. Therefore, after implementing the threshold resummation within the NLO CGC calculation, we should expect that the improved NLO calculation should be able produce satisfactory result as compared with the experimental data across the entire $p_T$ regime for hadron produced in the rapidity $y\geq 2.3$. As we can observe from Fig.\ref{fig:tau-plot}, our resummed results describe the data measured at RHIC and the LHC well and there is no issue of negativity across the whole $p_T$ regions.  

If we wish to push the envelope and extend the CGC calculation to the rapidity region $0<y\leq 2.3$ in $pA$ collision, we have to deal with the fact that there is no prescription for the dipole scattering amplitude beyond the boundary when $x_g >0.01$. To estimate the impact of the events in the large $x_g$ region, we adopt two prescriptions as follows.
\begin{itemize}
\item \textbf{Fixed boundary condition}: By adopting this boundary condition, we set the $k_\perp$ dependent gluon distribution to zero when $x_g>0.01$ since this region is beyond the applicable window of the CGC calculation. This prescription is equivalent to removing all the events with $x_g>0.01$ in our calculation. 
\item \textbf{Frozen boundary condition}: In this case, to extend the dipole gluon distribution in the large $x_g$ region, we freeze it at $x_g=0.01$. That is to say, when $x_g >0.01$, the input dipole scattering amplitude simply retains its value at the initial condition at $x_g=0.01$. 
\end{itemize}

In the first prescription, by removing all the events with large $x_g$, we underestimate the gluon distribution in the large-$x$ region. In comparison, with the frozen boundary condition, we overestimate the dipole gluon distribution for the region $x_g >0.01$. The numerical difference of these two prescriptions starts to show up in the large $p_T$ region as we see in Figs.~\ref{fig:freeze1}-\ref{fig:freeze3}, which can be viewed as the sign that the contribution from large $x_g$ regions starts to become important. In the high $p_T$ region of Fig.~\ref{fig:freeze4}, as compared with the LHC data measured in $pPb$ collisions by both the ATLAS and ALICE collaborations in the middle rapidity region, our results of these two prescriptions both start to deviate from the data when $p_T > 15 \text{GeV}$. In this case, we believe that our NLO formalism based on the dilute-dense factorization starts to breakdown, since both the projectile and the target are now rather dilute and the collinear framework certainly becomes the more appropriate theory to describe the high $p_T$ data. Nevertheless, we see that in the low $p_T$ region, these two boundary conditions give consistent descriptions of the data in the low $p_T$ region.

\end{widetext}

\end{document}